\documentclass[times,article]{elsarticle}

\usepackage{framed,multirow}

\usepackage{amssymb}
\usepackage{latexsym}

\usepackage[bookmarks=true,bookmarksnumbered=false,bookmarksopen=false,breaklinks=false,pdfborder={0 0 0},pdfborderstyle={},backref=false,colorlinks=true]{hyperref}
\usepackage{color}
\usepackage{amsfonts}
\usepackage{amsmath}
\usepackage{amssymb}
\usepackage{multirow}
\usepackage{subfig}
\usepackage{graphicx}
\usepackage{rotating}
\usepackage[shortlabels]{enumitem}

\newcommand{\bff}{\mathbf{f}}
\newcommand{\bfg}{\mathbf{g}}
\newcommand{\bfk}{\mathbf{k}}

\newcommand{\bfw}{\mathbf{w}}
\newcommand{\bfx}{\mathbf{x}}
\newcommand{\bfy}{\mathbf{y}}

\newcommand{\bfv}{\mathbf{v}}

\newcommand{\bfF}{\mathbf{F}}
\newcommand{\bfG}{\mathbf{G}}
\newcommand{\bfH}{\mathbf{H}}

\newcommand{\bfK}{\mathbf{K}}

\newcommand{\pd}[2]{\frac{\partial #1}{\partial #2}}

\newcommand{\mm}[1]{\rm mm}

\newcommand{\beq}{\begin{equation}}
\newcommand{\eeq}{\end{equation}}
\newcommand{\bea}{\begin{eqnarray}}
\newcommand{\eea}{\end{eqnarray}}

\newcommand{\bit}{\begin{itemize}}
\newcommand{\eit}{\end{itemize}}
\newcommand{\ben}{\begin{enumerate}}
\newcommand{\een}{\end{enumerate}}

\newcommand{\bF}{\mathbf{F}}

\newcommand{\bK}{\mathbf{K}}
\newcommand{\bk}{\mathbf{k}}

\newcommand{\bM}{\mathbf{M}}
\newcommand{\bP}{\mathbf{P}}

\newcommand{\bU}{\mathbf{U}}
\newcommand{\bV}{\mathbf{V}}

\newcommand{\bw}{\mathbf{w}}

\newcommand{\bg}{\mathbf{g}}
\newcommand{\bh}{\mathbf{h}}

\newcommand{\bx}{\mathbf{x}}

\newcommand{\inv}[1]{\frac{1}{#1}}

\newcommand{\brk}[1]{\left [ #1 \right ]}

\newcommand\scalemath[2]{\scalebox{#1}{\mbox{\ensuremath{\displaystyle #2}}}} 

\usepackage{url}
\usepackage{xcolor}
\definecolor{newcolor}{rgb}{.8,.349,.1}

\journal{Journal of Computational Physics}

\begin{document}
\begin{frontmatter}

\title{A variable high-order shock-capturing finite difference method with GP-WENO}

\author[1]{Adam {Reyes}}
\ead{acreyes@ucsc.edu}
\author[2]{Dongwook {Lee}\corref{cor1}}
\cortext[cor1]{Corresponding author: Tel.: +1-831-502-7708}
\ead{dlee79@ucsc.edu}
\author[4]{Carlo {Graziani}}
\ead{carlo@mcs.anl.gov}
\author[3]{Petros {Tzeferacos}}
\ead{petros.tzeferacos@flash.uchicago.edu}

\address[1]{Department of Physics, The University of California, Santa Cruz, CA, United States}
\address[2]{Department of Applied Mathematics, The University of California, Santa Cruz, CA, United States}
\address[3]{Flash Center for Computational Science, Department of Astronomy \& Astrophysics, The University of Chicago, IL, United States}
\address[4]{Mathematics and Computer Science, Argonne National Laboratory, Argonne, IL, United States}

\begin{abstract}
We present a new finite difference shock-capturing scheme for hyperbolic equations on static uniform grids.
The method provides selectable high-order accuracy by employing a kernel-based Gaussian Process (GP) data
prediction method which is an
extension of the GP high-order method originally introduced in a finite volume framework by the same authors.
The method interpolates Riemann states to high order,
replacing the conventional polynomial interpolations with polynomial-free GP-based interpolations.
For shocks and discontinuities, this GP interpolation scheme
uses a nonlinear shock handling strategy similar to
Weighted Essentially Non-oscillatory (WENO), with a novelty consisting in the fact that
nonlinear smoothness indicators are formulated in terms of
the Gaussian likelihood of the local stencil data, replacing the conventional $L_2$-type smoothness indicators
of the original WENO method.
We demonstrate that these GP-based smoothness indicators play a key role in
the new algorithm, providing significant improvements in delivering
high -- and selectable -- order accuracy in smooth flows, while successfully delivering non-oscillatory solution behavior in discontinuous flows.
\end{abstract}

\end{frontmatter}



\section{Introduction}
\label{sec:introduction}
High-order discrete methods for hyperbolic conservative equations comprise an
important research area in computational fluid dynamics (CFD). The rapid growth in development of 
high-order methods has  been to a great extent driven by a radical change in the balance between
computation and memory resources in modern high-performance computing (HPC) architectures.
To efficiently use computing resources of modern HPC machines CFD algorithms need to
adapt to hardware designs in which memory per compute core has become progressively more
limited \cite{Attig2011,Dongarra2012future,Subcommittee2014top}.
A computationally efficient numerical algorithm should
exploit greater availability of processing
power, while keeping memory use low. 
This goal can be achieved by 
discretizing the CFD governing equations to high order, thus providing
the desired solution accuracy at a higher cost in processor power, but with a smaller memory requirement \cite{Hesthaven2007,LeVeque2002,leveque2007finite}.
Of course, a practical consideration in designing such high-order numerical algorithms is that 
time-to-solution at a given grid resolution should not increase due to
the additional floating point operations.

The most popular approach to designing high-order methods for shock-capturing is
based on implementing highly accurate approximations to partial differential equations (PDEs) 
using piecewise local polynomials. 
By and large, polynomial approaches fall into three
categories: finite difference (FD) methods, 
finite volume (FV) methods, 
and discontinuous Galerkin (DG) methods.
These three formulations all interpolate or reconstruct fluid states using 
Taylor series expansions, with 
accuracy controlled by the number of expansion terms retained
in the interpolation or reconstruction.
Below, we briefly summarize the aspects of those schemes that are most
relevant to this paper.

The DG method, first proposed by Reed and Hill \cite{reed1973triangular} in 1973 for solving
neutron transport problems, approximates a conservation law 
by first multiplying a given PDE by a test function $v$ and then integrating it in each cell
to express the governing dynamics in integral form \cite{shu2009high}.
The method approximates both the numerical solution $u(x,t)$ and the test function $v(x)$
by piecewise polynomials of chosen degree $k$ in each cell. These polynomials are permitted to be
discontinuous at each cell interface, allowing flexibility in achieving high-order accuracy
in smooth regions, while achieving non-oscillatory shock capturing at discontinuities. 
Solutions are typically integrated with a $k$-stage Runge-Kutta (RK) method, 
in which case the scheme is referred to as RKDG. 
The advantages of RKDG are that it is well-adapted to complicated geometries \cite{cockburn2001runge};
it is easy to parallelize due to data locality \cite{beck2014high,atak2015discontinuous};
it lends itself to GPU-friendly computing implementations \cite{klockner2009nodal};
accommodates arbitrary $h-p$ adaptivity \cite{baccouch2015asymptotically,cao2015superconvergence};
it permits designs that preserve given structures in local approximation spaces 
\cite{cockburn2004locally,li2005locally}.
The weaknesses of the method include the fact that
it is significantly more complicated in terms of algorithmic design;
it potentially features less robust solution behaviors at strong shocks and discontinuities \cite{shu2009high};
its stability limit for timestep size becomes progressively more restrictive 
with increasing order of accuracy \cite{cockburn2001runge,zhang2005analysis,liu20082}.
For more discussions and an extended list of references, see also Cockburn and Shu 
\cite{cockburn1998runge,cockburn2001runge}; 
Shu \cite{shu2009high,shu2016high}; Balsara \cite{balsara_higher-order_2017}.
As an alternative to RKDG, another line of DG development called ADER-DG was
studied and first introduced by Dumbser
and Munz in 2005 \cite{dumbser2005ader,dumbser2005arbitrary}. 
The main advantage of ADER-DG over RKDG is its computational
efficiency achieved by
using a one-step time integration method of the ADER (Arbitrary DERivative in space and time) 
approach \cite{toro2001towards} instead of multi-stage RK methods. 
In \cite{dumbser2005ader,dumbser2005arbitrary},
ADER-DG was used to solve linear hyperbolic systems with constant
coefficients and linear systems with variable coefficients in conservative form.
Extensions and improvements of the ADER-DG approach over the last decade have been reported
in \cite{dumbser2006building,kaser2006arbitrary,dumbser2006arbitrary,
kaser2007arbitrary,de2007arbitrary,dumbser2007arbitrary,zanotti2015solving,castro2007high}.

The finite volume method (FVM) also uses the
governing equations in integral form, making use of volume-averaged 
conservative variables.
The discrete formulation of FVM provides a natural way of maintaining 
conservation laws \cite{toro2013riemann}. 
This inherently conservative property of the scheme makes FVM a very popular algorithmic
choice for application problems where shocks and discontinuities are dominant.
Historically, work on high-order FVM methods began with the
seminal works
by van Leer \cite{van1974towards, van1977towards, vanleer1979}, 
which overcame the Godunov Barrier theorem \cite{godunov1959difference} 
to produce effective second-order accurate methods.
More advanced numerical methods with higher solution accuracies than second-order
became available soon thereafter,
including the piecewise parabolic method (PPM) \cite {colella1984piecewise},
the essentially non-oscillatory (ENO) schemes \cite{harten1987uniformly,shu_efficient_1988},
and the weighted ENO (WENO) schemes
\footnote{Although the original WENO-JS scheme introduced in \cite{jiang1996efficient} is in FDM,
the key ideas of WENO-JS have been reformulated and studied in FVM by numerous practitioners.} 
\cite{liu1994weighted,jiang1996efficient} which improved ENO by means of nonlinear weights.
Most of these early schemes focused on obtaining high-order accuracy in 1D, and
their naive extension to multidimensional problems 
using a dimension-by-dimension approach \cite{buchmuller2014improved}
resulted in a second-order solution accuracy bottleneck
due to
inaccuracies in both space and time associated with computing a face-averaged flux function \cite{shu2009high,buchmuller2014improved,zhang2011order,mccorquodale2011high}.
Zhang \textit{et al.} \cite{zhang2011order} studied the effect of second-order versus 
high-order quadrature approximations combined with the 1D 5th order accurate WENO 
reconstruction as a baseline spatial formulation.
They demonstrated that  for nonlinear test problems in 2D,
the simple and popular dimension-by-dimension approach only converges to
second-order, despite the fact that the order of accuracy of the baseline
1D algorithm (e.g., 5th order accuracy for WENO) does in fact carry over to linear 2D problems.
See also a recent work on the piecewise cubic method (PCM) proposed by 
Lee \textit{et al.} \cite{lee2017piecewise}.
The spatial aspect of the problem is addressable by using multiple quadrature points per cell-face \cite{shu2009high,titarev_finite-volume_2004} or
quadrature-free flux integration through freezing the Riemann fan
along the cell interfaces \cite{dumbser2007quadrature}, while
high-order temporal accuracy is achievable by using multi-stage Runge-Kutta (RK) methods 
\cite{zhang2011order,buchmuller2014improved,mccorquodale2011high}
or single-step ADER-type formulations 
\cite{titarev2002ader,toro2006derivative,Titarev2005, Dumbser2008,balsara2009efficient,dumbser2013ader,balsara2013efficient,qian_generalized_2014}.

In the conservative FDM approach, one evolves
pointwise quantities, thus avoiding
the complexities of FVM associated with the need for dealing with volume-averaged quantities.
For this reason, FDM has also been a popular choice for obtaining
high-order solution accuracy in hyperbolic PDE solution, particularly when some of its well-known
difficulties such with geometry, AMR, and free-streaming preservation
are not considerations. See, e.g., 
\cite{jiang1996efficient,del_zanna_echo:_2007,mignone_high-order_2010,jiang_alternative_2013,chen_5th-order_2016,zhu2016new}.
Traditional FDM formulations directly \textit{reconstruct} high order numerical fluxes from cell-centered
point values \cite{shu1989,jiang1996efficient,mignone_high-order_2010}. In this approach, the
pointwise flux $F$ is assumed to be the volume average of the targeted high-order numerical flux $\hat{F}$,
thereby recovering the same reconstruction procedure of FVM (i.e., reconstructing
point values of Riemann states at each interface, given the volume-averaged quantities at cell centers) for
its flux approximation.
To ensure stability, upwinding is enforced through a splitting of the
FD fluxes into left and right moving fluxes. The most commonly used
flux splitting is the global Lax-Friedrichs splitting
\cite{shu2009high}, which, while able to maintain a high formal order
of accuracy in smooth flows, is known to be quite diffusive
\cite{chen_5th-order_2016}. 
Alternatively, an improvement can be achieved with 
the use of local characteristic field decompositions in the flux splitting in part of the flux reconstruction
\cite{mignone_high-order_2010}, in which the characteristic field calculation is heavily dependent on
a system of equations under consideration and adds an extra
computational expense.
Recently, some practitioners including Del Zanna \textit{et al.} and Chen \textit{et al.} 
studied another form of high-order FD flux formulation, called FD-Primitive
\cite{del_zanna_echo:_2007,chen_5th-order_2016} (referred to as FD-Prim hereafter), 
in which, instead of designing high-order numerical fluxes directly from cell pointwise values,
high-order fluxes are constructed first by solving Riemann problems, followed by a high-order
correction step where the
correction terms are derived from the so-called compact-like
finite difference schemes \cite{lele1992compact,deng2000developing,zhang2008development}.
In this way, FD-Prim is algorithmically quite analogous to FVM, in that it first interpolates (rather than reconstructs) 
high-order Riemann states at each cell interface, then uses them to calculate interface fluxes by solving 
Riemann problems using either exact or approximate solvers, and finally makes corrections
to the fluxes to deliver high-order-accurate FD numerical fluxes \cite{del_zanna_echo:_2007,chen_5th-order_2016}.
In this way the FD-Prim approach allows the added flexibility of choosing a Riemann solver
(e.g., exact \cite{saurel1994exact,delmont2009exact,takahashi2014exact,toro2013riemann},
HLL-types \cite{harten1983upstream,toro1994restoration,miyoshi2005multi,guo2016hlld}, or Roe
\cite{roe1981approximate}, etc.)  in a manner analogous to the FVM approach.

The aforementioned traditional polynomial-based high-order methods are complemented by 
a family of  ``non-polynomial'' (or ``polynomial-free'') 
methods called radial basis function (RBF) approximation methods. 
As a family of ``mesh-free'' or ``meshless'' method, RBF approximation methods
have been extensively studied (see \cite{buhmann2000radial}) to
provide more flexible approximations, in terms of approximating functions \cite{powell1985radial} 
as well as scattered data \cite{Franke1982}.
Unlike local polynomial methods, RBF has degrees of freedom that
disassociate the tight coupling between the stencil configuration 
and the local interpolating (or reconstructing) polynomials under consideration.
For this reason, interest has grown in meshfree methods based on RBF 
as means for designing numerical methods that achieve
high order convergence while retaining a simple (and flexible) algorithmic framework
in all spatial dimensions \cite{sarra2009multiquadric,safdari2018radial}.
Approximations based on RBF have been used
to solve hyperbolic PDEs 
\cite{Katz2009,morton2007,sonar1996,guo2017rbf,guo_non-polynomial_2016,bigoni_adaptive_2017},
parabolic PDEs \cite{moroney2006,moroney2007},
diffusion and reaction-diffusion PDEs \cite{shankar2015radial}, and
boundary value problems of elliptic PDEs \cite{liu2015kansa}, as well as for
interpolations on irregular domains \cite{chen2016reduced,heryudono2010radial,martel2016stability}, and for
interpolations on more general sets of scattered data \cite{Franke1982}.

In this paper, we develop a new high-order FDM in which the core interpolation formulation
is based on Gaussian Process (GP) Modeling \cite{bishop2007pattern,rasmussen2005,wahba1995}. This work
is an extension of our previous GP high-order method \cite{reyes_new_2016} introduced in a FVM framework.
Analogous to RBF, our GP approach is a non-polynomial method.
By being a meshless method, the proposed GP method features attractive properties similar to those of
RBF and
allows flexibility in code implementation and selectable orders of
solution accuracy in any number of spatial dimensions. An important feature of the GP
approach is that it comes equipped with a likelihood estimator for a
given dataset, which we have leveraged to form a new set of smoothness
indicators. Based on the probabilistic theory of 
Gaussian Process Modeling \cite{bishop2007pattern,rasmussen2005},
the new formulation of our smoothness indicators is
a key component of our GP scheme that we
present in Section~\ref{sec:handl-disc-1}. 
We call our new GP scheme with the GP-based smoothness indicators GP-WENO in this paper.
As demonstrated below in our numerical convergence study, the GP-WENO's formal accuracy
is $\mathcal{O}(\Delta^{2R+1})$ 
and is controlled by the parameter $R$, called the GP radius,
that determines the size of a GP stencil on which GP-WENO
interpolations take place. These numerical experiments also show that
the new GP-based smoothness indicators are better able to preserve
high order solutions in the presence of discontinuities than are
the conventional WENO smoothness indicators based on the $L_2$-like norm
of the local polynomials \cite{jiang1996efficient}.


\section{Finite Difference Method}
\label{sec:finite-diff-meth}

We are concerned with the solution of 3D conservation laws:
\begin{equation}
  \label{eq:cons_law}
  \pd{\bU}{t} + \pd{\bfF(\bU)}{x} + \pd{\bfG(\bU)}{y} + \pd{\bfH(\bU)}{z} = 0,
\end{equation}
where $\bU$ is a vector of conserved variables and $\bfF$, $\bfG$ and
$\bfH$ are the fluxes. For the Euler equations these are defined as
\begin{equation}
  \label{eq:euler}
  \bU =
  \begin{bmatrix}
    \rho \\
    \rho u \\
    \rho v \\
    \rho w \\
    E
  \end{bmatrix}, \;\;\;
  \bfF(\bU) =
  \begin{bmatrix}
    \rho u \\
    \rho u^2 + p \\
    \rho u v \\
    \rho u w \\
    u (E+p)
  \end{bmatrix}, \;\;\;
  \bfG(\bU) =
  \begin{bmatrix}
    \rho u \\
    \rho uv \\
    \rho v^2+p \\
    \rho vw \\
    v (E+p)
  \end{bmatrix}, \;\;\; 
  \bfH(\bU) =
  \begin{bmatrix}
    \rho u \\
    \rho uw \\
    \rho vw \\
    \rho w^2+p \\
    w (E+p)
  \end{bmatrix}.
\end{equation}

We wish to produce a conservative discretization of the pointwise
values of $\bU$, i.e., $\bU_{ijk} = \bU(\bfx_{ijk})$, and we write
Eq.~(\ref{eq:cons_law}) in the form:
\begin{equation}
  \label{eq:disc_fd}
  \pd{\bU_{ijk}}{t} = -\inv{\Delta x}(\hat{\bff}_{i+1/2,j,k}-\hat{\bff}_{i-1/2,j,k}) 
                         - \inv{\Delta y}(\hat{\bfg}_{i,j+1/2,k}-\hat{\bfg}_{i,j-1/2,k})
                         - \inv{\Delta z}(\hat{\bh}_{i,j,k+1/2}-\hat{\bh}_{i,j,k-1/2}).
\end{equation}
Here $\hat{\bff}_{i\pm 1/2,j,k}$, $\hat{\bfg}_{i,j\pm 1/2,k}$ and $\hat{\bh}_{i,j,k\pm
  1/2}$ are the $x$, $y$ and $z$ numerical fluxes evaluated at the
halfway point between cells in their respective directions. 
 The  numerical fluxes are defined so that their divided difference rule approximates
 the exact flux derivative with $q$-th order:
 \begin{equation}
   \label{eq:num_flux}
   \left .\pd{\bfF}{x} \right |_{\bfx=\bfx_{ijk}} = \inv{\Delta
     x}(\hat{\bff}_{i+1/2,j,k}-\hat{\bff}_{i-1/2,j,k}) +
   \mathcal{O}(\Delta x^q),
 \end{equation}
 and similarly for the $y$ and $z$ fluxes. 
 As a result, the overall finite difference scheme in Eq. (\ref{eq:disc_fd}) approximates the original
 conservation law in Eq. (\ref{eq:euler}) with the spatial accuracy of order $q$.
 The temporal part of Eq.~(\ref{eq:disc_fd}) can be discretized by a
 method-of-lines approach with a Runge-Kutta time discretization \cite{gottlieb_strong_2011}.

To determine the numerical fluxes $\hat{\bff}$ in Eq. (\ref{eq:disc_fd}), 
let us consider first the
pointwise $x$-flux, $\bfF_{ijk}$, as the 1D cell average of an
auxiliary function, $\hat{\bfF}$, in the $x$-direction. 
If we also define another function, $\bP(x)=\int_{-\infty}^x \hat{\bfF}(\xi)
d\xi$, we can write $\bfF_{ijk}$ as
\begin{equation}
  \label{eq:prim_F}
  \bfF_{ijk} = \inv{\Delta
    x}\int_{x_{i-1/2}}^{x_{i+1/2}}\hat{\bF}(\xi)d\xi = \inv{\Delta x}
  (\bP(x_{i+1/2}) - \bP(x_{i-1/2})).
\end{equation}
Differentiating Eq.~(\ref{eq:prim_F}) with respect to $x$ gives
\begin{equation}
  \label{eq:exct_num_flux}
  \left .\pd{\bfF}{x} \right |_{\bfx=\bfx_{ijk}} = \inv{\Delta
     x}(\hat{\bfF}_{i+1/2,j,k}-\hat{\bfF}_{i-1/2,j,k}),
\end{equation}
and comparing with Eq.~(\ref{eq:num_flux}) we can identify $\hat{\bfF}$ as the analytic flux function we wish to
approximate with the numerical flux $\hat{\bff}$.
This can be repeated in a similar fashion for the $y$ and $z$ fluxes, $\hat{\bfG}$ and $\hat{\bfH}$. 
The goal is
then to form a high order approximation to the integrand quantities $\hat{\bfF}$,
$\hat{\bfG}$ and $\hat{\bfH}$, 
knowing the mathematically cell-averaged integral quantities and physically pointwise fluxes $\bfF_{ijk}$,
up to some design accuracy of order $q+1$, e.g.,
\begin{equation}
  \label{eq:design}
  \hat{\bff}_{i+1/2,j,k} =
\hat{\bfF}_{i+1/2,j,k} + \mathcal{O}(\Delta x^{q+1}).
\end{equation}
Note that this is exactly the same reconstruction
procedure of computing high-order accurate Riemann states at cell interfaces
given the integral volume-averaged quantities at cell centers in 1D FVM.

In the high order finite difference method originally put forward by
Shu and Osher \cite{shu1989}, the problem of approximating
the numerical fluxes is accomplished by directly
\textit{reconstructing} the face-centered numerical fluxes from the
cell-centered fluxes on a stencil that extends from the points
$\bfx_{i-l,j,k}$ to $\bfx_{i+r,j,k}$. That is, $\hat{\bff}_{i+1/2,j,k} =
\bf{\mathcal{R}}(\bfF_{i-l,j,k},...,\bfF_{i+r,j,k})$ in complete analogy
to reconstruction in the context of finite volume schemes, where
$\bf{\mathcal{R}}(\cdot)$ is a high order accurate procedure to reconstruct
face-centered values from cell-averaged ones. Such flux-based finite
difference methods (or FD-Flux in short), as just described, are easily implemented using
the same reconstruction procedures as in 1D finite volume codes and
provide high order of convergence on multidimensional
problems. For this reason, they have been widely
adopted \cite{mignone_high-order_2010,shu_high-order_2003,jiang1996efficient}. One
pitfall of this approach is that proper upwinding is typically achieved
by appropriately splitting the fluxes into parts moving towards and away from the
interface of interest using the global Lax-Friedrichs splitting
\cite{mignone_high-order_2010,shu_high-order_2003} at the cost of
introducing significant diffusion to the scheme.

On the other hand, it can be readily seen from Eq.~(\ref{eq:prim_F})
that the naive use of the interface value of the flux $\bfF_{i+1/2,j,k}$ as the
numerical flux can provide at most a second order approximation 
and should be avoided for designing a high-order FDM, 
no matter how accurately $\bfF_{i+1/2,j,k}$ is computed, since
\begin{equation}
  \label{eq:flux_2nd}
  \bfF_{i+1/2,j,k} = \inv{\Delta x} \int_{x_{i}}^{x_{i+1}} \hat{\bfF}(\xi)d\xi = \hat{\bfF}_{i+1/2,j,k} + \mathcal{O}(\Delta x^2).
\end{equation}

Alternatively, in an approach originally proposed by Del Zanna
\cite{del_zanna_echo:_2007,del_zanna_efficient_2003}, upwinding is
provided by solving a  Riemann problem
at each of the face centers, $\bfx_{i+1/2,j,k}$,
$\bfx_{i,j+1/2,k}$ and $\bfx_{i,j,k+1/2}$ for the corresponding
face-normal fluxes, $\bfF_{i+1/2,j,k}$, $\bfG_{i,j+1/2,k}$ and
$\bfH_{i,j,k+1/2}$. The numerical flux is then viewed as being the
face-center flux from the Riemann problem, i.e., at the cell interface
$\bfx_{i+1/2,j,k}$,
\begin{equation}
\label{eq:riemann_prob}
\bfF_{i+1/2,j,k} = \mathcal{RP}(\bU^L_{i+1/2,j,k},\bU^R_{i+1/2,j,k}),
\end{equation}
plus a series of high order corrections using the
spatial derivatives of the flux evaluated at the face-center,
\begin{align}
  \label{eq:numFlux_cor}
  \begin{split}
    \hat{\bff}_{i+1/2,j,k}&= \bfF_{i+1/2,j,k} + {\alpha}{\Delta x^2} \bfF^{(2)}_{i+1/2,j,k}
                                                                +{\beta}{\Delta x^4}\bfF^{(4)}_{i+1/2,j,k}+ \dots, \\
    \hat{\bg}_{i,j+1/2,k}&= \bfG_{i,j+1/2,k} + {\alpha}{\Delta y^2} \bfG^{(2)}_{i,j+1/2,k}
                                                                +{\beta}{\Delta y^4}\bfG^{(4)}_{i,j+1/2,k}+ \dots, \\
    \hat{\bh}_{i,j,k+1/2}&= \bfH_{i,j,k+1/2} + {\alpha}{\Delta z^2} \bfH^{(2)}_{i,j,k+1/2}
                                                                +{\beta}{\Delta z^4}\bfH^{(4)}_{i,j,k+1/2}+ \dots,
  \end{split}
\end{align}
where parenthesized superscripts denote numerical derivatives in the corresponding dimension, and where $\alpha$ and $\beta$ are constants chosen so
Eq.~(\ref{eq:exct_num_flux}) holds up to the desired order of accuracy, e.g., 

\begin{equation}
  \label{eq:numF_cond}
  \left .\pd{\bfF}{x} \right |_{\bfx=\bfx_{ijk}} = \inv{\Delta
    x}(\hat{\bff}_{i+1/2,j,k}-\hat{\bff}_{i-1/2,j,k}) +\mathcal{O}(\Delta x^q).
\end{equation}
For the choice $q=5$ only the terms up to the fourth derivative in
Eq.~(\ref{eq:numFlux_cor}) need to be retained. The constants $\alpha$
and $\beta$ are determined by Taylor expanding the terms in
(\ref{eq:numFlux_cor}) and enforcing the condition in
(\ref{eq:numF_cond}). For this reason, the values of $\alpha$ and $\beta$ depend on
the stencil geometry used to approximate the derivatives. Del Zanna
\cite{del_zanna_echo:_2007} used the Riemann fluxes at neighboring
face-centers to calculate the derivatives and found $\alpha=-1/24$ and
$\beta = 3/640$. A disadvantage of this choice is that it requires
additional guard cells on which to solve the Riemann problem in order
to compute the high order correction terms near the domain boundaries. 
Chen \textit{et al.} \cite{chen_5th-order_2016} converted the cell-centered
conservative variables to cell-centered fluxes around the face-center
of interest to compute the flux derivatives. This leads to
$\alpha=-1/24$ and $\beta=1/480$. While this approach doesn't require
additional guard cells, the conversion from conservative (or
primitive) variables to flux variables needs to be performed at each grid point in the
domain, incurring additional computational cost. For this reason, we
adopt the face-center flux approach of Del Zanna. For example, the second and
fourth derivatives of the $x$-flux are then given by the finite difference formulas,
\begin{align}
  \label{eq:FD-ders}
  \begin{split}
    \bfF^{(2)}_{i+1/2,j,k} &= \frac{1}{\Delta x^2}\left(\bfF_{i-1/2,j,k} - 2\bfF_{i+1/2,j,k} + \bfF_{i+3/2,j,k}\right), \\
    \bfF^{(4)}_{i+1/2,j,k} &= \frac{1}{\Delta x^4}\left(\bfF_{i-3/2,j,k} - 4\bfF_{i-1/2,j,k} + 6\bfF_{i+1/2,j,k} -
                                                                                                        4\bfF_{i+3/2,j,k} + \bfF_{i+5/2,j,k} \right).
  \end{split}
\end{align}
The derivatives of the $y$-fluxes and $z$-fluxes are given in the same way.
These correction terms were originally derived in the context of 
compact finite difference interpolation \cite{lele1992compact,deng2000developing,zhang2008development}.
For instance,
the explicit formula in \cite{lele1992compact}
for the first derivative approximation $\left .\pd{\bfF}{x} \right |_{\bfx=\bfx_{ijk}}$ 
using six neighboring interface fluxes $\bfF_{i-5/2,j,k}, \cdots, \bfF_{i+5/2,j,k}$ 
reduces to the high order correction formula in Eqs.~(\ref{eq:numFlux_cor}) -- (\ref{eq:FD-ders})
(see also the Appendix in \cite{del_zanna_echo:_2007}).

In summary, the finite difference method described in this paper consists of the following steps:
\begin{enumerate}
\item \textit{Pointwise} values of either the primitive or
  conservative variables $\bU^n_{ijk}$ are given on a uniform grid at time $t^n$.
\item The Riemann states $\bU^L$ and $\bU^R$ as given in Eq.~(\ref{eq:riemann_prob}) 
  at the face-centers between grid points,
  $\bU_{i\pm 1/2,j,k}$, $\bU_{i,j\pm 1/2,k}$ and $\bU_{i,j,k\pm 1/2}$
   are interpolated from pointwise cell-centered values. These
  interpolations should be carried out in a non-oscillatory way with the desired spatial accuracy.
  A stepwise description of the interpolation procedure is given in Section~\ref{sec:gp_weno_stepwise}.
\item The face-center normal fluxes are calculated from the Riemann
  problem in Eq.~(\ref{eq:riemann_prob}) at the halfway points between grid points.
\item The second and fourth derivatives of the fluxes are calculated by following Eq.~(\ref{eq:FD-ders})
  using the Riemann fluxes from the previous step. In principle these
  should be carried out in a non-oscillatory fashion using a nonlinear slope limiter as was done in
  \cite{del_zanna_echo:_2007}. However, it was pointed out in
  \cite{chen_5th-order_2016} that the flux-derivatives' contribution
  to the numerical fluxes in Eq.~(\ref{eq:numFlux_cor}) is relatively
  small and will produce only small oscillations near
  discontinuities. 
  For this reason, the derivatives are finite differenced without any limiters in our implementation.
  The numerical fluxes are constructed as in
  Eq.~(\ref{eq:numFlux_cor}).
\item The conservative variables can then be updated $t^n\rightarrow
  t^{n+1}$ using a standard SSP-RK method
  \cite{gottlieb_strong_2011}. 
\end{enumerate}

So far, we have not yet described what type of spatial interpolation method is to be
used in Step 2 to compute high-order Riemann states at each cell interface. Typically,
non-oscillatory high-order accurate local polynomial schemes are adopted such as
MP5 \cite{suresh_accurate_1997} in \cite{chen_5th-order_2016} or 
WENO-JS \cite{jiang1996efficient} in  \cite{del_zanna_echo:_2007}.
In the next section, we will introduce our high-order polynomial-free interpolation scheme,
based on Gaussian Process Modeling.


\section{Gaussian Process Modeling}
\label{sec:gp-weno-interp}
In this section, we briefly outline the statistical theory underlying the
construction of GP-based Bayesian prior and posterior distributions (see Section~\ref{sec:gp-interpolation}). 
Interested readers are encouraged to refer to our previous paper
\cite{reyes_new_2016} for a more detailed discussion
in the context of applying GP for the purpose of achieving high-order algorithms for FVM schemes.
For a more general discussion of GP theory see \cite{rasmussen2005}.

\subsection{GP Interpolation}
\label{sec:gp-interpolation}
GP is a class of stochastic processes, i.e., processes that sample
functions (rather than points) from an infinite-dimensional function
space. The distribution over the space of functions $f$
is specified by the prior mean and covariance functions, which give rise to the GP \textit{prior}:
\begin{itemize}
  \item a mean function $\bar{f}(\mathbf{x}) = \mathbb{E}[f(\bfx)]$ 
over $\mathbb{R}^N$, and 
\item a covariance GP kernel function which is a symmetric and positive-definite 
integral kernel over $\mathbb{R}^{N}\times\mathbb{R}^{N}$ given by $K(\mathbf{x},\mathbf{y}) =
\mathbb{E}\left[
\left(f(\mathbf{x})-\bar{f}(\mathbf{x})\right)
\left(f(\mathbf{y})-\bar{f}(\mathbf{y})\right)
\right]$. 
\end{itemize}
The GP approach to interpolation is to regard the values of the
function $f$ at a series of points $\bx \in \mathbb{R}^{N}$ 
as samples from a function that is only known probabilistically in terms of the prior GP 
distribution, and to form a \textit{posterior}
distribution on $f(\bx)$ that is \textit{conditioned} on the observed values. One frequently refers to
the observed values as \textit{training points}, and to the passage from prior distribution to 
posterior predictive distribution as a training process.
We may use the trained distribution to predict the behavior of functions $f$
at a new point $\bx_*$. 
For example, in the current context, the fluid variables $U(\bfx)$
are assumed to be sample functions from GP distributions with prescribed mean and covariance functions, 
written as $U(\bfx)\sim
\mathcal{GP}(\bar{f}(\bfx),K(\bfx,\bfy))$. We then train the GP on the known values of
the fluid variables at the cell centers, $U_{ijk}$, to predict the
Riemann states at cell-face centers, e.g., $U_{i\pm 1/2,j,k}$ with $\bx_*=\bx_{i\pm1/2,j,k}$.

The mean function is often
taken to be constant, $\bar{f}(\bfx) = f_0$. We have found a choice of zero
mean, $f_0 = 0$ works well, and we adopt this choice in this paper. For the
kernel function, $K(\bfx, \bfy)$, we will use the ``Squared
Exponential'' (SE) kernel,
\begin{equation}
  \label{eq:se}
  K(\bfx, \bfy) = K_{\text{SE}}(\bfx,\bfy) = \Sigma^2\exp \brk{-\frac{(\bfx-\bfy)^2}{2\ell^2}}.
\end{equation}
For other choices of kernel functions and the related discussion in the context of
designing high-order approximations for numerical PDEs, 
readers are referred to \cite{reyes_new_2016}.

The SE kernel has two free parameters, $\Sigma$ and $\ell$, called
\textit{hyperparameters}. We will see below that $\Sigma$ plays no role in the calculations that are
presented here, and may as well be chosen to be $\Sigma=1$. However, $\ell$ is a
length scale that controls the characteristic scale on which the
GP sample functions vary. As was demonstrated in
\cite{reyes_new_2016}, $\ell$ plays a critical role in the solution
accuracy of a GP interpolation/reconstruction scheme and ideally should match the length
scales of the problem that are to be resolved.

Formally, a GP is a collection of random variables, any finite
collection of which has a joint Gaussian distribution
\cite{bishop2007pattern,rasmussen2005}. We consider the
function values $f(\bfx_i)$ at points $\bfx_i$, $i=1,\dots,N$, as our
$N$ ``training'' points. Introducing the data vector $\bff$ with components
$[\bff]_i=f(\bfx_i)$, the likelihood, $\mathcal{L}$, of $\bff$ given
a GP model (i.e., $f\sim \mathcal{GP}(\bar{f},K)$) is given by
\begin{equation}
  \label{eq:likely}
  \mathcal{L} =
  (2\pi)^{-\frac{N}{2}}|\det\bfK|^{-\inv{2}}\exp \brk{-\inv{2}(\bff-\bar{\bff})^T\bfK^{-1}(\bff-\bar{\bff}) },
\end{equation}
where $[\bar{\bff}]_i = \bar{f}(\bfx_i)$ and $[\mathbf{K}]_{ij}\equiv
K(\mathbf{x}_{i},\mathbf{x}_{j})$. The likelihood $\mathcal{L}$ is a measure
of how compatible the data $\bff$ is with the GP model specified by
the mean $\bar{f}(\bfx)$ and the covariance $K(\bfx,\bfy)$.

Given the function value samples $\bff$, the GP theory furnishes the
posterior predictive distribution over the value $f_*=f(\bfx_*)$ of the unknown function
$f\sim \mathcal{GP}(\bar{f},K)$ at any new point $\bfx_*$. 
The mean of this distribution is the \textit{posterior mean function},
\begin{equation}
  \label{eq:post_mean}
  \tilde{f}_* = \bar{f}(\bfx_*) + \bfk_*^T\bfK^{-1}(\bff-\bar{\bff}),
\end{equation}
where $[\bfk_{*}]_i = K(\bfx_*,\bfx_i)$. Taking a zero mean GP, $\bar{f}(\bfx) =\bff= 0$, 
Eq.~(\ref{eq:post_mean}) reduces to
\begin{equation}
  \label{eq:zero-mean}
  \tilde{f}_* = \bfk_*^T\bfK^{-1}\bff.
\end{equation}
According to Eqs.~(\ref{eq:post_mean}) and (\ref{eq:zero-mean}), the GP posterior mean is a linear
function of $\bff$, with a weight vector $\bfk_*^T\bfK^{-1}$
specified entirely by the choice of covariance kernel
function, the stencil points, and the prediction point. We take this posterior mean $\tilde{f}_*$ of the
distribution in Eq.~(\ref{eq:zero-mean}) as the interpolation $f_*$ of the
function $f$ at the point $\bfx_*\in\mathbb{R}^\text{D}$,
$\text{D}=1,2,3$, where $f$ is any one of the fluid variables in
primitive, conservative or characteristic form, which we will denote
as $q$. Note that had we retained the multiplicative scale factor $\Sigma$ 
as a model hyperparameter, it would have canceled
out in Eq.~(\ref{eq:zero-mean}). This justifies our choice of $\Sigma=1$.

\subsection{GP Interpolation for FD-Prim}
\label{sec:gp-interpolation-cfd}
Hereafter, we restrict ourselves to describe our new multidimensional GP high-order interpolation scheme
in the framework of FD-Prim, which only requires us to consider 1D data interpolations as typically
done in the dimension-by-dimension approach for FDM.
The notation and the relevant discussion will therefore be formulated in 1D.

We wish to interpolate the Riemann states
$\bU_{i\pm 1/2}$ from the pointwise cell centered values $\bU_i$. We
consider a (fluid) variable $q$ on a 1D stencil of  $R$ points on either side of the
central point $x_i$ of the $i$-th cell $I_i=[x_{i-1/2},x_{i+1/2}]$, and write
\begin{equation}
  \label{eq:2R1-stencil}
  S_R = \bigcup_{k=i-R}^{k=i+R} I_k.
\end{equation}
We seek
a high-order interpolation of $q$ at $\bfx_*=x_*=x_{i\pm1/2}$,
\begin{equation}
  \label{eq:intrp}
  q_{i\pm 1/2} = \mathcal{I}_{\text{GP}} (q_{i-R},\dots,q_{i+R}),
\end{equation}
where $\mathcal{I}_{\text{GP}}(\cdot)$ is the GP interpolation given
in Eq.~(\ref{eq:zero-mean}). We define the data vector on $S_R$ by
\begin{equation}
  \label{eq:f-vec}
  \bff = [q_{i-R},\dots,q_{i+R}]^T,
\end{equation}
and we define a vector of
weights $\bfw^T_* = \bfw^T_{i\pm 1/2} \equiv \bfk_{i\pm 1/2}^T\bfK^{-1}$, so that the interpolation
in Eq.~(\ref{eq:zero-mean})
can be cast as a product between the (row) vector of weights $\bfw^T_*$ and the
data $\bff$,
\begin{equation}
  \label{eq:q-intrp}
  q_{i\pm 1/2} = \bfw^T_{i\pm 1/2} \bff.
\end{equation}
Note here that $\bfK$ is a covariance kernel matrix of size $(2R+1)\times(2R+1)$ whose
entries are defined by
\begin{equation}
\label{eq:kmat}
[\bfK]_{jk}=K(x_j,x_k)=\exp\left[-\frac{(x_j - x_k)^2}{2\ell^2}\right], \;\;\; i-R \le j, k \le i+R,
\end{equation}
and
$\bfk_*\equiv\bfk_{i\pm1/2}$ is a vector of length $(2R+1)$
with entries are defined by
\begin{equation}
\label{eq:kstar}
[\bfk_{*}]_{k}=K(x_*,x_k)=\exp\left[-\frac{(x_* - x_k )^2}{2\ell^2}\right], \;\;\; i-R \le k \le i+R.
\end{equation}

The weights $\bfw_*$ are independent of the
data $\bff$ and depend only on the locations of the data points $x_j, x_k$,
and the interpolation point $x_*$. Therefore, for cases where the grid
configurations are known in advance, the weights can be computed and
stored a priori for use during the simulation.


\subsection{Handling Discontinuities: GP-WENO}
\label{sec:handl-disc-1}
The above GP interpolation procedure works well for smooth flows
without any additional modifications. For non-smooth flows, however, it requires
some form of limiting to avoid numerical oscillations at discontinuities
that can lead to numerical instability. To this end, we adopt
the approach of the Weighted Essentially Non-Oscillatory (WENO)
methods \cite{jiang1996efficient}, where the effective stencil size
is adaptively changed to avoid interpolating through a discontinuity,
while retaining high-order convergence in smooth regions. 
In the work by Del Zanna \textit{et al.}
\cite{del_zanna_echo:_2007}, 
a high-order Riemann state is constructed by considering the conventional
WENO's weighted combination of interpolations  
from a set of candidate
sub-stencils. The weights are chosen based on $L_2$-norms of the derivatives of polynomial reconstructions 
on each local stencil (e.g., $S_m$, see below) 
in such a way that a weight is close to zero when the
corresponding stencil contains a discontinuity, while weights are \textit{optimal} in
smooth regions in the sense that they reduce to the interpolation over
a global\footnote{The term \textit{global} here is to be understood
in a sense that the desired order of accuracy, e.g., 5th-order in WENO-JS, 
is to be optimally achieved in this ``larger'' or ``global'' stencil, rather than the global entire
computational domain.} stencil (e.g., $S_R$, see below).

For the proposed GP-WENO scheme, we
 introduce a new GP smoothness indicator inspired by the probabilistic
 interpretation of GP, replacing
the standard $L_2$-norm-based formulations of WENO.
The GP-WENO scheme will be fully specified by combining
the linear GP interpolation in Section~\ref{sec:gp-interpolation-cfd} 
and the nonlinear GP smoothness local indicators in this section.

We begin with considering the global stencil, $S_R$,
in Eq.~(\ref{eq:2R1-stencil}) with $2R+1$ points centered at the cell
$I_i$ and the $R+1$ candidate sub-stencils $S_m \subset S_R$, each
with $R+1$ points,
\begin{equation}
  \label{eq:sub-stencil}
  S_m = \{I_{i-R+m-1},\dots ,I_i,\dots,I_{i+m-1}\}, \;\;\; m=1, \dots ,R+1,
\end{equation}
which satisfy
\beq
\begin{array}{cccc}
\bigcup\limits_{m=1}^{R+1}S_m = S_R, & \;
                                       \bigcap\limits_{m=1}^{R+1}S_m =
                                       I_i, &\; \mbox{ and } & x_*=x_{i\pm1/2} \in I_m, \forall m.
\end{array}
\eeq
Eq.~(\ref{eq:q-intrp}) can then be evaluated to give a GP
interpolation at the location $x_*=x_{i\pm 1/2}$ from the $m$-th candidate stencil
$S_m$,
\begin{equation}
  \label{eq:candidate_interp}
  {q}^m_{*} = \bfw_m^T  \bff_m.
\end{equation}
We now take the weighted combination of these candidate GP
approximations as the final interpolated value,
\begin{equation}
  \label{eq:weight_sum}
  q_* = \sum _{m=1}^{R+1} \omega_m q_*^m.
\end{equation}

As in the traditional WENO approach, the nonlinear weights,
$\omega_m$, should reduce to some optimal weights $\gamma_m$ in
smooth regions, so that the approximation in
Eq.~(\ref{eq:candidate_interp}) reduces to the GP approximation
(Eq.~(\ref{eq:q-intrp})) over the global $2R+1$ point stencil
$S_R$. 
The $\gamma_m$'s then should satisfy,
\begin{equation}
  \label{eq:gammas_0}
  q_* = \sum_{m=1}^{R+1} \gamma_m q_*^m,
\end{equation}
or equivalently,
\begin{equation}
  \label{eq:gammas}
  \bfw_*^T  \bff = \sum_{m=1}^{R+1} \gamma_m\bfw_m^T  \bff_m.
\end{equation}
We then seek $\boldsymbol{\gamma} = [\gamma_1,\dots,\gamma_{R+1}]^T$ as
  the solution to the $(2R+1)\times(R+1)$ overdetermined system
  \begin{equation}
    \label{eq:gamm-system}
    \textbf{M} \boldsymbol{\gamma} = \bfw_*,
  \end{equation}
  where the $n$-th column of $\textbf{M}$ is given by $\bfw_n$ for
  $(R+1)$ row entries and zeros for the rest:
  \begin{equation}
    \label{eq:M-mat}
    [\textbf{M}]_{mn} = \left \{
      \begin{array}{cl}
        \mathrm{w}_{m-n+1,n}& \text{if } I_m \in S_n, \\
        0 & \text{otherwise},
      \end{array}
    \right.
  \end{equation}
  where $\mathrm{w}_{m,n}\equiv[\bfw_n]_m$.
  For example, in the case of $R=2$ the above system reduces to the
  $5\times 3$ overdetermined system,
  \beq
\gamma_1 
\left[ 
\begin{array}{ccccc}
{\mathrm{w}_{1,1} }  \\ 
{\mathrm{w}_{2,1}} \\
{\mathrm{w}_{3,1}} \\
0 \\
0\\
\end{array}
\right]
\begin{array}[c]{@{}l@{\,}l}
   \left. 
   \begin{array}{c} 
   \vphantom{\vdots}\\ 
   \vphantom{0} 
   \end{array} 
   \right\} 
   & 
   {\bfw_1} \\ \\ \\
\end{array}
+
\gamma_2
\left[ 
\begin{array}{ccccc}
0\\
{\mathrm{w}_{1,2} }  \\ 
{\mathrm{w}_{2,2}} \\
{\mathrm{w}_{3,2}} \\
0\\
\end{array}
\right]
\begin{array}[c]{@{}l@{\,}l}
   \left. 
   \begin{array}{c} 
   \vphantom{\vdots}\\ 
   \vphantom{0} 
   \end{array} 
   \right\} 
   & 
   {\bfw_2} \\
\end{array}
+
\gamma_3
\left[ 
\begin{array}{ccccc}
0\\
0\\
{\mathrm{w}_{1,3} }  \\ 
{\mathrm{w}_{2,3}} \\
{\mathrm{w}_{3,3}} \\
\end{array}
\right]
\begin{array}[c]{@{}l@{\,}l}\\\\
   \left.
   \begin{array}{c}
   \vphantom{\vdots}\\ 
   \vphantom{0} 
   \end{array} 
   \right\} 
   &
   {\bfw_3}
\end{array}
=
\left[ 
\begin{array}{ccccc}
{\mathrm{w}_{1} }  \\ 
{\mathrm{w}_{2}} \\
{\mathrm{w}_{3}} \\
{\mathrm{w}_{4}}\\ 
{\mathrm{w}_{5}} \\
\end{array}
\right]
\begin{array}[c]{@{}l@{\,}l}
   \left. 
   \begin{array}{c} 
   \vphantom{0}  \\ 
   \vphantom{\vdots}\\ 
   \vphantom{\vdots}\\    
   \vphantom{0} 
   \end{array} 
   \right\} 
   & 
   {\bfw_*} \\
\end{array},
\eeq
or in matrix form, $\bM \boldsymbol{{\gamma}} = \bfw_*$,
\beq
\left[ 
\begin{array}{ccccc}
{\mathrm{w}_{1,1} } &\;\;& 0           &\;\;& 0   \\ 
{\mathrm{w}_{2,1}}  &\;\;& \mathrm{w}_{1,2}  &\;\;& 0\\
{\mathrm{w}_{3,1}}  &\;\;& \mathrm{w}_{2,2}  &\;\;& \mathrm{w}_{1,3}\\
0             &\;\;& \mathrm{w}_{3,2}  &\;\;& \mathrm{w}_{2,3}\\
0             &\;\;& 0           &\;\;& \mathrm{w}_{3,3} \\
\end{array}
\right]
\left[ 
\begin{array}{ccccc}
\gamma_1\\
\gamma_2\\
\gamma_3\\
\end{array}
\right]
=
\left[ 
\begin{array}{ccccc}
{\mathrm{w}_{1} }  \\ 
{\mathrm{w}_{2}} \\
{\mathrm{w}_{3}} \\
{\mathrm{w}_{4}}\\ 
{\mathrm{w}_{5}} \\
\end{array}
\right].
\eeq

The optimal weights, $\gamma_m$, then depend only on the choice of
kernel (Eq.~(\ref{eq:se})) and the stencil $S_R$, and as with the
weights $\bfw_*$ and $\bfw_m$, the $\gamma_m$'s need only be computed
once and used throughout the simulation. We take $\gamma_m$ as the
least squares solution to Eq.~(\ref{eq:gamm-system}), which can be
determined numerically.

All that remains to complete GP-WENO is to specify the nonlinear weights $\omega_m$ in
Eq.~(\ref{eq:weight_sum}). These should reduce to the optimal weights
$\gamma_m$ in smooth regions, and more importantly, they need to serve as an
indicator of the continuity of data on the candidate stencil $S_m$,
becoming small when there is a strong discontinuity on $S_m$. We first
adopt the weighting scheme of the WENO-JS schemes
\cite{jiang1996efficient},
\begin{equation}
  \label{eq:weights}
  \omega_m^{} = \frac{\tilde{\omega}_m^{}}{ \sum_{s}\tilde{\omega}_s^{}}, \mbox{ where }
\tilde{\omega}_m^{} = \frac{\gamma_m^{}}{(\epsilon + \beta_m)^p},
\end{equation}
where we have set $p=2$ and $\epsilon=10^{-36}$ in our tests. 
The quantity $\beta_m$ is the so-called smoothness indicator of the data
$\bff_m$ on the stencil $S_m$. In WENO schemes the smoothness
indicators are taken as the scaled sum of the square $L_2$ norms of
all the derivatives $\frac{d^{l}}{dx^{l}}p_m(x)$, $l=1, \dots, k$, 
of the local $k$-th degree reconstruction polynomials $p_m(x)$ over the cell
$I_i$  
where the interpolating points $x_{i\pm 1/2}$ are located.

In our GP formulation, however, there is no polynomial to use for $\beta_m$, and hence
a non-polynomial approach is required. The GP theory furnishes the concept of the data likelihood function,
which measures how likely the data is to have been sampled from the chosen GP distribution. The
likelihood function is very well-adapted to detecting departures from smoothness, because
the SE kernel (Equation \ref{eq:se}) is a covariance over the space of smooth ($C^\infty$) functions \cite{rasmussen2005},
so that non-smooth functions are naturally assigned smaller likelihood by the model.
As in \cite{reyes_new_2016} we construct the smoothness indicators
$\beta_m$ within the GP framework as the negative $\log$ of the GP likelihood in
Eq.~(\ref{eq:likely}),
\begin{equation}
  \label{eq:log-like}
  -\log [\mathcal{L}] = \frac{N}{2}\log[2\pi] +\inv{2}\log \left|\det \bfK_m\right|
  + \inv{2}(\bff_m-\bar{\bff})^T\bfK_m^{-1}(\bff_m-\bar{\bff}),
\end{equation}
which is non-negative.
The three terms on the right hand side of Eq.~(\ref{eq:log-like}) can
be identified as a normalization, a complexity penalty and a data fit
term, respectively
\cite{bishop2007pattern,rasmussen2005}. The GP covariance
matrix, $\bfK_m$, on each of the sub-stencils $S_m$ are identical here in
the uniform grid geometry, causing the first two terms in
Eq.~(\ref{eq:log-like}) -- the normalization and complexity penalty terms -- to
be the same on each candidate stencil regardless of the data
$\bff_m$. For this reason, we use only the data fit term in our GP
smoothness indicators. 
With the choice of zero mean $\bar{\bff}=0$ 
the GP-based smoothness indicator becomes
\begin{equation}
  \label{eq:gp-smooth}
  \beta_m = \bff_m^{\ T}(\bfK_m^{-1})\bff_m.
\end{equation}
Let us consider a case in which 
the data on $S_j$ is discontinuous,
while the other  sub-stencils $S_m$ ($m = 1, \cdots, R+1, m \ne j$) contain smooth data.
The statistical interpretation
of Eq.~(\ref{eq:gp-smooth}) is that the short length-scale 
variability (i.e., the short shock width ranging over a couple of grid spacing $\Delta$) in
the data makes $\bff_j$ unlikely (i.e., low probability) according to the smoothness of the
model represented by $\bfK_j$, in which case $\beta_j \sim
-\log[\mathcal{L}(\bff_j,\bfK_j)]$ is relatively larger than the other $\beta_m$, $m \ne j$. 
On the other hand, for smooth $\bff_j$
where the data is likely (i.e., high probability), $\beta_j$ becomes relatively smaller
than the other $\beta_m$, $m \ne j$.

As in the standard WENO schemes, the nonlinear GP-WENO interpolation
relies on the ``relative ratio'' of each individual $\beta_m$ to the others.
For this reason, the choice of $\Sigma=1$ for $\bfK_m$ 
in Eq.~(\ref{eq:gp-smooth})
can also be justified due to cancellation.

We note that, with the use of zero mean, 
$\beta_m$ does not reduce to zero
on a sub-stencil where
the data $\bff_m$ is non-zero constant.
In this case, the value of $\beta_m$ could be any non-zero value 
proportional to $\bff_m^2$ which could be arbitrarily large depending on the constant value of $\bff_m$.
One resolution to this issue to guarantee
$\beta_m = 0$ in this case is to use a non-zero mean $\bar{\bff}$.
In our numerical experiments, the use of non-zero mean helps to improve
some under- and/or over-shoots adjacent to constant flow regions.
However, away from such constant regions, the GP solution becomes
more diffusive than with zero mean function. 
In some multidimensional problems where there is an assumed flow symmetry, 
the GP solutions with non-zero mean failed to preserve the symmetry during the course of
evolution.
For this reason, we use zero mean function in this paper,
leaving a further investigation of this issue to a future study.

The calculation of $\beta_m$ in Eq.~(\ref{eq:gp-smooth}) can be speeded
up by considering the eigenvalues $\lambda_i$ and eigenvectors
$\bfv_i$ of the square matrix $\bfK_m$, which allow $\beta_m$ to be
expressed as (see \cite{reyes_new_2016} for derivation),
\begin{equation}
  \label{eq:beta-eigen}
  \beta_m = \sum_{i=1}^{R+1}\inv{\lambda_i}(\bfv_i\cdot \bff_m)^2.
\end{equation}
As previously mentioned, for the uniform grids considered here the
$\bfK_m$'s are the same for every candidate stencil. Hence, like
$\gamma_m$ and $\bfw_m$, the combination $\bP_i =
{\lambda_i}^{-1/2}\bfv_i$ need only be computed once before starting the
simulation and then used throughout the simulation.

It is worthwhile to note that our smoothness indicators $\beta_m$
in Eq.~(\ref{eq:beta-eigen}) are written compactly as a sum of perfect squares, 
which is an added advantage recently 
studied by Balsara \textit{et al.} \cite{balsara_efficient_2016} for their WENO-AO formulations.
In addition, all eigenvalues $\lambda_i$ of the symmetric, positive-definite
matrix $\bfK_m$ are positive-definite, so that the
smoothness indicators $\beta_m$ are always positive by construction.


\subsection{The Length Hyperparameter}
\label{sec:length-hyperp}

As mentioned in Section~\ref{sec:gp-interpolation} the SE kernel in Eq.~(\ref{eq:se}) 
used in this paper contains a length
hyperparameter $\ell$ that controls the characteristic length scale
of the GP model. Alternatively, the GP prediction given in
Eq.~(\ref{eq:post_mean}), using the SE kernel, can be viewed as a
linear smoother of the data $\bff$ over the length scale $\ell$. For
this reason, $\ell$ should be comparable to, if not larger than the
size of the given stencil, $\ell/\Delta \gtrsim R$. We
demonstrate in Section~\ref{sec:accuracy-results} how this length
scale can be tuned to obtain better accuracy for smooth flows, as was
also seen in \cite{reyes_new_2016}.

It is important to note that the length hyperparameter serves a second, logically separate purpose when used to
compute the smoothness indicators according to the GP likelihood in Eq.~(\ref{eq:gp-smooth}). 
In this application we are not 
using the GP model to smooth the data over the given sub-stencil but
rather to determine whether there is a discontinuity in any of the
candidate sub-stencils. 
In general, these two applications have different purposes and requirements.
We therefore introduce a second length hyperparameter $\sigma$ for determining the 
GP smoothness indicators $\beta_m$ in Eq.~(\ref{eq:gp-smooth})
so that we compute $\bfK_m$ using $\sigma$ instead of $\ell$, thus
treating $\sigma$ separately
from the ``smoothing'' length hyperparameter $\ell$.
This modification allows us to obtain greater stability in the
presence of discontinuities by considering length scales comparable to
the grid spacing $\Delta$, $\sigma \gtrapprox \Delta$, based on the fact 
that the typical shock width in high-order shock capturing codes is of order $\Delta$.
Viewed as a whole, the method essentially first
attempts to detect discontinuities on the (shorter) scale of  $\sigma$, and
then smooths the data on the (larger) scale of $\ell$.

We have found that using $\ell/\Delta = 12$ and $\sigma/\Delta = 3$
works well on a wide range of problems, especially for the $R=2$
GP-WENO scheme. Additional stability for problems with strong shocks
using larger stencil radii can be achieved by using lower values of
$\sigma$, down to $\sigma/\Delta=1.5$.
Larger values of $\ell$ can also
lead to marginal gains in stability. 
However, we observe that setting $\sigma \le \Delta$ makes the GP-WENO
scheme unstable regardless of any choice of $\ell$ values.

\section{A Stepwise Description of GP-WENO}
\label{sec:gp_weno_stepwise}
Before we present the numerical results of the GP interpolation algorithms outlined so far,
we give a quick summary of the step-by-step procedures of the GP-WENO algorithm as below:

\begin{enumerate}
\item \textbf{Pre-Simulation:} The following steps are carried out
  before starting a simulation, and any calculations therein need only
  be performed {\it once, stored, and used} throughout the actual simulation.
  \begin{enumerate}[(a)]
  \item \textbf{Configure computational grid:} Determine a GP stencil
    radius $R$ and choose the size of the hyperparameter
    $\ell$. This configuration determines the SE kernel function in
    Eq.~(\ref{eq:se}) as well as the global and candidate stencils in
    Eqs.~(\ref{eq:2R1-stencil}) and (\ref{eq:sub-stencil}).
  \item
    \label{it:gp_wgt}
    \textbf{Compute GP weights:} Compute the covariance
    matrices, $\bK$ and $\bK_m$, according to Eq.~(\ref{eq:kmat}) on the stencils $S_R$ and
    each of $S_m$, respectively. Compute the prediction vectors, $\bk_*$ and $(\bk_*)_m$ respectively on
    $S_R$ and each of $S_m$ using Eq.~(\ref{eq:kstar}). The GP weight
    vectors, $\bw^T=\bk_*\bK^{-1}$ on $S_R$ and $\bw^T_m=(\bk_*)_m \bK_m^{-1}$
    on each of $S_m$, can then be stored for use in the GP interpolation.
    See also \cite{reyes_new_2016} for a detailed discussion to circumvent a potential singularity issue
    in computing the covariance matrices for large values of $\ell/\Delta$.
    %
    %
  \item
    \label{it:lin_wgt}
    \textbf{Compute linear weights:} Use the GP weight vectors $\bw$ and $\bw_m$ from Step \ref{it:gp_wgt} to
    calculate and store the optimal linear weights $\gamma_m$ according to Eq.~(\ref{eq:gamm-system}).
  \item
    \label{it:eigen}
    \textbf{Compute kernel eigensystem:} Choose a length
    hyperparameter, $\sigma/\Delta$, for the smoothness indicators. The eigensystem for the
    covariance matrices $\bK_m$ used in GP-WENO are calculated
    using Eq.~(\ref{eq:kmat}) using $\sigma$ instead of $\ell$. Since the matrices $\bK_m$
    are the same on each of the candidate stencils in the GP-WENO
    scheme presented here, only one eigensystem (e.g., $m=1$) needs to be
    determined. This eigensystem is then used to calculate and store the vectors
    $\bfv_i/\sqrt{\lambda_i}$ in Eq.~(\ref{eq:beta-eigen}) for use in determining the
    smoothness indicators in the interpolation Step \ref{it:interpolation_gp} below.
  \end{enumerate}
  %
  %
\item \textbf{Riemann state interpolation:} \label{it:interpolation_gp}
  Start a simulation. 
  At each cell $x_i$, calculate the updated posterior mean function
  $q_*^m$ according to Eq. (\ref{eq:candidate_interp}) 
  as a high-order GP interpolator to compute 
  high-order Riemann state values at $x_*=x_{i\pm 1/2}$
  using each of the candidate sub-stencils $S_m$. 
  The
  smoothness indicators (Eq.~(\ref{eq:beta-eigen})),  calculated using
  the eigensystem from Step \ref{it:eigen} in conjunction with the linear weights
  from Step \ref{it:lin_wgt}, form the nonlinear weights
  (Eq.~(\ref{eq:weights})). Then take the convex combination of  $q_*^m$ to get  $q_*$ 
  according to Eq.~(\ref{eq:weight_sum}).
  \item \textbf{Calculate numerical fluxes:} 
  \label{it:num_fluxes}
  Solve Riemann problems at
    cell interfaces, $x_{i\pm1/2}$, using 
  the high-order GP Riemann states in Step \ref{it:interpolation_gp} as inputs. Use the
  interface fluxes to calculate a numerical flux at each interface according to Eq.~(\ref{eq:numFlux_cor}).
  \item \textbf{Temporal update:} 
  Using the high-order Godunov fluxes from Step \ref{it:num_fluxes}, 
  update the pointwise solutions $q_i$ 
  from $t^n$ to $t^{n+1}$ according to Eq.~(\ref{eq:disc_fd}).
\end{enumerate}

\section{GP-WENO Code Implementation and Distribution}
The implementation of the GP-WENO FD-Prim scheme is parallelized using Coarray Fortran (CAF)
\cite{garain_comparing_2015}. CAF is made to work with the GNU Fortran
compiler through the OpenCoarrays library
\cite{eachempati_open-source_2010}, and IO is handled with the HDF5 library. 
The GP-WENO FD-Prim 
source code described in this paper is available at 

\begin{center}
\texttt{https://github.com/acreyes/GP-WENO\_FD-Prim}
\end{center} 

\noindent under a Creative Commons Attribution 4.0 International License. 

\section{Accuracy and Performance Results on Smooth Test Problems}
\label{sec:accuracy-results}
In this section we assess the accuracy and performance of the proposed 
GP-WENO scheme on smooth advection problems in 1D and 2D.
As demonstrated below, the high-order GP-WENO solutions converge linearly in 
$\mathcal{O}(\Delta^{2R+1})$ and target solution errors are reached 
faster, in terms of CPU-time, than the 5th-order WENO-JS we chose
as the polynomial-based counterpart algorithm for comparison.
A suite of nonlinear discontinuous problems in 1D, 2D and 3D are outlined in
Section~\ref{sec:numer-test-probl} to display the shock-capturing capability
of the GP-WENO scheme.

For smooth-flow problems where there is no shock or discontinuity,
treating $\sigma$ differently from $\ell$ is not required 
because the associated nonlinear smoothness indicators
all become equally small in magnitude and do not
play an important role. 
We performed the smooth advection problems in Section~\ref{sec:accuracy-results}
by setting $\sigma/ \Delta \sim 3$ to follow the same convention we use for all discontinuous
problems in Section~\ref{sec:numer-test-probl} (i.e., $\ell/\Delta \sim 12$ and  $\sigma/\Delta \sim 3$).
Alternatively, one can set $\sigma = \ell$ in all smooth flow problems, which does not 
qualitatively change the results reported in this section.

\subsection{1D Smooth Gaussian Advection}
\label{sec:1d-smooth-advection}
The test presented here considers the passive advection of a Gaussian
density profile in 1D. The problem is set up on a computational domain $[0,1]$ 
with periodic boundary conditions. The initial condition is
given by a density profile of  
$\rho(x) = 1 + e^{-100(x-x_0)^2}$ with $x_0=0.5$, with constant velocity and
pressure, $u=1$ and $P=1/\gamma$. The ratio of specific heats is
chosen to be $\gamma=5/3$. The profile is propagated for one period
through the boundaries until $t=1$ where the profile returns to its
initial position at $x=x_0$. At this point, any deformation to the initial
profile is solely due to phase errors and/or numerical diffusion of the algorithm under consideration,
serving as a metric of the algorithm's accuracy.

\begin{figure}[h!]
  \centering
  \includegraphics[width=0.55\textwidth]{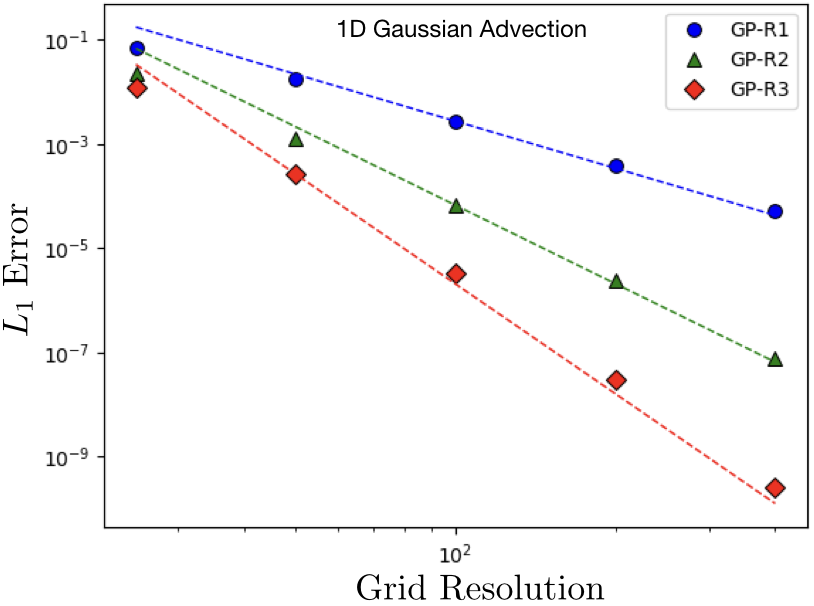}
  \caption{Plot of the $L_1$ errors for GP-WENO using R=1,2,3 on the
    1D smooth Gaussian advection problem. Dotted lines show the corresponding
    $(2R+1)$ convergence rates.}
  \label{fig:1d_smth}
\end{figure}

We perform this test for the GP-WENO method using values of $R=1,2,3$, 
a length hyperparameter of $\ell=0.1$ (the choice of this value becomes apparent below),
and $\sigma/\Delta = 3$. We employ RK4 for time integration, adjusting the time step
to match the temporal and spatial accuracy of the scheme as the resolution is
increased (e.g., see \cite{mignone_high-order_2010-1}).
The results are summarized in Fig.~\ref{fig:1d_smth} and Table~\ref{tab:1d_smth}. All three choices
of $R$ demonstrate a $(2R+1)$ convergence rate, as shown in
\cite{reyes_new_2016} for the same problem using the GP-WENO finite volume scheme
reported therein.

\begin{table}[h!]
  \centering
  \caption{L1 errors for the 1D smooth Gaussian advection
    problem on five different grid resolutions. All simulations use $\ell=0.1$, $\sigma/\Delta=3$, 
    and RK4 for time
    integration with an appropriately limited CFL condition to match
    the temporal and spatial accuracy.}  
  \footnotesize
  \begin{tabular}{|c|lc|lc|lc|} 
    \hline 
   \multirow{2}{*}{Grid $\Delta$}& \multicolumn{2}{c|}{GP-R1} & \multicolumn{2}{c|}{GP-R2} & \multicolumn{2}{c|}{GP-R3} \\ \cline{2-7}
    & $L_1$ & $L_1$ Order & $L_1$ & $L_1$ Order & $L_1$ & $L_1$ Order \\ \hline
    1/25 & $ 7.03 \times 10^{-2} $ & --  & $ 2.25 \times 10^{-2} $ & --  &$ 1.19 \times 10^{-2} $ & --  \\
1/50 & $ 1.74 \times 10^{-2} $ & 2.02  & $ 1.30 \times 10^{-3} $ & 4.11  &$ 2.64 \times 10^{-4} $ & 5.49\\
1/100 & $ 2.75 \times 10^{-3} $ & 2.66  & $ 6.70 \times 10^{-5} $ & 4.28 & $ 3.22 \times 10^{-6} $ & 6.36 \\
1/200 & $ 4.01 \times 10^{-4} $ & 2.78  & $ 2.48 \times 10^{-6} $ & 4.75 & $ 2.97 \times 10^{-8} $ & 6.76  \\
1/400 & $ 5.14 \times 10^{-5} $ & 2.96  & $ 7.84 \times 10^{-8} $ &
                                                                      4.99 & $ 2.51 \times 10^{-10} $ & 6.88  \\ \hline
  \end{tabular}
  \label{tab:1d_smth}
\end{table}

\begin{figure}[h!]
  \centering
  \includegraphics[width=0.6\textwidth]{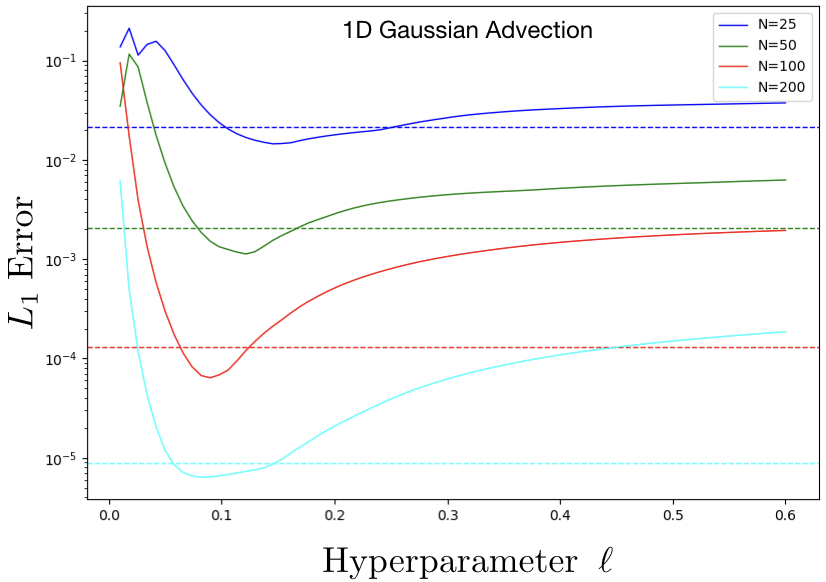}
  \caption{$L_1$ errors for different values of $\ell$ for the 1D
    Gaussian smooth advection. Solid lines show the GP-R2 scheme and
    dashed lines $L_1$ errors for the WENO-JS scheme using RK3 with a CFL=0.8.}
  \label{fig:ell_1d}
\end{figure}

The length hyperparameter $\ell$ provides an additional knob to tune
the solution accuracy. Fig.~\ref{fig:ell_1d} shows how the $L_1$ error
changes with the choice of $\ell$ on the 1D Gaussian advection
problem for different resolutions using the 5th-order GP-R2 scheme
compared to the 5th-order WENO-JS scheme (denoted with dotted
lines). The dependence of the $L_1$ errors on $\ell$ is qualitatively
similar for all resolutions. At larger values of $\ell$ (e.g., $\ell > 0.3$) the errors
begin to asymptotically plateau out and are generally higher than the corresponding WENO-JS
simulation. This can be explained by the nature of the GP's kernel-based data prediction, 
in which larger values of $\ell$ results in the GP model under-fitting the data.
On the other hand, we see that all errors diverge at small $\ell$ (e.g., $\ell < 0.1$)
due to the fact that the GP model over-fits the data
(i.e., large oscillations between the data points).
The errors of GP-WENO reach a local minimum at $\sim\ell=0.1$, roughly the full-width
half-maximum (FWHM) of the initial Gaussian density profile. In all cases
this local minimum in the $L_1$ error is lower than the errors
obtained using the WENO-JS scheme. This behavior is similar to that
observed for radial basis function (RBF) methods for CFD
\cite{bigoni_adaptive_2017}, for
the RBF shape parameter $\epsilon$ which represents an
inverse length scale, i.e., $\epsilon \sim 1/\ell$. 
Nonetheless, the connection between the optimal $\ell$ and
the length scales of the problem has only been made in the context of
Gaussian process interpolations/reconstructions in our recent work
\cite{reyes_new_2016}. This suggests that, to best
resolve the ``smallest'' possible features in a simulation for a given grid
spacing $\Delta$, the choice of $\ell \gtrsim \Delta$ may be optimal.

\subsection{2D Isentropic Vortex}
\label{sec:2d-isentropic-vortex}

Next, we test the accuracy of the GP-WENO schemes using the
multidimensional nonlinear isentropic vortex problem, initially
presented by Shu \cite{shu_essentially_1998}. The problem consists of
the advection of an isentropic vortex along the diagonal of a
Cartesian computational box with periodic boundary conditions. We set
up the problem as presented in \cite{spiegel_survey_2015}, where the
size of the domain is doubled to be $[0,20]\times[0,20]$ compared to
the original setup in \cite{shu_essentially_1998} to prevent self-interaction 
of the vortex across the periodic domain. The problem is
run for one period of the advection through the domain until the
vortex returns to its initial position, where the solution accuracy can
be measured against the initial condition.

\begin{figure}[h!]
  \centering
  \includegraphics[width=0.55\textwidth]{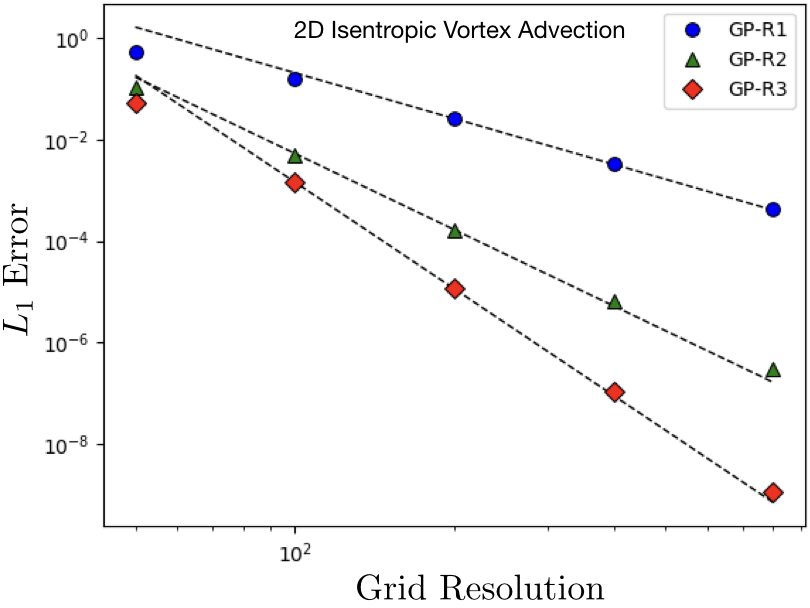}
  \caption{Plot of the $L_1$ errors for GP-WENO using R=1,2,3 on the
    isentropic vortex problem. Dotted lines show the corresponding
    $(2R+1)$ convergence rates. All simulations use $\ell=1.0$ and
    $\sigma/\Delta=3$. Temporal integration is done using RK4 with a
    suitably limited time step to match the temporal and spatial
    errors between different resolutions.}
  \label{fig:vortex}
\end{figure}

Our $L_1$ error results are shown in Fig.~\ref{fig:vortex} and summarized in
Table~\ref{tab:vortex} using $\ell=1$, $\sigma/\Delta = 3$, and
RK4 for time integration, utilizing once more the appropriate reduction in time
step to  match the spatial and temporal accuracies.
As with the 1D smooth advection in the previous section, the GP-WENO
method obeys a $(2R+1)$ order of convergence rate.

\begin{table}[h!]
  \caption{$L_1$ errors for the 2D isentropic vortex
    problem. All simulations use $\ell=1.0$, $\sigma/\Delta=3$, and RK4 for time
    integration with an appropriately limited CFL condition to match
    temporal and spatial accuracy.}
  \footnotesize
  \centering
  \begin{tabular}{|c|cc|cc|cc|}
    \hline 
   \multirow{2}{*}{$\Delta$}& \multicolumn{2}{c|}{GP-R1} & \multicolumn{2}{c|}{GP-R2} & \multicolumn{2}{c|}{GP-R3} \\ \cline{2-7}
    & $L_1$ & $L_1$ Order & $L_1$ & $L_1$ Order & $L_1$ & $L_1$ Order \\ \hline
   2/5 & $ 5.34 \times 10^{-1} $ & --  & $ 1.33 \times 10^{-1} $ & --  & $ 6.46 \times 10^{-2} $ & --   \\
1/5 & $ 1.60 \times 10^{-1} $ & 1.74  & $ 4.71 \times 10^{-3} $ & 4.82  & $ 1.14 \times 10^{-3} $ & 5.82   \\
1/10 & $ 2.60 \times 10^{-2} $ & 2.62  & $ 1.54 \times 10^{-4} $ & 4.93  & $ 1.11 \times 10^{-5} $ & 6.68   \\
1/20 & $ 3.38 \times 10^{-3} $ & 2.94  & $ 5.73 \times 10^{-6} $ & 4.75  & $ 1.03 \times 10^{-7} $ & 6.75   \\
1/40 & $ 4.25 \times 10^{-4} $ & 2.99  & $ 2.34 \times 10^{-7} $ & 4.61  & $ 1.02 \times 10^{-9} $ & 6.66   \\ \hline
  \end{tabular}
  \label{tab:vortex}
\end{table}

\begin{figure}[h!]
  \centering
  \includegraphics[width=0.55\textwidth]{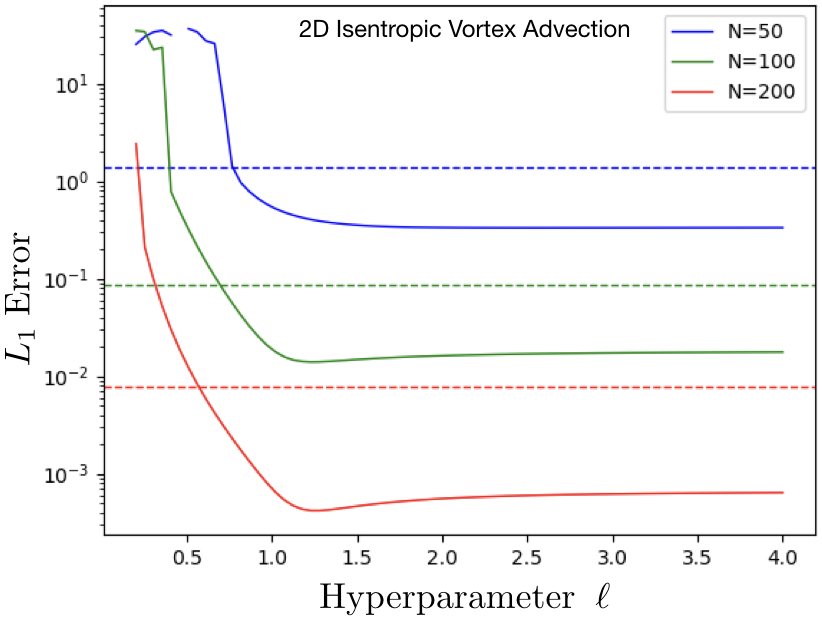}
  \caption{$L_1$ errors for different values of $\ell$ for the 2D
    isentropic vortex problem using $\sigma/\Delta=3$. Dotted lines
    show the WENO-JS $L_1$ errors at the same resolutions.}
  \label{fig:vortex_ell}
\end{figure}

We also repeat the test of the dependence of the errors on the length
hyperparameter $\ell$ and show the results in
Fig.~\ref{fig:vortex_ell}. Similar to the 1D case in Fig.~\ref{fig:ell_1d} there is a
minimum for the error at higher resolution around $\ell \gtrapprox 1$.  
The errors diverge at small values of $\ell$ and plateau at large $\ell$. Shown as dotted lines in
Fig.~\ref{fig:vortex_ell} are the errors for the WENO-JS
interpolation. For all resolutions, the minimum of the error for the
GP-WENO scheme is significantly smaller than the errors of the WENO-JS
scheme. This can also be seen by comparing the GP-R2 $L_1$ column of Table~\ref{tab:vortex}
to the WENO-JS $L_1$ column of Table~\ref{tab:weno-vortex}. Also, the order of convergence for the WENO-JS is
smaller than that of GP-R2 method.

\begin{table}[h!]
  \caption{$L_1$ errors or the 2D isentropic vortex using WENO
    interpolations comparing the choice of smoothness
    indicators. WENO-JS indicates the use of the classical polynomial-based
    smoothness indicators and WENO-GP indicates the use of the new
    GP-based smoothness indicators.}
\footnotesize
  \centering
  \begin{tabular}{|c|cc|cc|}
    \hline 
   \multirow{2}{*}{$\Delta$}& \multicolumn{2}{c|}{WENO-JS} &
                                                             \multicolumn{2}{c|}{WENO-GP}
    \\ \cline{2-5}
                            & $L_1$ & $L_1$ Order & $L_1$ & $L_1$ Order \\ \hline
    2/5 & $ 8.68 \times 10^{-2} $ & --  & $ 8.10 \times 10^{-2} $ & --  \\
1/5 & $ 3.28 \times 10^{-3} $ & 4.72  & $ 4.83 \times 10^{-3} $ & 4.07  \\
1/10 & $ 5.81 \times 10^{-4} $ & 2.50  & $ 1.73 \times 10^{-4} $ & 4.80  \\
1/20 & $ 3.83 \times 10^{-5} $ & 3.92  & $ 7.41 \times 10^{-6} $ & 4.55  \\
1/40 & $ 1.69 \times 10^{-6} $ & 4.50  & $ 2.69 \times 10^{-7} $ & 4.78  \\ \hline
  \end{tabular}
  \label{tab:weno-vortex}
\end{table}

We attribute the significant improvement in errors for the
GP-R2 scheme over the classical WENO-JS to the use of the GP-based
smoothness indicators in Eq.~(\ref{eq:beta-eigen}), which seem to
better suit the adaptive stencil process in WENO-type methods for
nonlinear problems like the isentropic vortex test. It is known that
the original weighting scheme of the WENO-JS method
\cite{jiang1996efficient} suffers from reduced accuracy in the
presence of inflection points 
that lowers the scheme's formal order of accuracy. This has been
addressed with such schemes as the Mapped WENO
\cite{henrick_mapped_2005} or the WENO-Z \cite{borges2008improved} methods.
All of these methods use
the same smoothness indicators as in WENO-JS and acquire improved
behavior by modifying the way nonlinear weights
are formulated (see Eq.~(\ref{eq:weights})). The 
GP-WENO method uses \textit{exactly the same} weighting as in the
classical WENO-JS scheme and the observed improvement originates from the new
GP-based smoothness indicators.

This suggests that the GP-based smoothness indicators could also be applied in 
conventional polynomial-based WENO interpolations/reconstructions to achieve
improved accuracy in smooth solutions. More specifically, a WENO polynomial-interpolation
is used for $\bff_m$ on the candidate stencils $S_m$ (Eq.~(\ref{eq:weight_sum})), while
the GP-based smoothness indicators are adopted in Eq.~(\ref{eq:beta-eigen}).
We refer to such a scheme as the WENO-GP weighting scheme.
\footnote{On a static, uniform grid configuration, WENO-GP 
requires a one-time formulation and storage of the GP covariance
matrix $\bfK_m$ on a sub-stencil $S_m$, followed by the computation
of its eigenvalues $\lambda_i$ and eigenvectors $\bfv_i$.   
The GP-based smoothness indicators are then computed using $\bff_m$ on each $S_m$ via
Eq.~(\ref{eq:beta-eigen}), and applied to an any polynomial-based WENO scheme.}
In Table~\ref{tab:weno-vortex}, we compare $L_1$ errors
for the WENO-JS and WENO-GP schemes.  The latter outperforms the former,
without changing the formulation of the weights $\tilde{\omega}_m$.
\begin{figure}[htpb!]
  \centering
  \includegraphics[width=0.55\textwidth]{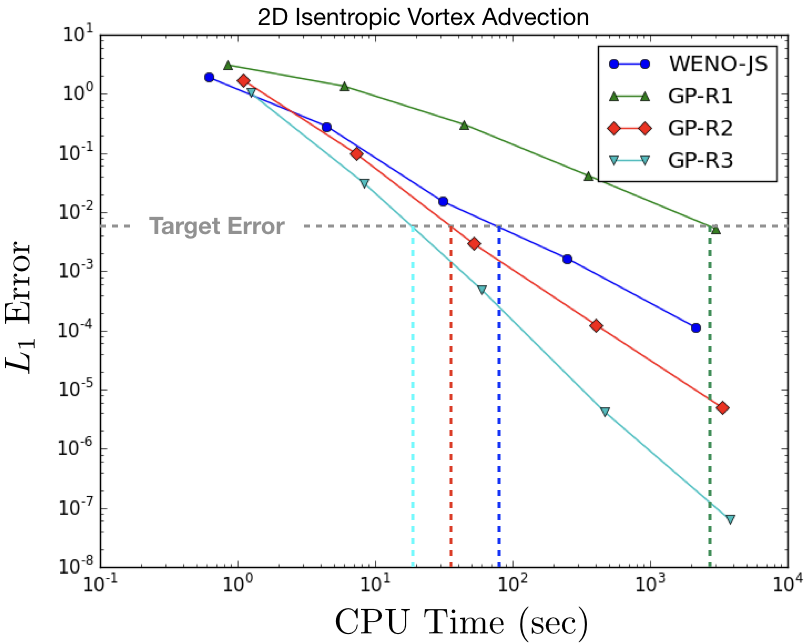}
  \caption{$L_1$ errors vs. CPU time (sec) for the 2D
    isentropic vortex problem on $[0,20]\times[0,20]$, parallelized on four compute cores 
    (2.7 GHz 12-Core Intel Xeon E5).
    The results were obtained using CFL=0.4 and temporally integrated using RK4 without 
    any use of timestep reduction. The two hyperparameters were set as $\ell = 1.2$ and $\sigma/\Delta=3$.
    An HLLC Riemann solver was used in all cases. The dotted horizontal line is the target $L_1$ error
    of $5\times 10^{-3}$.}
  \label{fig:vortex_timing}
\end{figure}

In Fig.~\ref{fig:vortex_timing} we show the CPU efficiency of WENO-JS and of GP-WENO with $R=1,2$, and $3$
as the $L_1$ error versus the CPU time for the isentropic vortex problem.
The 5th-order GP-R2 and 7th-order GP-R3 schemes yield faster time-to-solution accuracy when
compared to the 5th-order WENO-JS scheme. The comparison is quantitatively 
summarized in Table~\ref{tab:performace-time-to-error}.
\begin{table}[ht!]
  \caption{Relative CPU time for the four schemes
  to reach the target $L_1$ error of $5\times 10^{-3}$, 
  represented by the dotted horizontal line in Fig.~\ref{fig:vortex_timing}.
  All CPU times are normalized to that of WENO-JS.}
  \label{tab:performace-time-to-error}
  \centering
  \begin{tabular}{||c|c||}
    \hline
    Scheme & Relative time-to-error \\ \hline
    GP-R1   & 35.81 \\
    GP-R2   & 0.43 \\
    GP-R3   & 0.22 \\
    WENO-JS  & 1.0 \\ \hline 
  \end{tabular}
\end{table}


\section{Numerical Test Problems with Shocks and Discontinuities}
\label{sec:numer-test-probl}

In this section we present test problems using the GP-WENO interpolation method
described in Section~\ref{sec:gp-weno-interp} and applied to the compressible
Euler equations in 1, 2, and 3D. The  
GP-WENO with $R=2$ (or GP-R2) scheme is chosen as the default method and is
compared to the 5th order WENO method \cite{chen_5th-order_2016,jiang1996efficient,del_zanna_echo:_2007}
that is nominally of the same order of accuracy.
All interpolations are carried out on characteristic variables to minimize
unphysical oscillations in the presence of discontinuities \cite{qiu_construction_2002} .
A 3rd-order TVD Runge-Kutta method (RK3)
\cite{gottlieb_strong_2011} for temporal integration and an HLLC
\cite{li_hllc_2005,toro1994restoration} Riemann solver are used
throughout unless otherwise specified. The two hyperparameters $\ell$ and $\sigma$
are chosen to have values so that $\ell/\Delta\sim 6-12$ and $\sigma/\Delta \sim 1.5-3$.

\subsection{1D Scalar Advection}
\label{scal-adv}

\begin{figure}
    \centering
    \subfloat[]{
    \includegraphics[width=.45\textwidth]{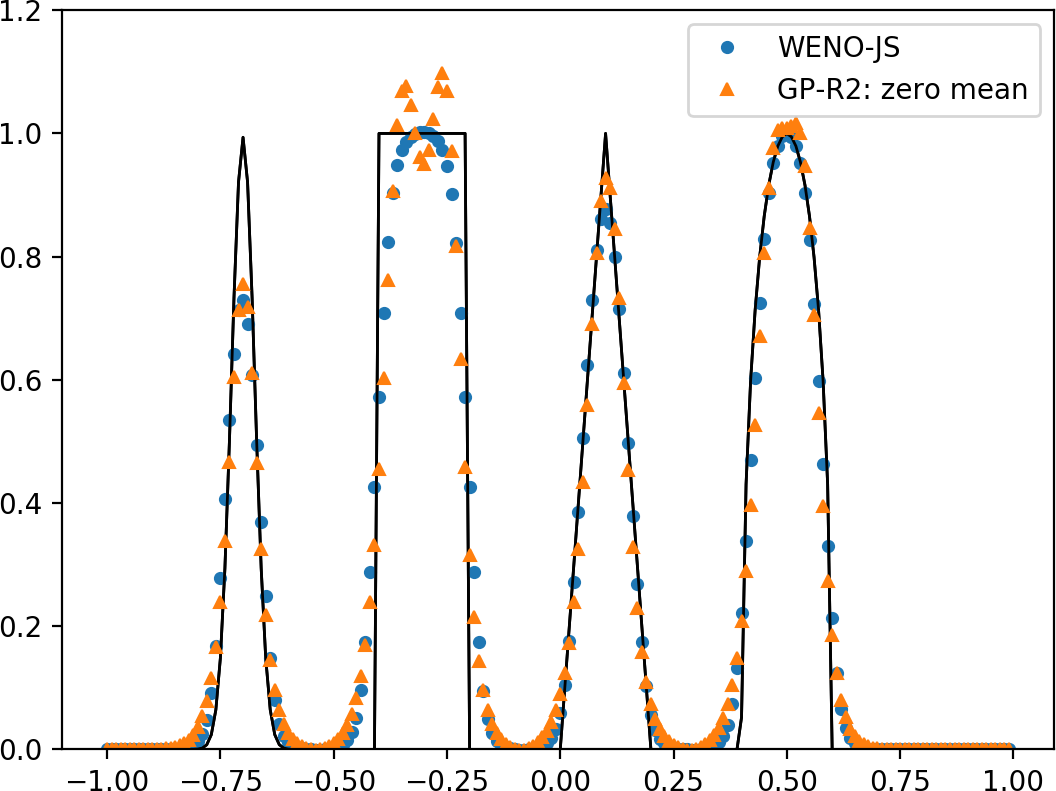}
    \label{subfig:scal_adv_zero}
    }
    \subfloat[]{
    \includegraphics[width=.45\textwidth]{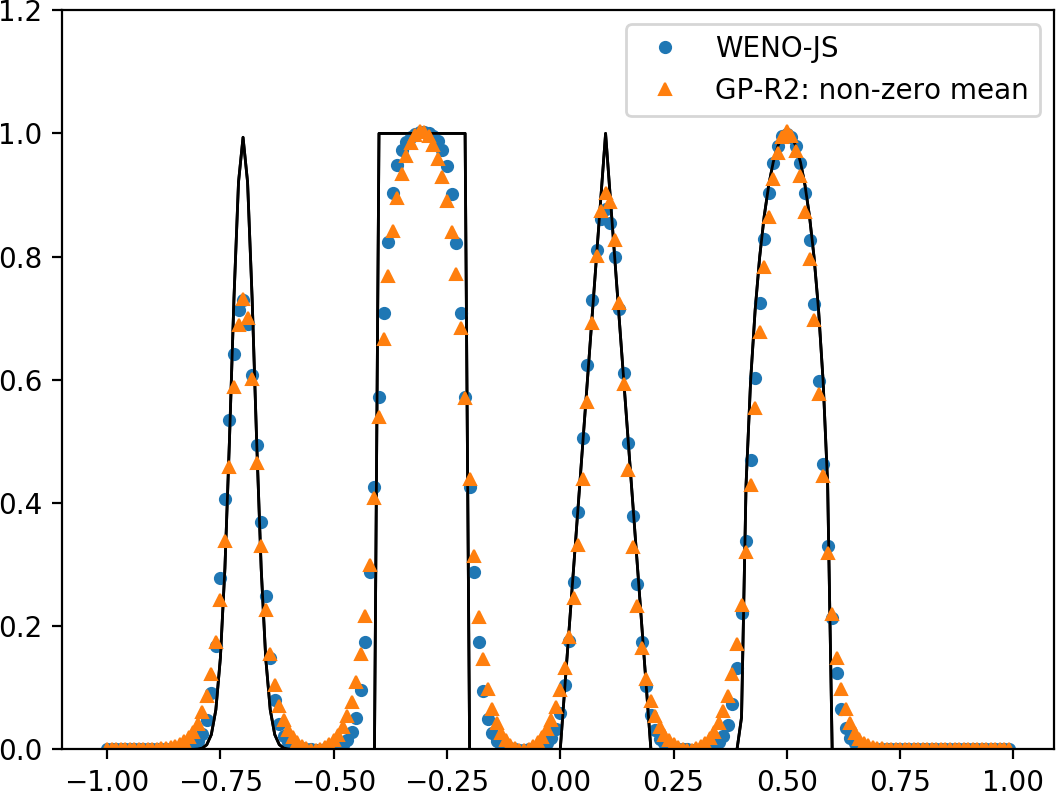}
    \label{subfig:scal_adv_nonzero}
    } 
    \caption{WENO-JS and GP-WENO solutions for the scalar advection
      problem obtained on a 200 point grid up to $t=8$ with a CFL of
      0.4. GP-WENO solutions used values of $\ell/\Delta=12$ and
      $\sigma/\Delta=3$. The left panel \ref{subfig:scal_adv_zero} shows a GP-WENO solution
      using a zero mean in the calculation of the smoothness
      indicators, and the right panel \ref{subfig:scal_adv_nonzero} shows a GP-WENO solution
      obtained using a non-zero mean.}
    \label{fig:scal_adv}
\end{figure}

Our first test is the scalar advection problem of
\cite{jiang1996efficient}, which consists of passive advection of
several shapes through a periodic domain. The profiles consist of
a combination of (i) a Gaussian, (ii) a square pulse, (iii) a sharply
peaked triangle and (iv) a half-ellipse from left to right. 
We follow \cite{jiang1996efficient} to set up the
problem on a grid with 200 points and evolve the solution up to a final time of
$t=8$, which corresponds to 4 periods through the domain. Results
comparing the WENO-JS and GP-R2 methods for this problem are shown
in Fig.~\ref{subfig:scal_adv_zero} using the GP-WENO scheme with the
smoothness indicators outlined in Eq.~(\ref{eq:beta-eigen}). 
The GP-R2 solution captures the amplitudes of the Gaussian and
triangle waves more effectively than does the WENO-JS solution. 
However, for the square wave the GP-R2 solution
introduces some Gibbs phenomena at the discontinuities. 
Similar oscillations associated
with flat discontinuities like those shown here have been also observed 
in shock-tube problems in our previous finite volume version of GP-WENO \cite{reyes_new_2016}. 
We find that this issue stems from the use of the zero-mean prior in the calculation of the
smoothness indicators, which biases the stencil choice towards those that
are more compatible with the zero-mean prior (i.e., data that are
closer to zero). The issue can be alleviated by setting a non-zero
mean using the cell-centered data $\bff_i$ on the cell $I_i$ in
Eq.~(\ref{eq:beta-eigen}), changing from 
$\bff_m$ to $(\bff_m-\bff_i)$. 
A solution obtained using these smoothness indicators
for GP-R2 is shown in Fig.~\ref{subfig:scal_adv_nonzero}. 
While this greatly improves GP-WENO's performance on this scalar advection
problem, the non-zero mean can have some unintended consequences on
multidimensional problems. Because the smoothness indicators with non-zero means $\bff_i$ are 
computed using the residuals of the data, $\bff_m - \bff_i$,
the stencil adaptation could become sensitive to machine precision differences in the data on
a stencil. Such differences could potentially lead to an instability in the stencil selection. 
In problems that contain a specific symmetry, such as the 2D Riemann Problem presented in
Sec.~\ref{subsec:2drp}, machine precision asymmetries in the initial
conditions can be observed to be amplified significantly in a runaway
process due to an asymmetry in the stencil adaptation in the WENO
procedure. In practice, for all nonlinear problems tested in this paper,
the oscillations from the zero mean choice are relatively small and so
the choice of zero mean is used in the remainder of the results presented.

\subsection{1D Shu-Osher Problem}
\label{shu-osher}

The Shu-Osher problem \cite{shu1989} is a compound test of a method
accuracy and stability.  The goal is to resolve small scale features
(i.e., high-frequency waves) 
in a post-shock region and concurrently capture a Mach 3 shock in a stable and
accurate fashion.
\begin{figure}[hb!]
  \centering
  \subfloat[]{
    \includegraphics[width=0.45\textwidth]{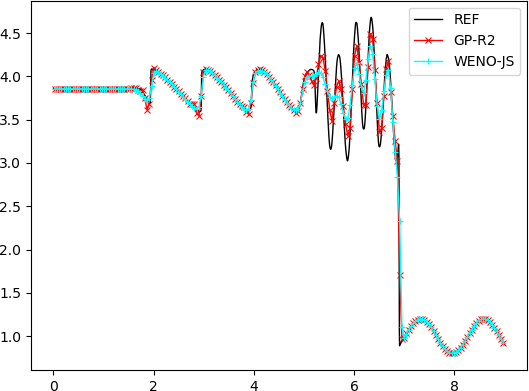}
    \label{subfig:shu-full}
  }
  \subfloat[]{
    \includegraphics[width=0.458\textwidth]{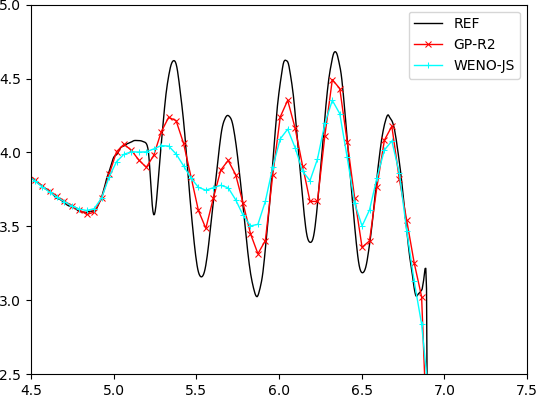}
    \label{subfig:shu-zoom}
  }
  \caption{(Left) The Shu-Osher problem at $t=1.8$ using RK3 and the
    HLLC Riemann solver. The GP-R2 scheme is shown in red, using
    $\ell/\Delta = 6$ and $\sigma/\Delta=3$. The WENO-JS scheme is shown in cyan.
    Both schemes are resolved on 200 grid points using a CFL of 0.8.
    The reference solution (black) is obtained using WENO-JS on a resolution of 2056 grid points. 
    (Right) Close-up of the frequency-doubled oscillations.}
  \label{fig:shu-full}
\end{figure}
In this problem, a (nominally) Mach 3 shock wave propagates into a constant density field
with sinusoidal perturbations. As the shock advances, two sets of
density features develop behind the shock: one that has the same
spatial frequency as the initial perturbation; one that has
twice the frequency of the initial perturbations and is closer to the 
shock. The numerical method must correctly capture the dynamics and the amplitude
of the oscillations behind the shock, and be compared against a reference solution obtained using
much higher resolution.

\begin{figure}[h!]
  \centering
  \subfloat[]{

    \includegraphics[width=5in, trim= 0.0in 1.5in 0.2in 2.0in,clip=true]
    {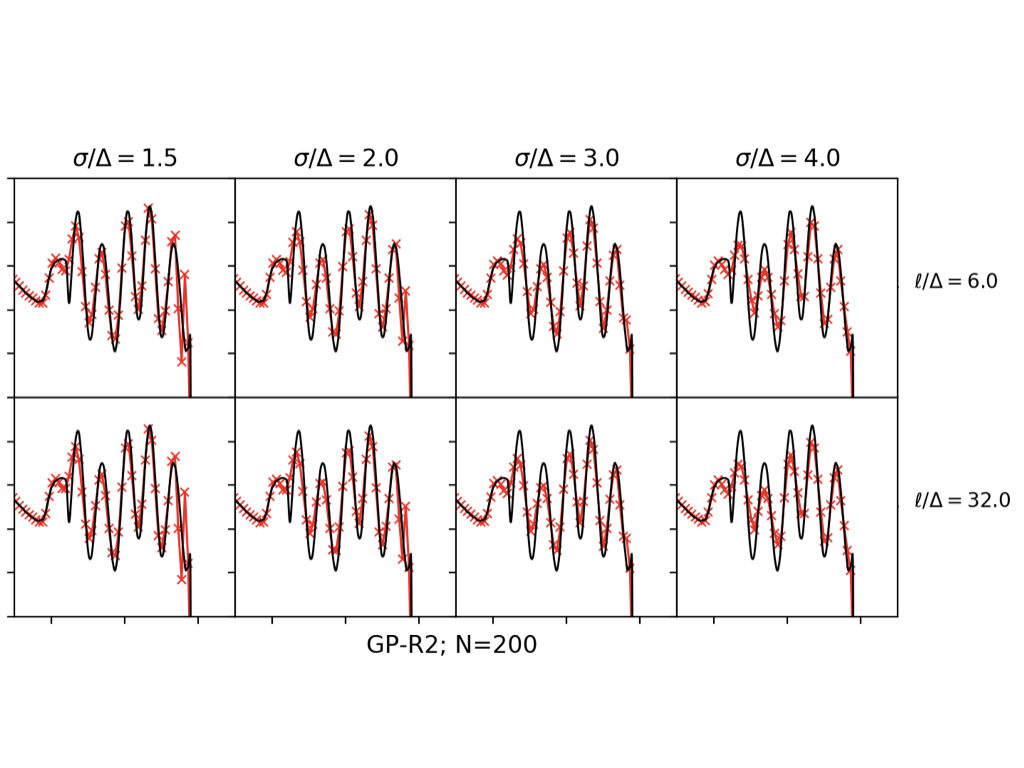}
    \label{subfig:shu-R2}
  } \\
  \subfloat[]{
    \includegraphics[width=5in, trim= 0.0in 1.5in 0.2in 2.0in,clip=true]
    {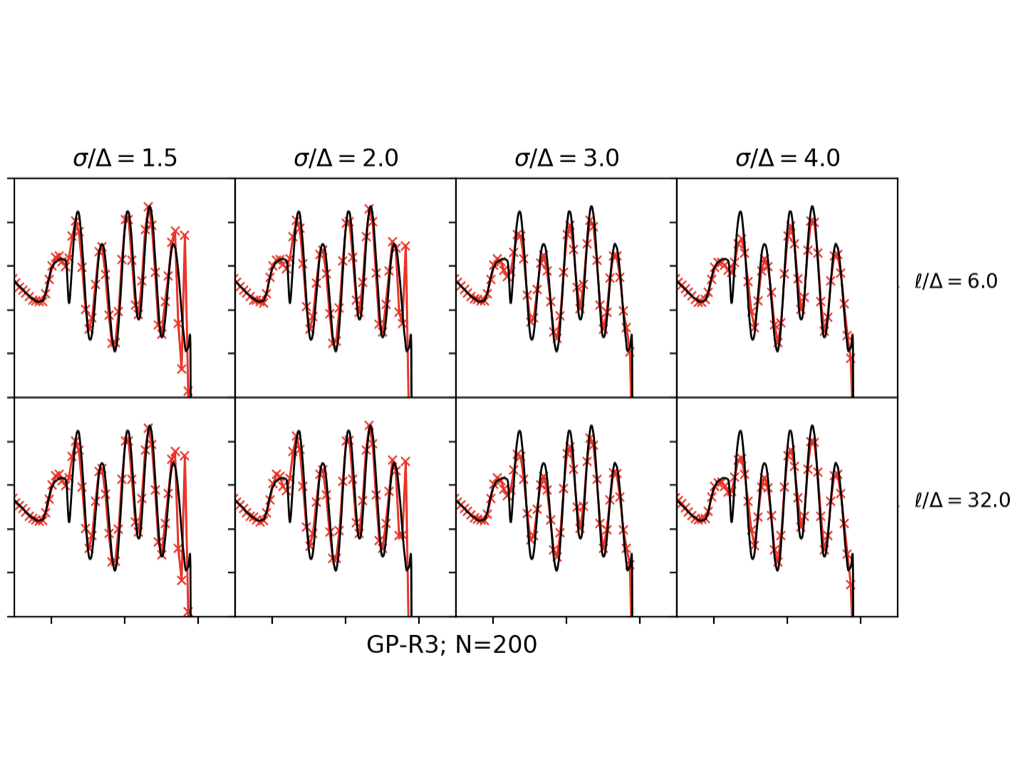}
    \label{subfig:shu-R3}
  }
  \caption{Comparison for different values of $\ell$ and $\sigma$ of
    the GP-R2 (\ref{subfig:shu-R2}) and GP-R3 (\ref{subfig:shu-R3})
    schemes on the Shu-Osher problem on 200 grid points. The 
    reference solution is shown in black.}
  \label{fig:shu-comp}
\end{figure}

\begin{figure}[h!]
  \centering
  \subfloat[][]{
    \includegraphics[width=.45\textwidth]{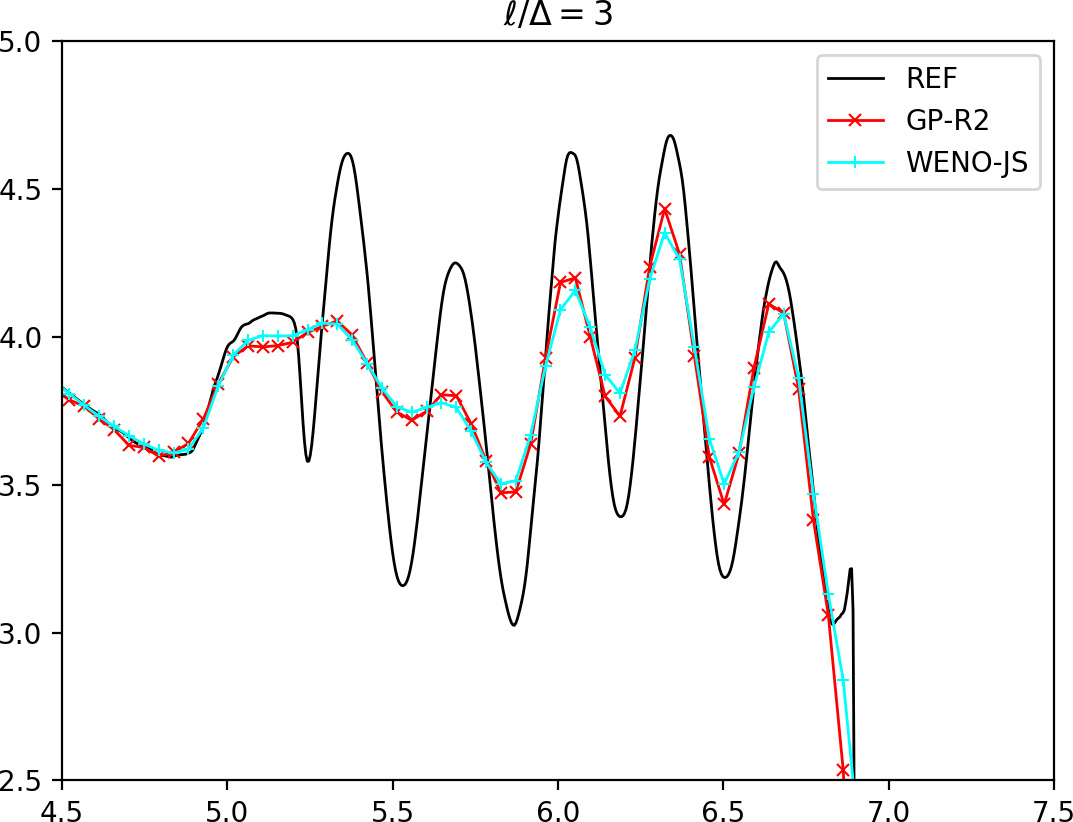}
    \label{subfig:shuW3}
  }
  \subfloat[][]{
    \includegraphics[width=.45\textwidth]{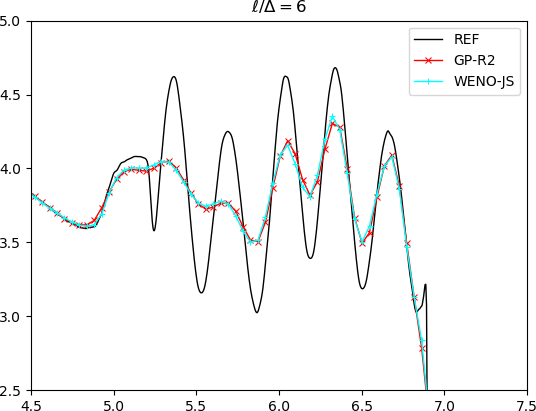}
    \label{subfig:shuW6}
    }
  \caption{Close-ups of the Shu-Osher problem using GP interpolation
    with the WENO-JS smoothness indicators instead of the default GP-based
    smoothness indicators.}
  \label{fig:shu-weno}
\end{figure}

The results for this problem are shown in Fig.~\ref{subfig:shu-full} for the
whole domain (left). A close-up of the frequency doubled oscillations is shown in
Fig.~\ref{subfig:shu-zoom}. 
We compare the GP-R2 method, with $\ell/\Delta=12$ and $\sigma/\Delta=3$,
to the WENO-JS method. The GP-R2 scheme clearly captures the amplitude of the
oscillations better than the 5th order WENO-JS. Again, the improvement
over the WENO-JS scheme is attributed to the use of the GP smoothness
indicators.

Fig.~\ref{fig:shu-comp} shows the Shu-Osher problem for
the 5th order GP-R2 and the 7th order GP-R3 schemes, for different
values of $\ell$ and $\sigma$. Changing $\ell$ results in small
changes in the amplitude of the oscillations, while the variation of
$\sigma$ has a more significant impact. Smaller values of $\sigma/\Delta$
result in more oscillations, while larger values better match
the reference solution. From this parameter study we conclude that
$\sigma/\Delta=3$ is fairly a robust choice for this shock tube problem.
Further, we found that $\sigma/\Delta$ can be further reduced closer to $\sim 1.5$
on higher resolution runs and in problems with stronger shocks.

In Fig.~\ref{fig:shu-weno} we show the combination of the GP-R2 interpolation
with the classical WENO smoothness indicators. The FWHM of the post shock
oscillations is $\sim 3-4$ times the grid spacing $\Delta$. 
This suggests that, following the FWHM discussion in Section~\ref{sec:accuracy-results},
a choice of $\ell/\Delta = 3$ for GP-R2 is optimal. This is confirmed in 
Fig.~\ref{fig:shu-weno}, where the solution in panel (a) with $\ell/\Delta = 3$  
better captures the amplitude of the oscillations, when compared to WENO-JS and
the GP-R2 solution with $\ell/\Delta = 6$. Notwithstanding, the 
default combination of GP-based smoothness indicators with GP-WENO yields
much better results overall (Fig.~\ref{subfig:shu-zoom}).

\subsection{1D Two Blast Wave Interaction}
\label{sec:1d-two-blast}
This problem was introduced by Woodward and Colella
\cite{woodward_numerical_1984} and consists of two strong shocks interacting with one another.
\begin{figure}[h!]
  \centering
  \subfloat[][]{
    \includegraphics[width=.45\textwidth]{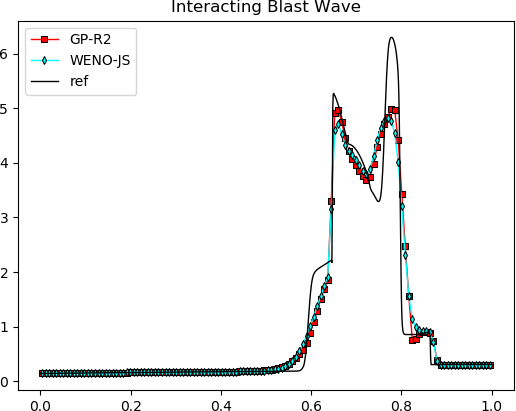}
    \label{subfig:2blast}
  }
  \subfloat[][]{
    \includegraphics[width=.469 \textwidth]{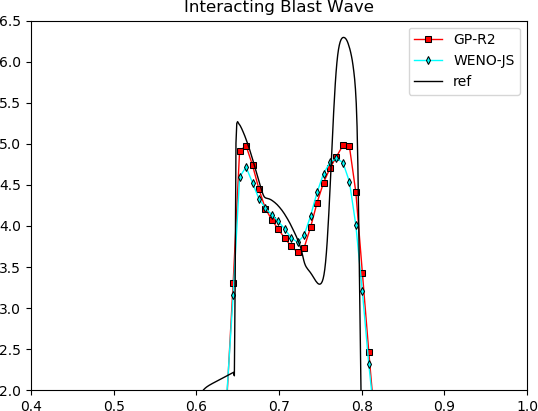}
    \label{subfig:2blast-zoom}
    }
  \caption{ (a) Two blast interaction problem showing GP-R2 (red) with
    $\ell/\Delta = 12$ and $\sigma/\Delta=3$ and WENO-JS (cyan) on 128 points using a CFL of
    0.8. (b) Zoom-in of the shock-interaction region. }
  \label{fig:2blast}
\end{figure}
We follow the
original setup of the problem and compare the GP-R2 scheme with the
5th order WENO-JS schemes on a computational domain of $[0,1]$,
resolved onto 128 grid points. 
We set $\ell/\Delta=12$ and $\sigma/\Delta=3$.
Fig.~\ref{fig:2blast} shows the density profiles for GP-R2 and WENO-JS 
at $t=0.038$, compared against a highly-resolved WENO-JS solution (2056 grid points). 
As shown in Fig.~\ref{subfig:2blast}, both methods yield acceptable solutions. 
However, the close-ups in Fig.~\ref{subfig:2blast-zoom} reveals that the
GP-R2 scheme better resolves the peaks and is closer to
the reference solution.

\subsection{2D Sedov}
\label{sec:2d_sedov}

Next, we consider the Sedov blast test \cite{sedov1993similarity}.
This problem studies the methods ability to maintain the symmetry of the
self-similar evolution of a spherical shock, generated by a high
pressure point-source at the center of the $[-0.5, 0.5] \times [-0.5,0.5]$ domain.
We follow the setup of \cite{fryxell2000flash}. 
\begin{figure}[h!]
  \centering
  \subfloat[]{
    \includegraphics[width=.4\textwidth]{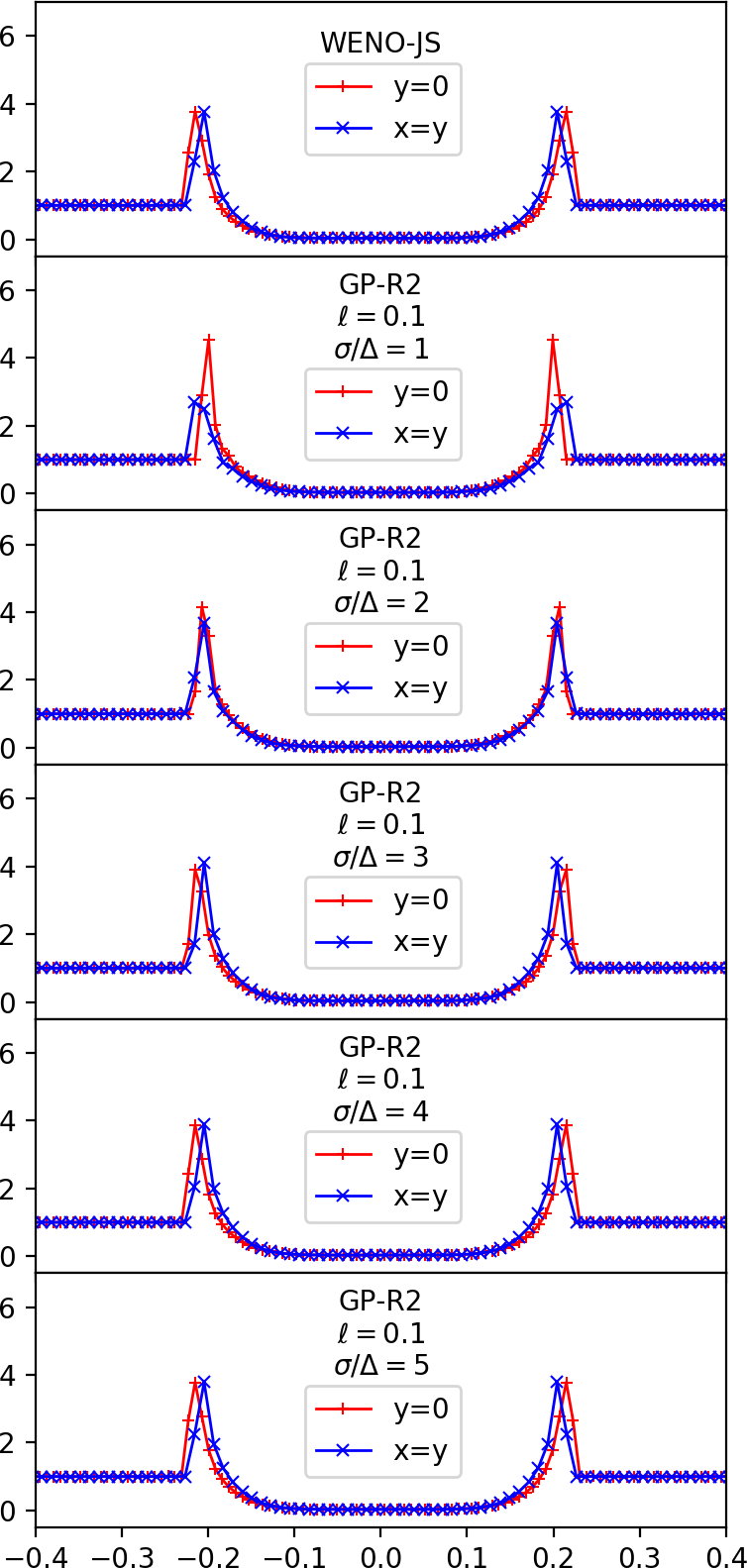}
    \label{subfig:sedov-128}
  }
  \subfloat[]{
    \includegraphics[width=.4\textwidth]{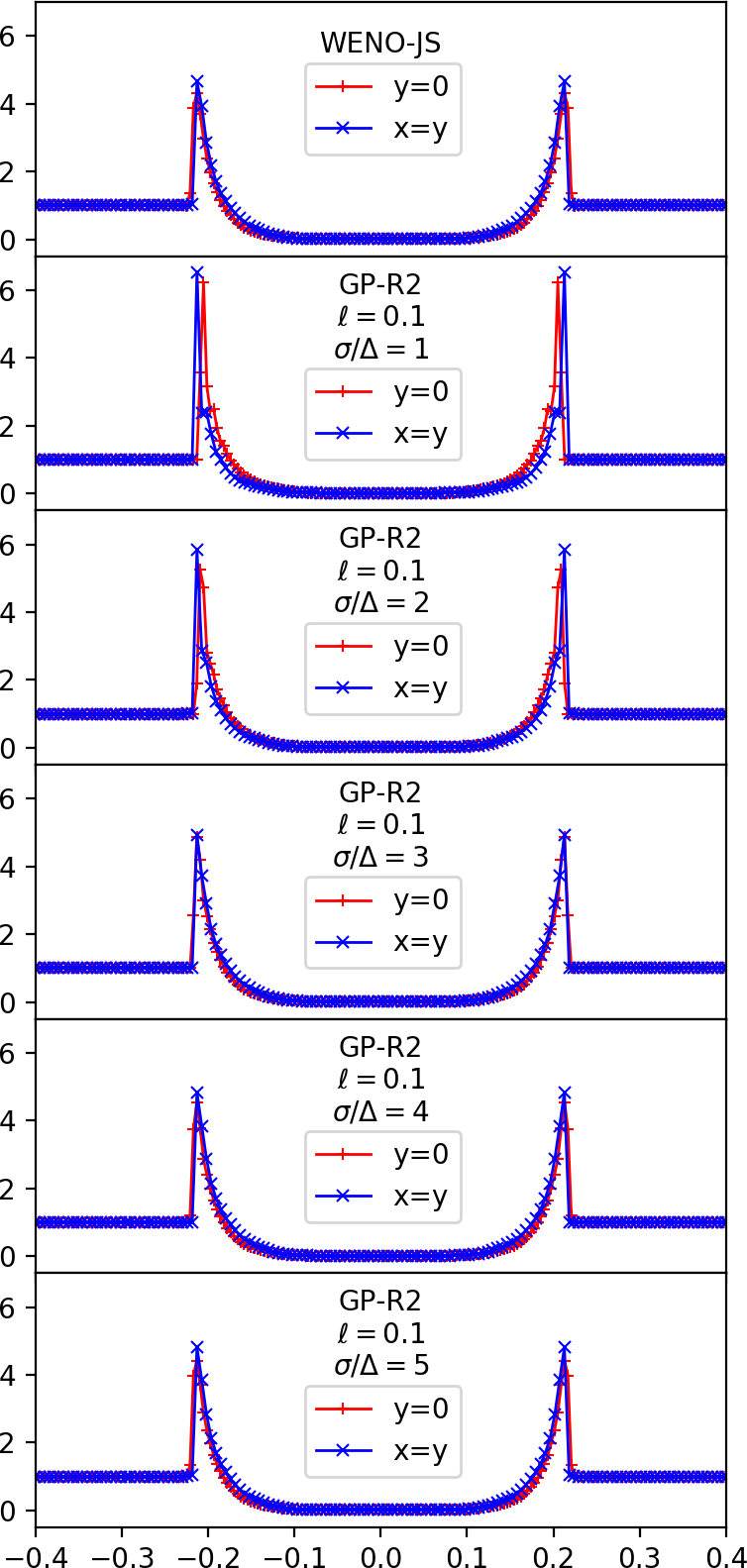}
    \label{subfig:sedov-256}
  }
  \caption{Comparison of $\sigma/\Delta$ values on two resolutions,
    $128\times128$ (\ref{subfig:sedov-128}) and $256\times256$ (\ref{subfig:sedov-256}),
    using GP-R2 for the Sedov problem. Shown are the density profiles
    along the $x$-axis (red) and along the diagonal $y=x$ (blue). }
  \label{fig:sedov}
\end{figure}

Fig.~\ref{fig:sedov} shows density profiles along 
$y=x$ (i.e., the diagonal) and $y=0$ (i.e., the $x$-axis) with 
GP-R2, at two different resolutions ($128\times 128$ and $256\times 256$) and different  
choices of $\sigma/\Delta$. All runs used a value of $\ell=0.1$ to perform 
a parameter scan on $\sigma/\Delta$.
The top two panels show the solutions obtained with WENO-JS. Again, as in the
Shu-Osher problem, small values of $\sigma/\Delta$ introduce oscillations
and lead to asymmetric shock evolution, whereas a choice of $\sigma/\Delta=3$
gives a good balance.

\subsection{2D Mach 3 Wind Tunnel with a Step}
\label{sec:mach-3-windtunnel}
The next 2D shock problem
consists of a Mach 3 wind tunnel setup with a forward facing
reflecting step, originally 
proposed by Woodward and Colella \cite{woodward_numerical_1984}.
We initialize the problem as in \cite{woodward_numerical_1984} with an
entropy fix for the region immediately above the corner of the step.
After the bow shock reflects on the step, the shock progressively reaches
the top reflecting wall of the tunnel at around $t=0.65$.
A triple point is formed due to the reflections and interactions of the shocks,
from which a trail of vortices travels towards the right boundary.
\begin{figure}[h!]
  \centering
  \subfloat[]{
    \includegraphics[width=4.5in, trim= 0.5in 1.8in 0.4in 1.8in,clip=true]
    {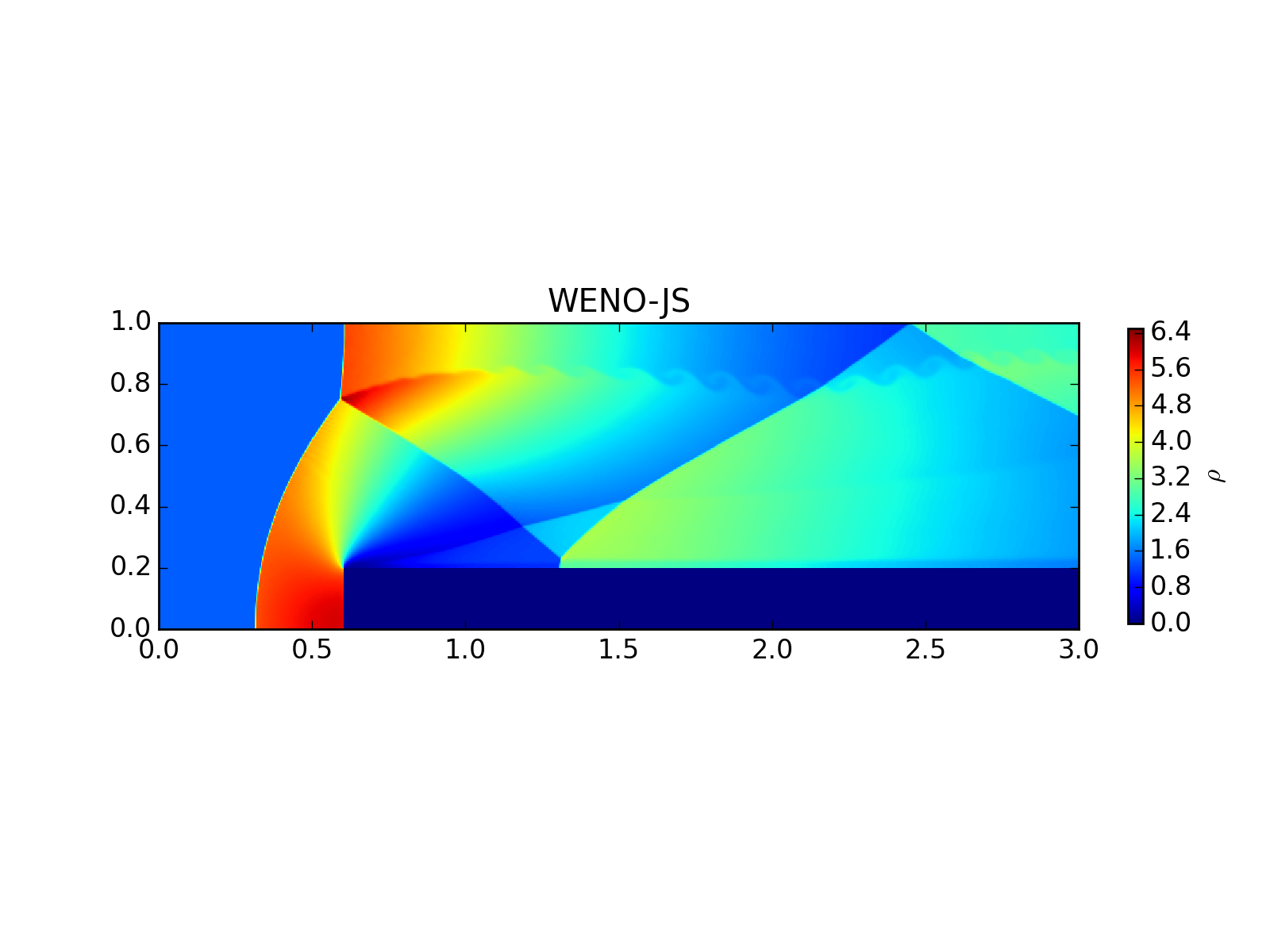} 
    \label{subfig:wind-gp}
  } \\
  \subfloat[]{
    \includegraphics[width=4.5in, trim= 0.5in 1.8in 0.4in 1.8in,clip=true]
    {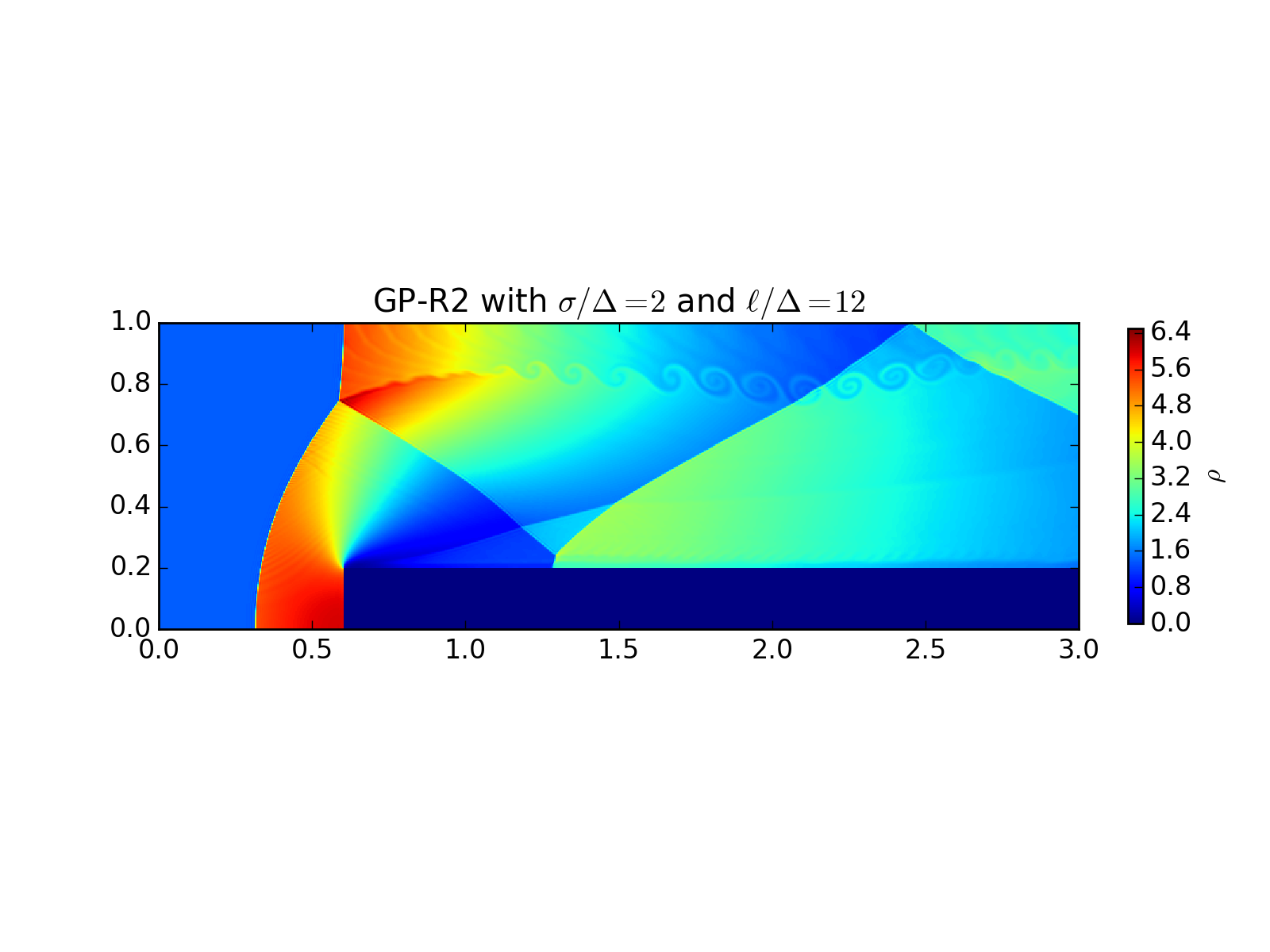}
    \label{subfig:wind-gpr2}
  } \\
  \subfloat[]{
    \includegraphics[width=4.5in, trim= 0.5in 1.8in 0.4in 1.8in,clip=true]
    {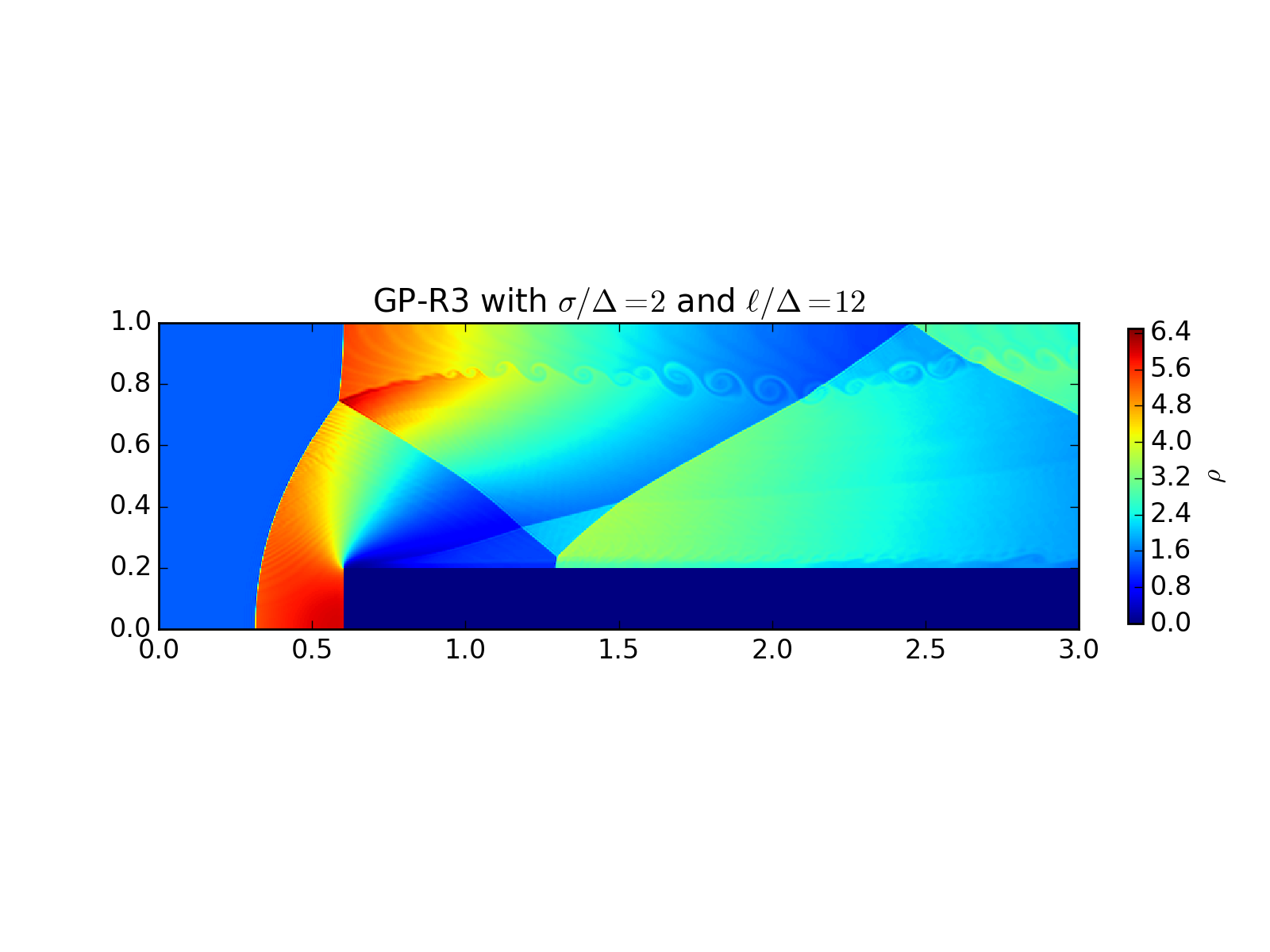}
    \label{subfig:wind-gpr3}
  } 
  \caption{Three density profiles for the Mach 3 wind tunnel problem with a step are shown:
  (a) the WENO-JS scheme 
  (b) the GP-R2 scheme;
  (c) the GP-R3 scheme. 
   The domain is resolved onto a $768\times 256$ grid. We use an HLL Riemann solver and RK3,
   with a CFL of $0.4$. The temporal evolution is followed up to $t=4$.  
   The two GP runs in (b) and (c) used $\sigma/\Delta=2$ and $\ell/\Delta=12$.      
    }
  \label{fig:windtunnel}
\end{figure}

Shown in Fig.~\ref{fig:windtunnel} are the results computed using the
GP-R2 and GP-R3 schemes, along with the WENO-JS solution on a $768\times256$ grid
at the final time $t=4$. Using the HLLC Riemann solver, we noticed that
the GP and WENO-JS schemes converged to different solutions, on account of the 
singularity at the corner of the step and despite the entropy fix.
To compare the two schemes, we ran our simulations using
an HLL Riemann solver, for which the two schemes
converged to similar solutions. Both methods are able to capture the
main features of the flow but the GP schemes produce more well-developed
Kelvin-Helmholtz roll-ups that originate from the triple point.

\subsection{2D Riemann Problem -- Configuration 3}
\label{subsec:2drp}
Next, we consider a specific Riemann Problem configuration 
that is drawn from a class of two dimensional
Riemann problems that have been studied extensively 
\cite{zhang1990conjecture,schulz-rinne_classification_1993}
and have been applied to code verification efforts
\cite{buchmuller2014improved,balsara2010,lax1998solution,schulz-rinne_numerical_1993, don_hybrid_2016,lee2017piecewise}.
Specifically, we look at the third configuration of the 2D Riemann problem presented
in \cite{don_hybrid_2016,lee2017piecewise}.

\begin{figure}[h!]
  \centering
    \includegraphics[width=2.3in, trim= 0.8in 0.0in 0.8in 0.2in,clip=true]
    {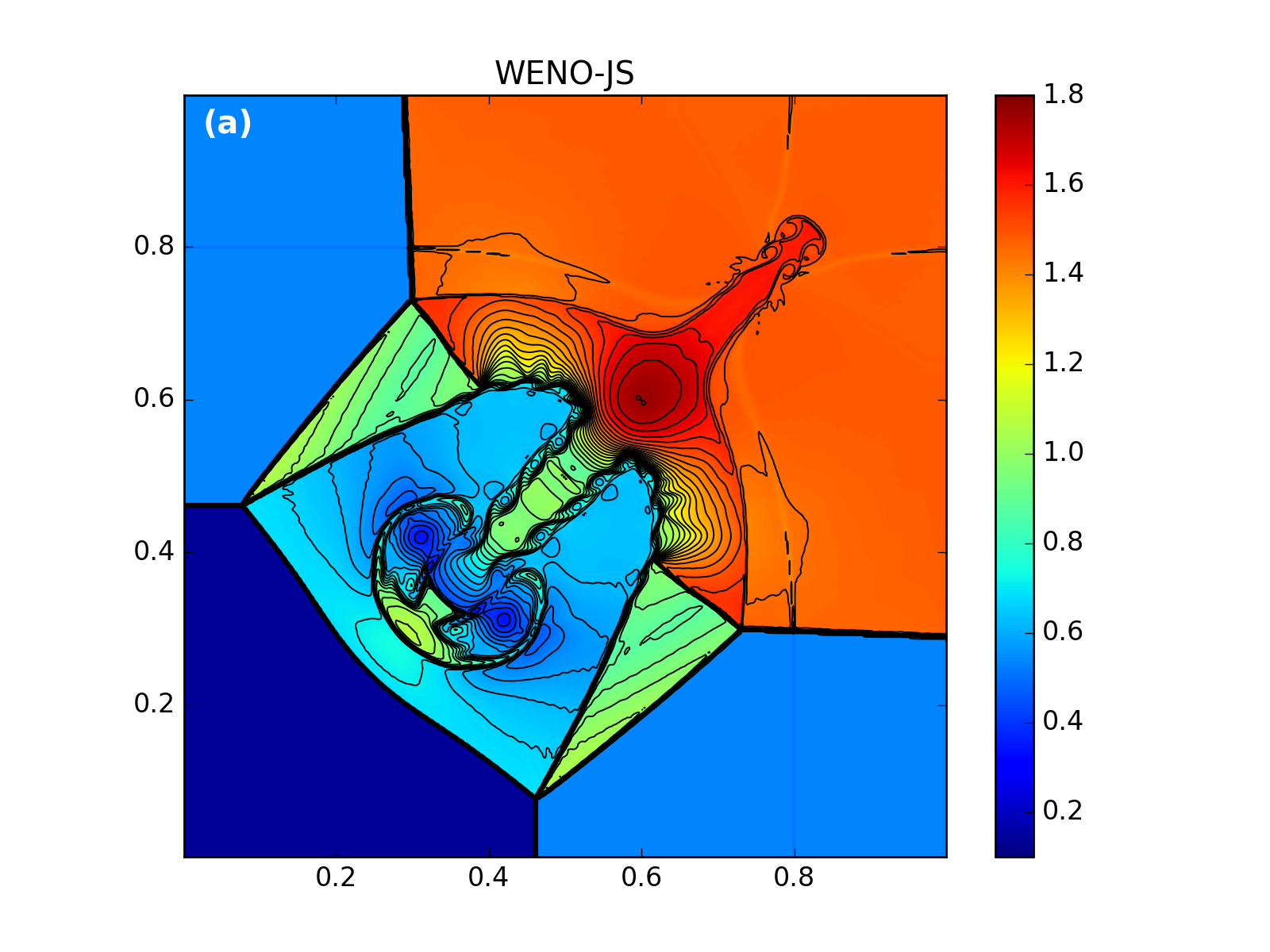}
    \label{subfig:2drp-weno}
    \includegraphics[width=2.3in, trim= 0.8in 0.0in 0.8in 0.2in,clip=true]
    {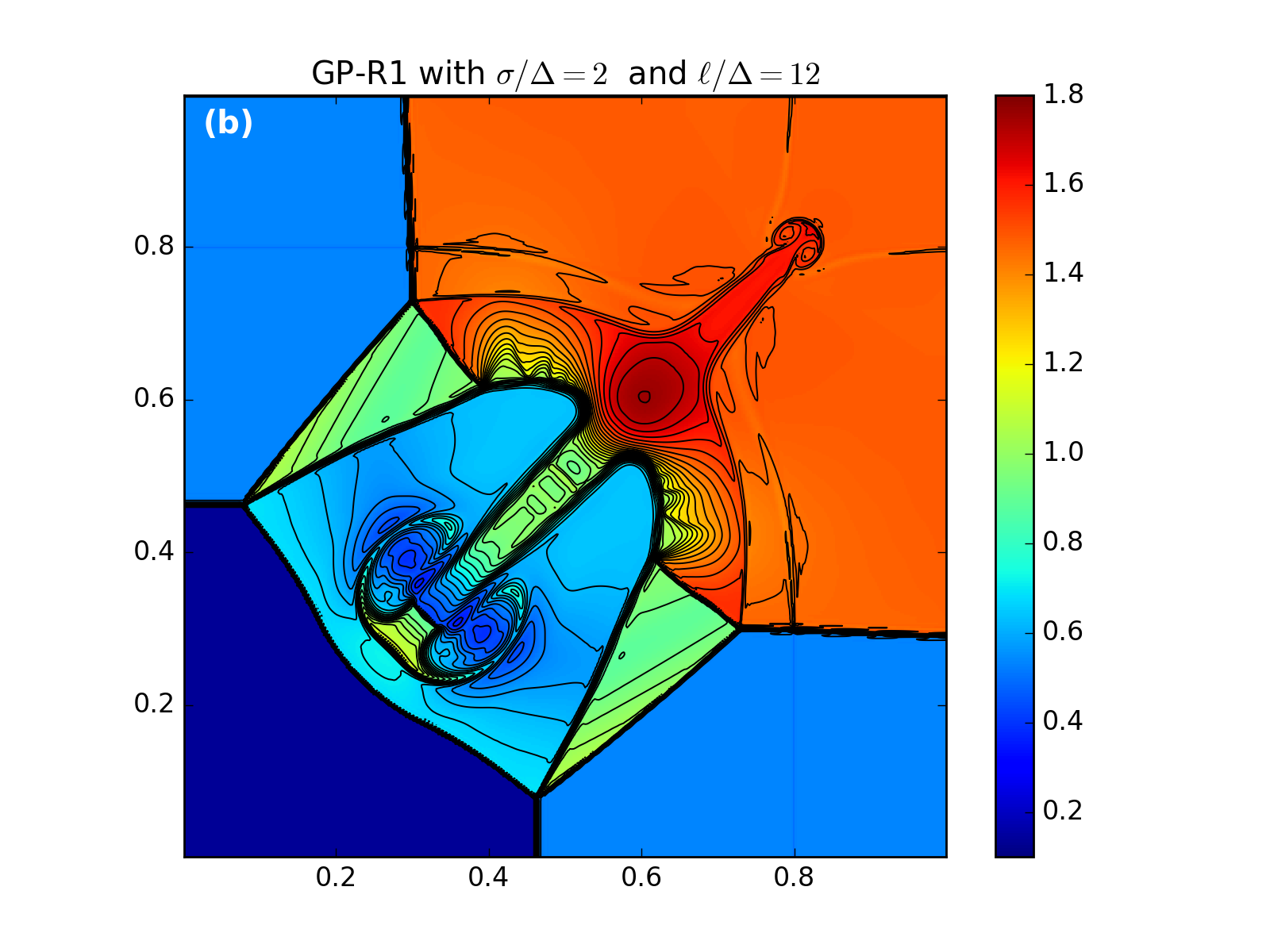}
    \label{subfig:2drp-gpr1}
    
    \includegraphics[width=2.3in, trim= 0.8in 0.0in 0.8in 0.2in,clip=true]
    {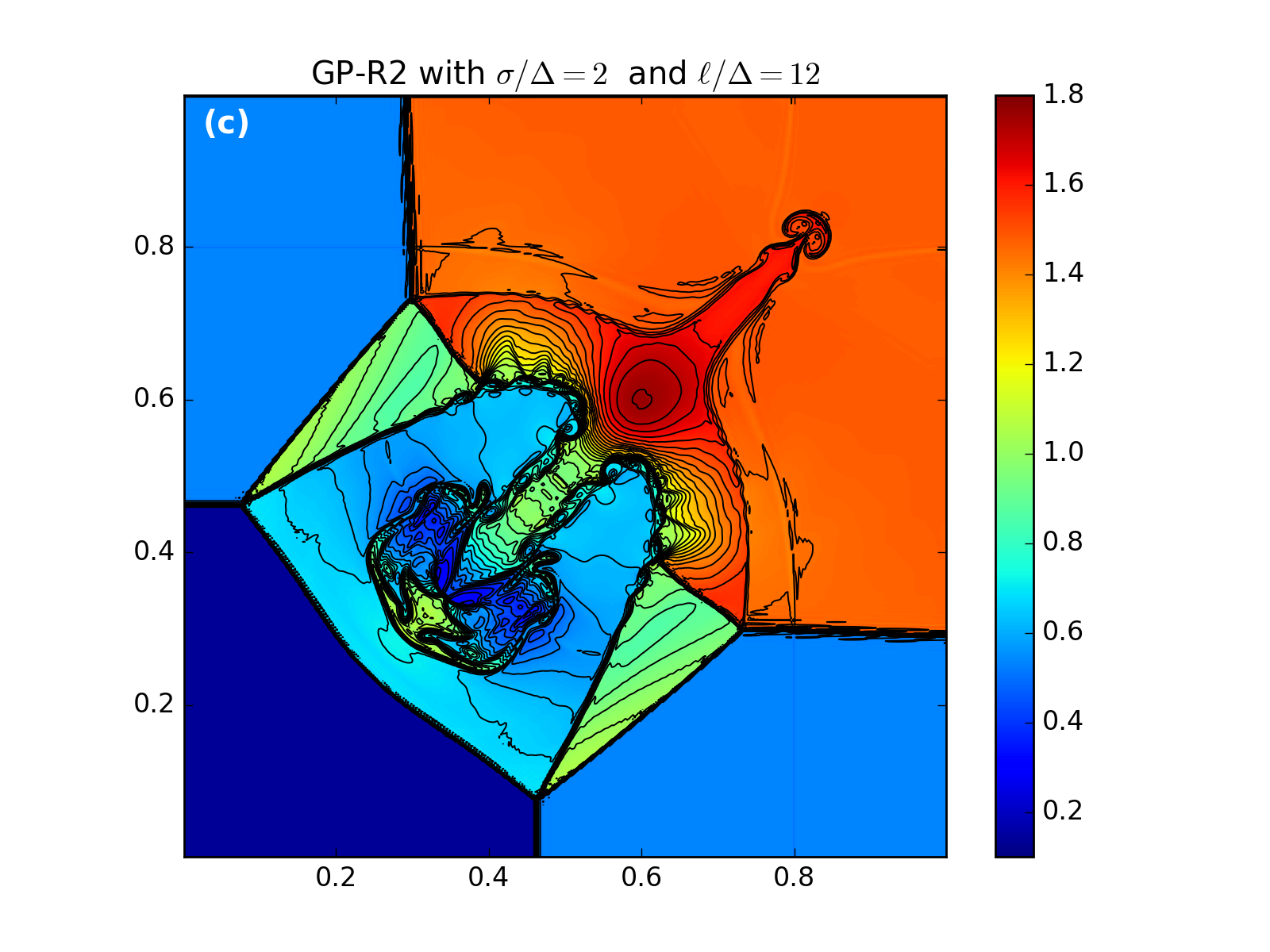}
    \label{subfig:2drp-gpr2}
   %
    \includegraphics[width=2.3in, trim= 0.8in 0.0in 0.8in 0.2in,clip=true]
    {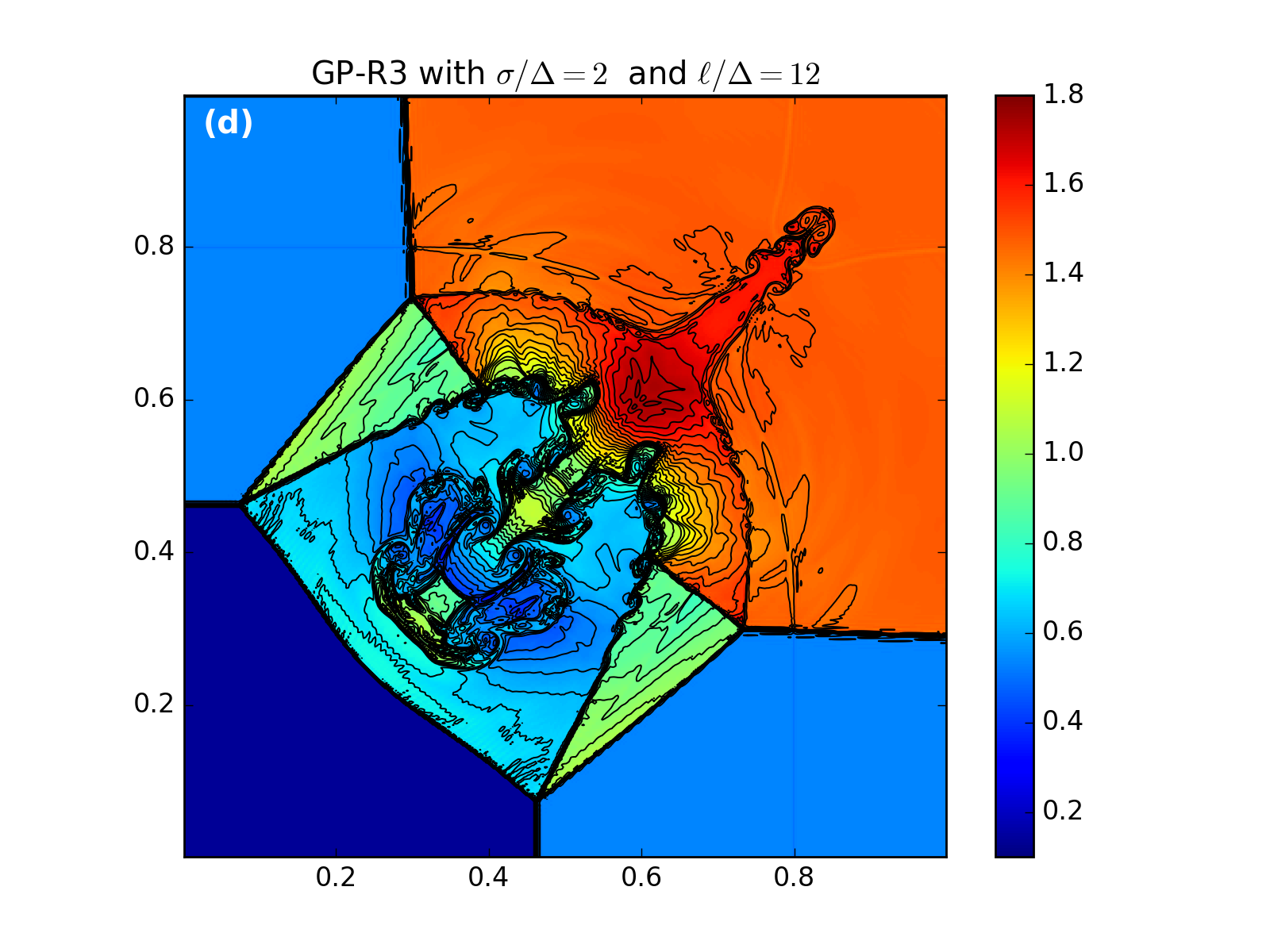}
    \label{subfig:2drp-gpr3}
    
    \includegraphics[width=2.3in, trim= 0.8in 0.0in 0.8in 0.2in,clip=true]
    {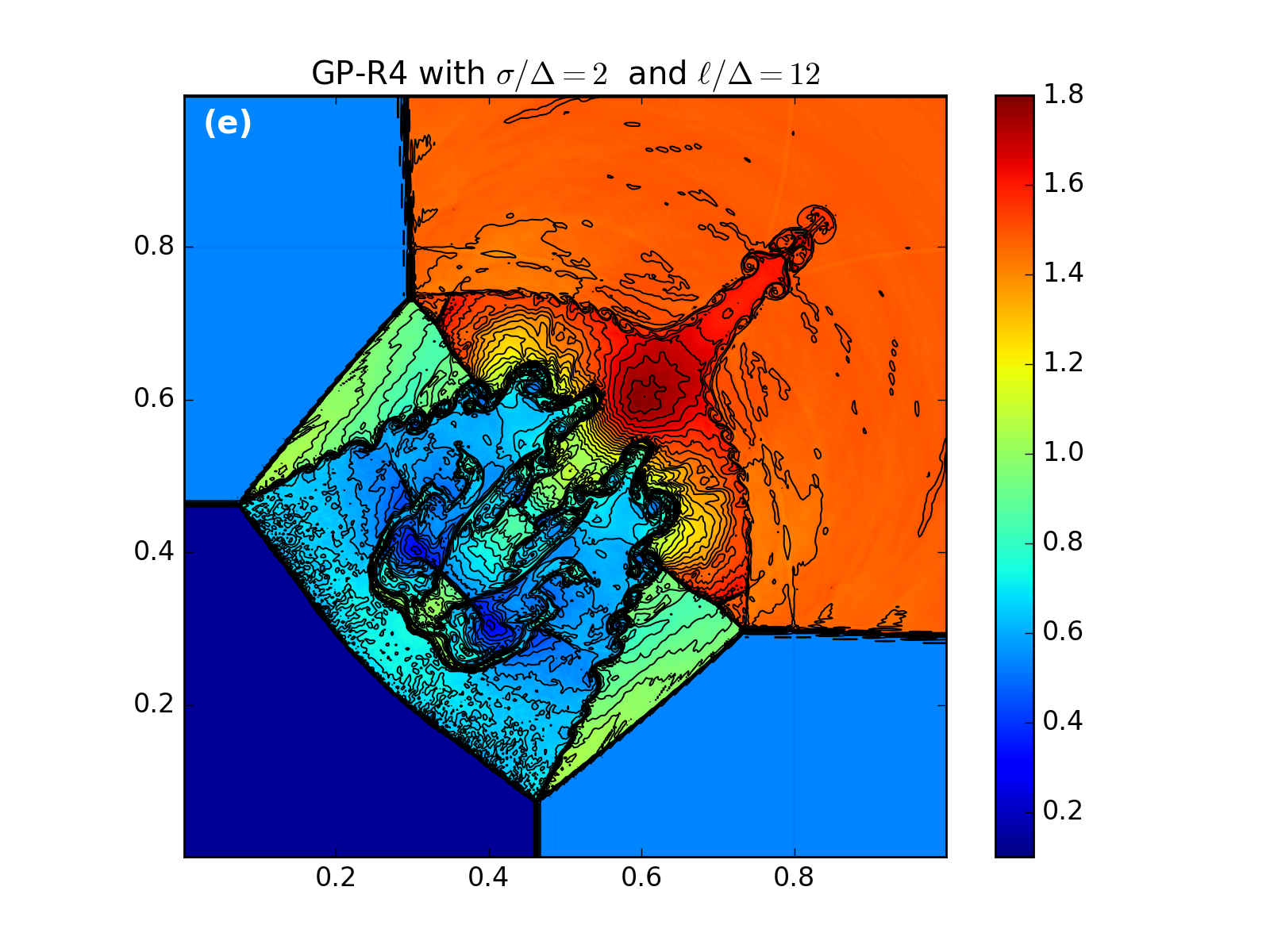}
    \label{subfig:2drp-gpr4}
    \includegraphics[width=2.3in, trim= 0.8in 0.0in 0.8in 0.2in,clip=true]
    {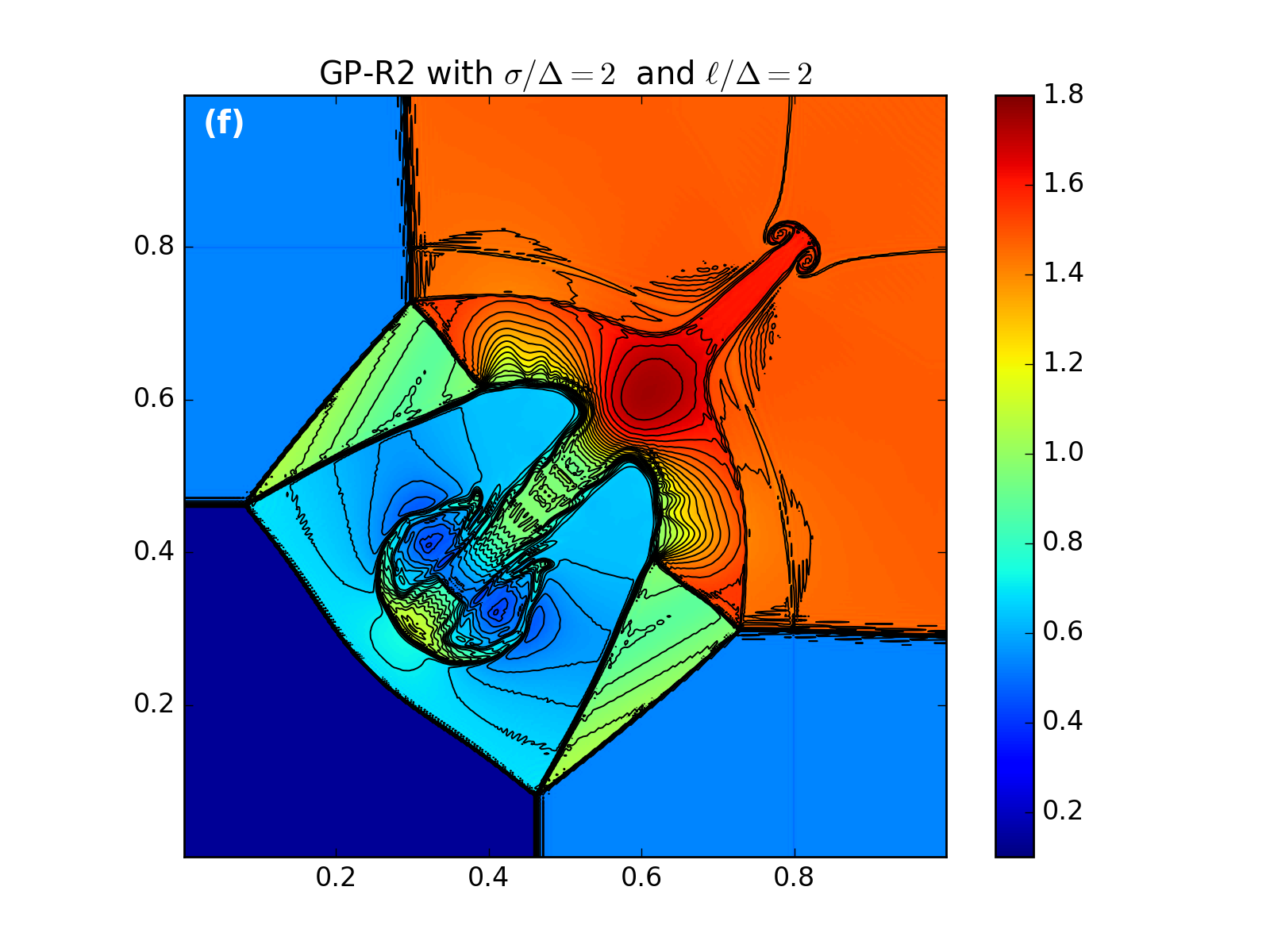}
    \label{subfig:2drp-gpr2_sig_eq_ell}
  \caption{Configuration 3 of the 2D Riemann problem using four different methods:
      (a) WENO-JS, (b) GP-R1, (c) GP-R2, (d) GP-R3, (e) GP-R4, 
      where (b) -- (e) have $\sigma/\Delta \ne \ell/\Delta$,
      and finally,
      (f) GP-R2 with $\sigma/\Delta = \ell/\Delta$. 
      Each panel shows the density values at $t=0.8$ between $[0.1, 1.8]$ in linear scale,
      computed on a $400\times 400$ grid. We over-plot 40 contour lines.
      All GP calculations used $\ell/\Delta=12$ except for (f). 
      An HLLC Riemann solver and RK3 were employed in all calculations, with CFL=0.4.}
  \label{fig:2DRP}
\end{figure}

Panels in Fig.~\ref{fig:2DRP} show density
profiles at $t=0.8$, along with 40 contour lines, for different choices
of GP radii, $R=1,2,3,4$ on a $400\times 400$ grid resolution. 
All GP methods correctly capture the expected features of the problem. 
In this experiment, we see that the increase of $R$ results in a
sharper representation of the flow features. 
Notably, the 5th-order GP-R2 solution in (c) captures significantly more features
when compared to the 5th-order WENO-JS in (a), as evinced by the formation of
more developed Kelvin-Helmholtz vortices along the slip lines (i.e., the
interface boundaries between the green triangular regions and the sky blue
areas surrounding the mushroom-shaped jet).

\subsection{Double Mach Reflection}
\label{sec:double-mach-refl}

\begin{sidewaysfigure}
  \centering 
    \includegraphics[width=3.7in]
    {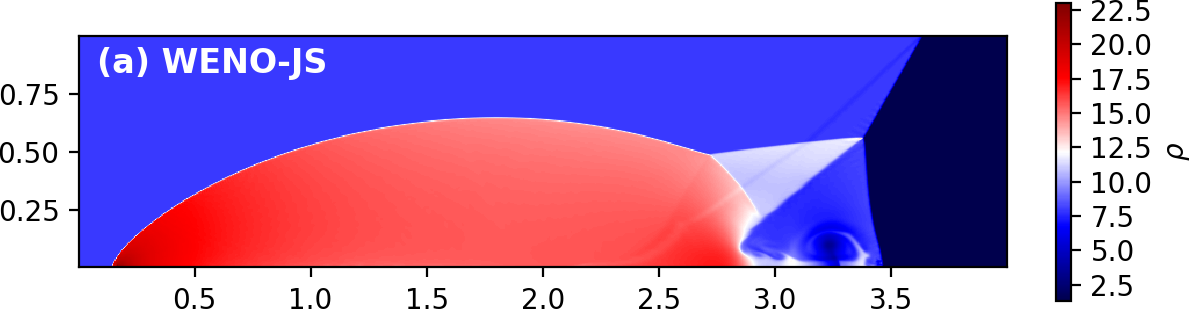}
    \includegraphics[width=3.7in]
    {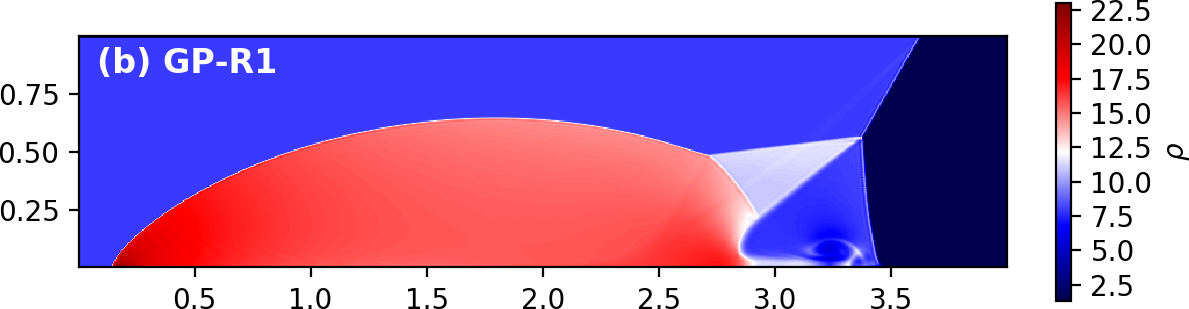}
    \includegraphics[width=3.7in]
    {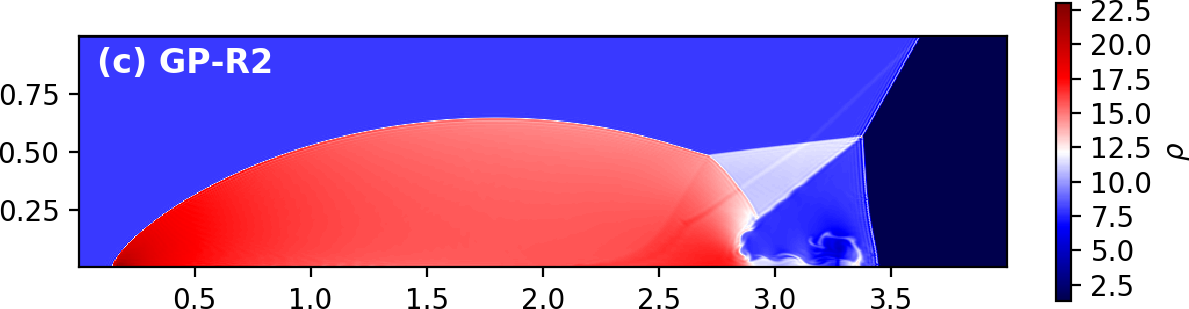}
    \includegraphics[width=3.7in]
    {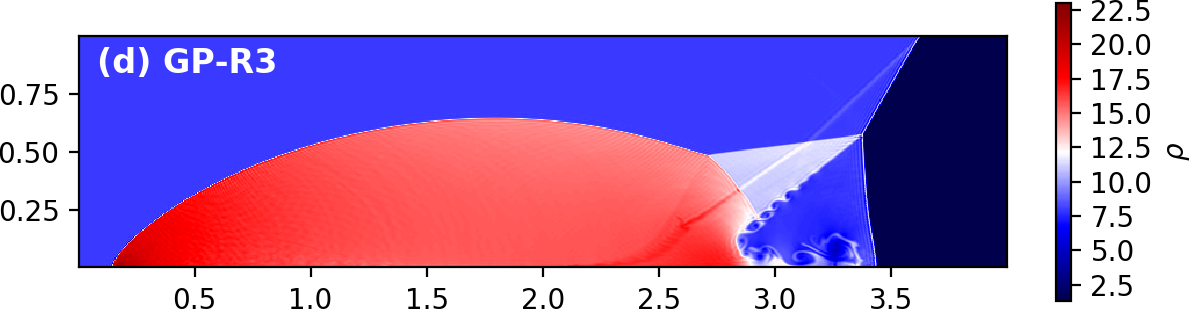}
    \includegraphics[width=3.7in]
    {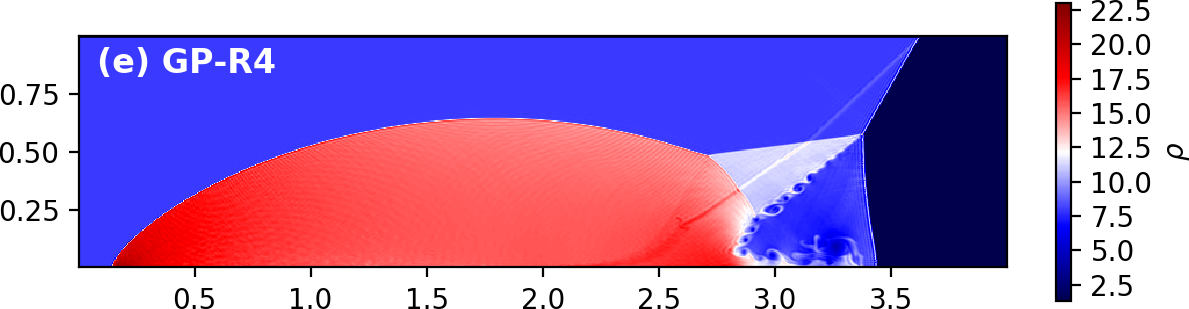}
    \includegraphics[width=3.7in]
    {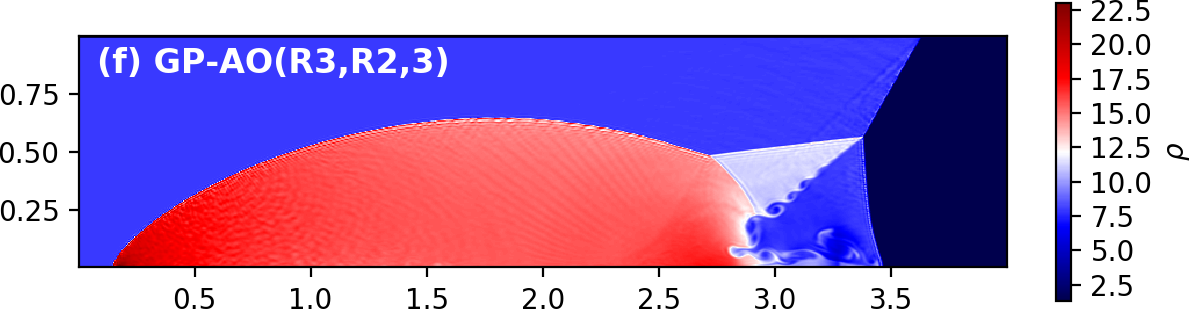}
\caption{The double Mach reflection (DMR) problem solved using six different methods,
 (a) WENO-JS, (b) GP-R1, (c) GP-R2, (d) GP-R3, (e) GP-R4, and (f) GP-AO($R3$,$R2$,3).
 We show density plots at $t=0.25$.
The calculations were performed on a $800\times 200$ grid, with $\ell/\Delta=12$ and $\sigma/\Delta=1.8$. 
For all cases we used the HLLC Riemann solver, RK3, and CFL=0.4.
}
  \label{fig:DMR}
\end{sidewaysfigure}

\begin{figure}[h!]
  \centering    
  \includegraphics[width=2.1in]
  {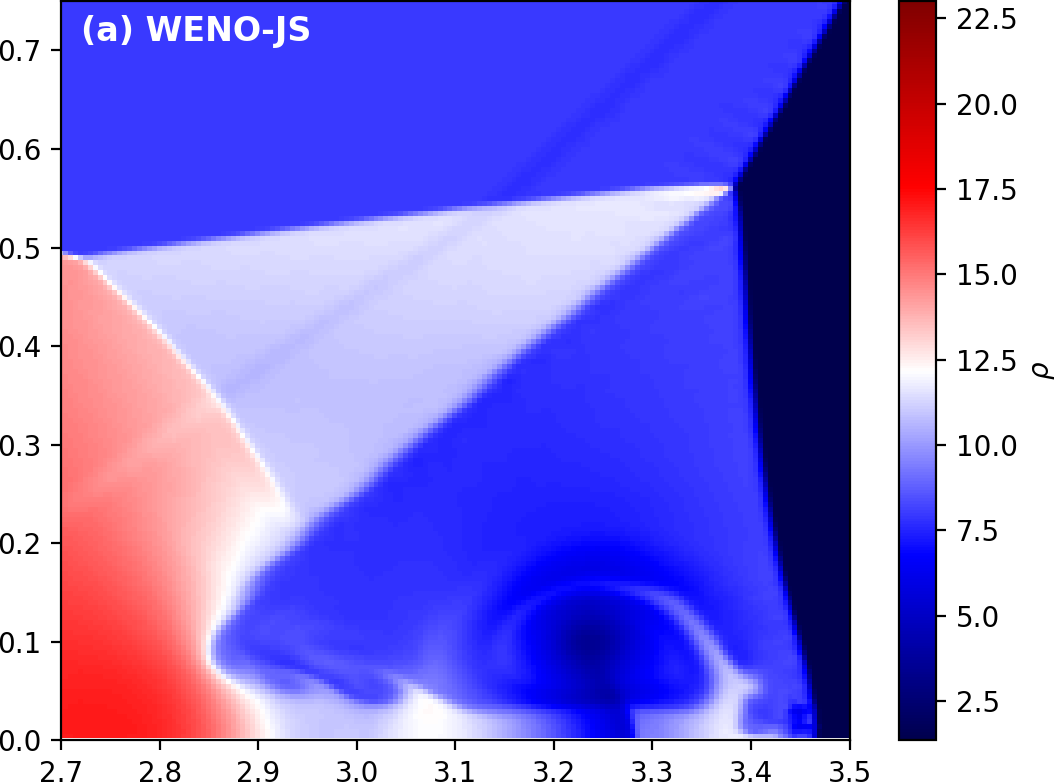} 
  \includegraphics[width=2.1in]
  {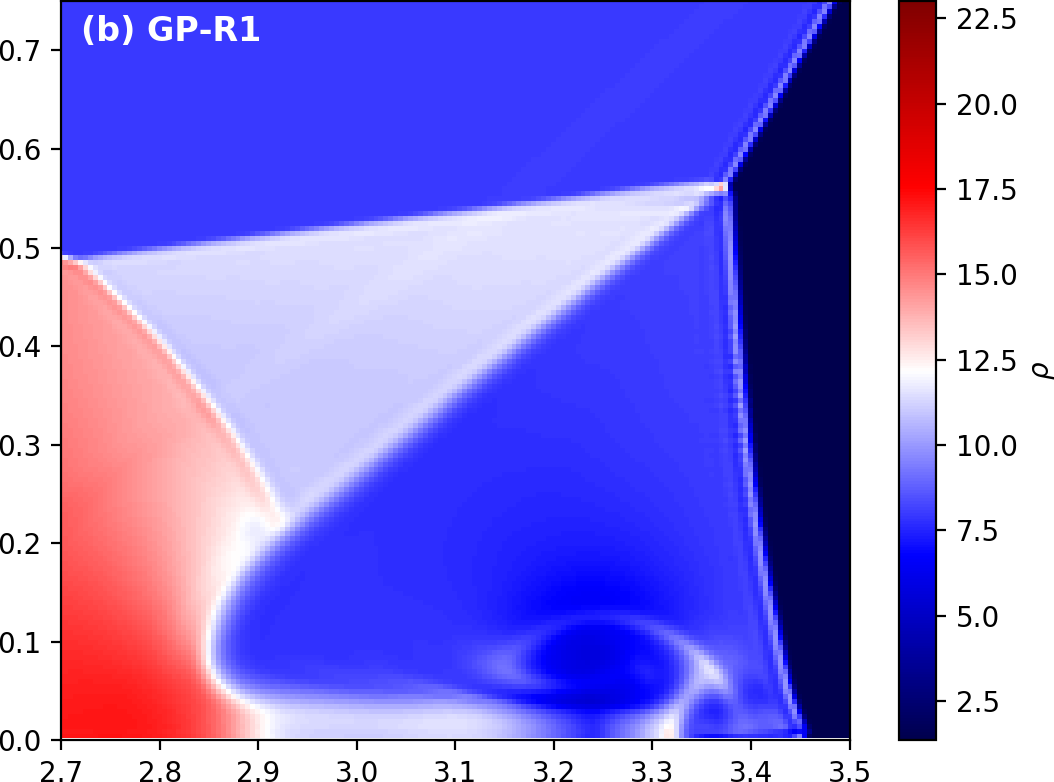} \\ \vspace{0.2in}
  \includegraphics[width=2.1in]
  {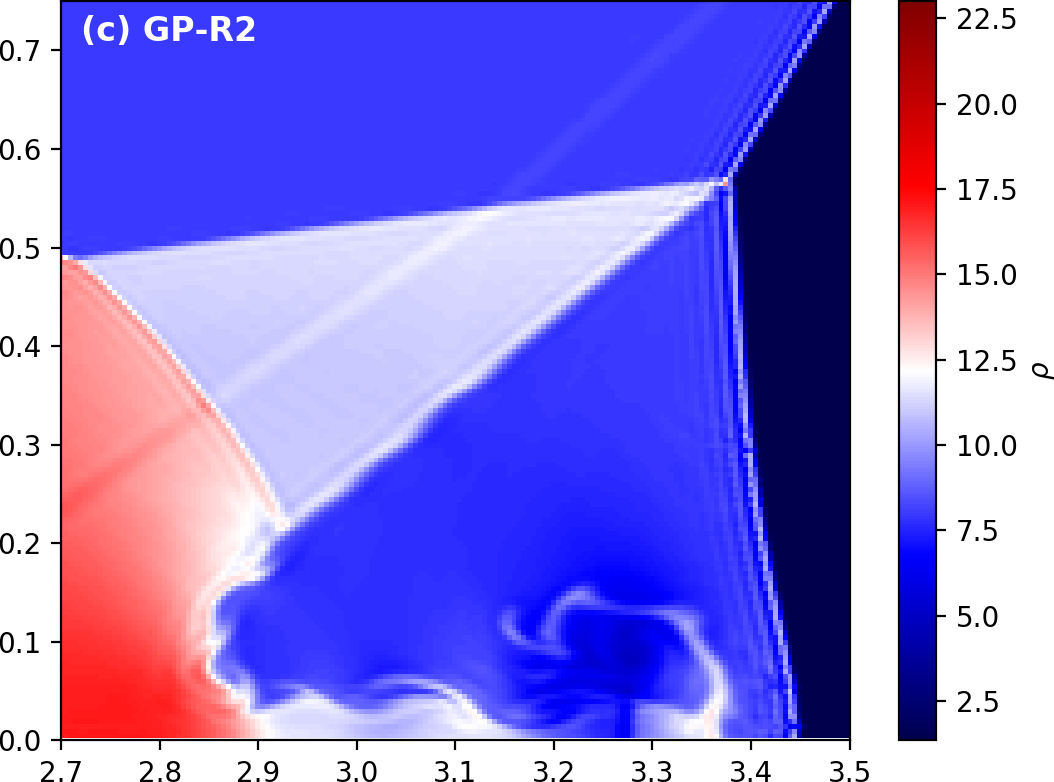} 
  \includegraphics[width=2.1in]
  {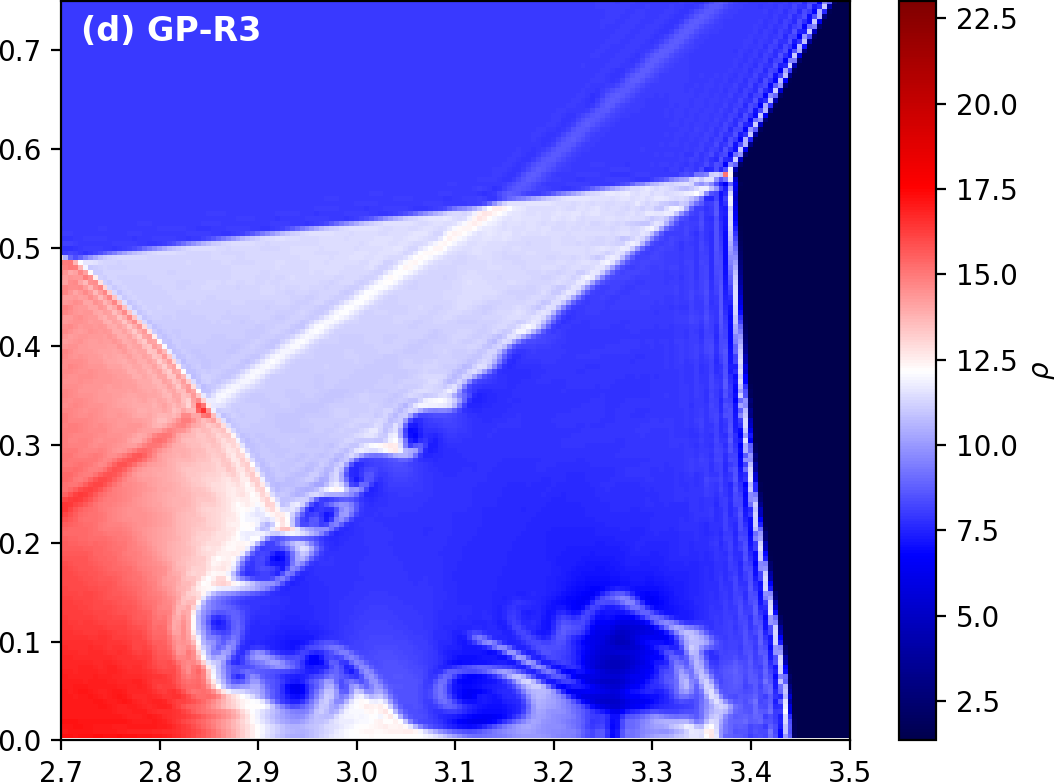} \\ \vspace{0.2in} 
  \includegraphics[width=2.1in]
  {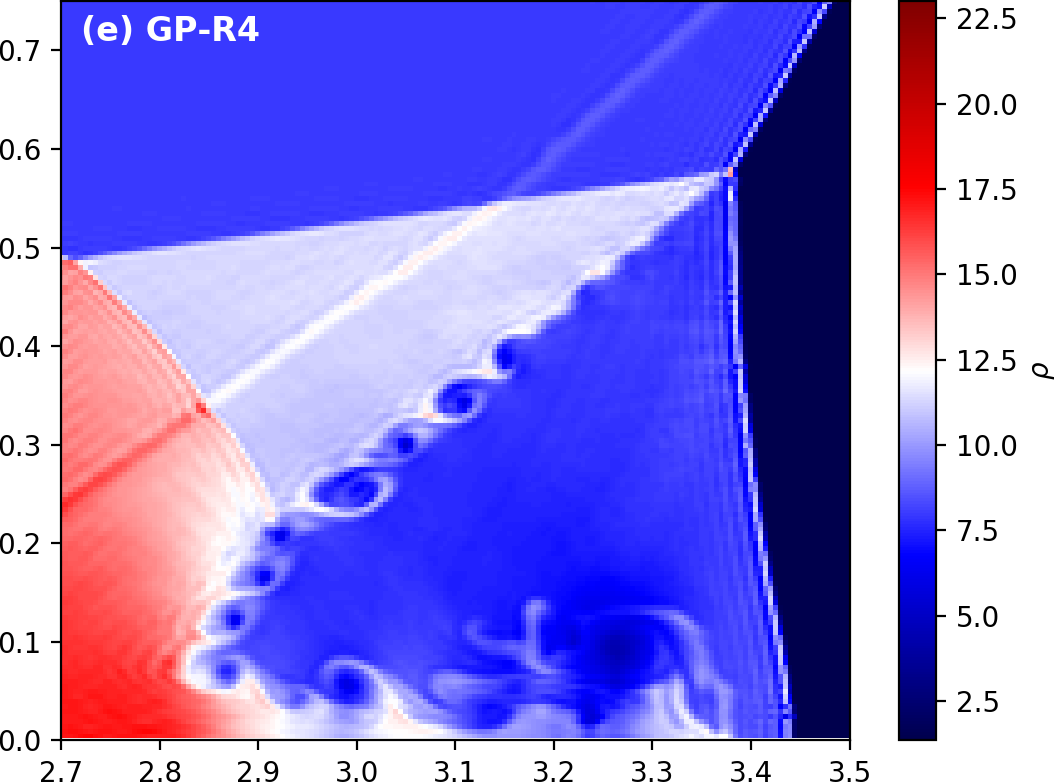} 
  \includegraphics[width=2.1in]
  {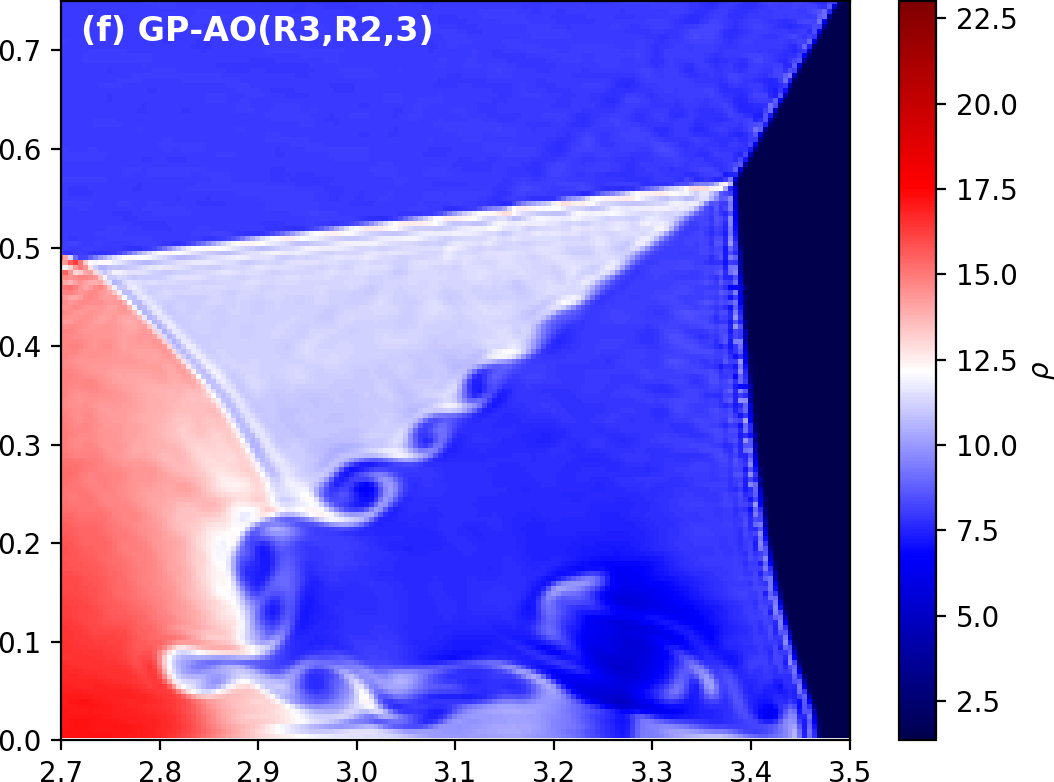}
  \caption{Close-ups near the triple point of the density profiles of Fig. ~\ref{fig:DMR}. 
  }
  \label{fig:DMR_zoomed}
\end{figure}

\begin{sidewaysfigure}
  \centering 
    \includegraphics[width=3.7in]
    {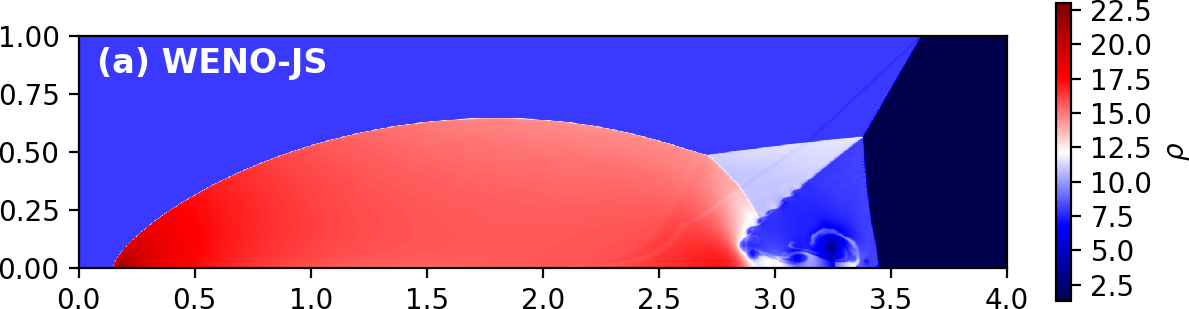}
    \includegraphics[width=3.7in]
    {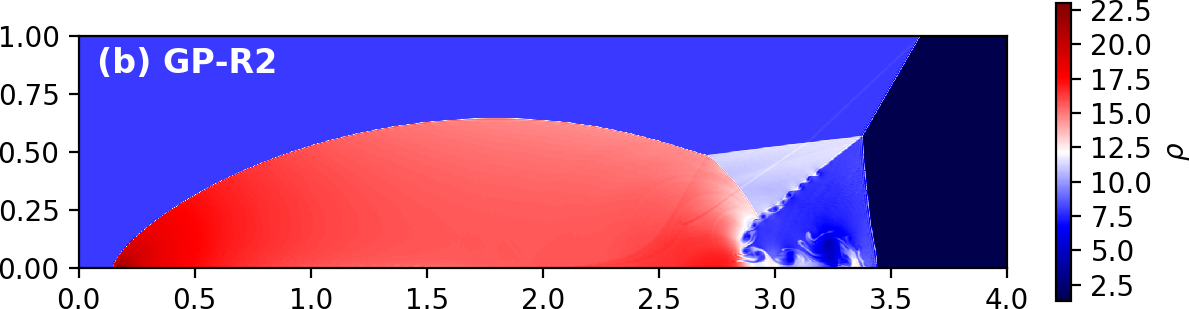}
    \includegraphics[width=3.7in]
    {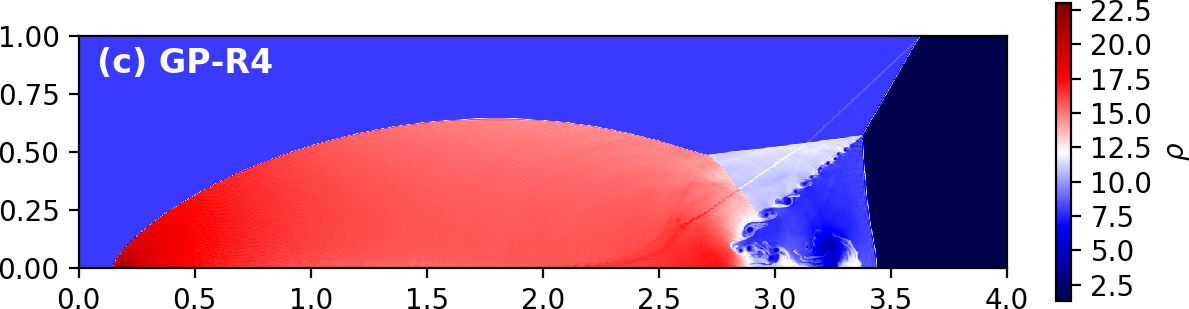}
    \includegraphics[width=3.7in]
    {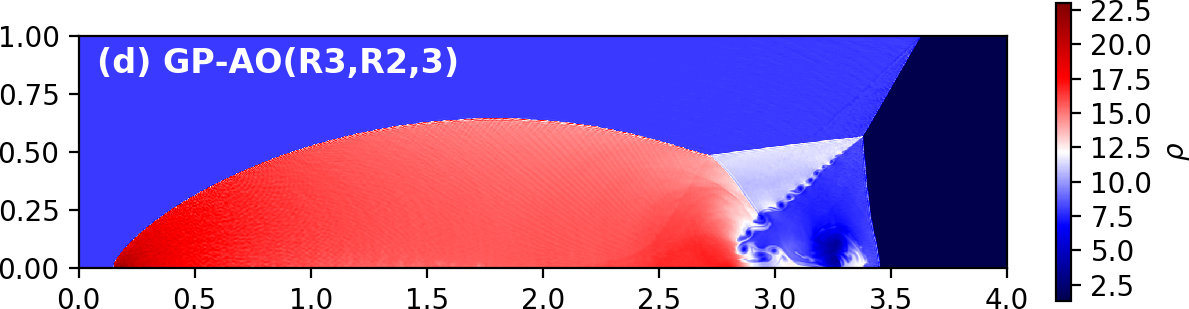}
    \\
    \vspace{0.5in}
    \includegraphics[width=1.82in]
    {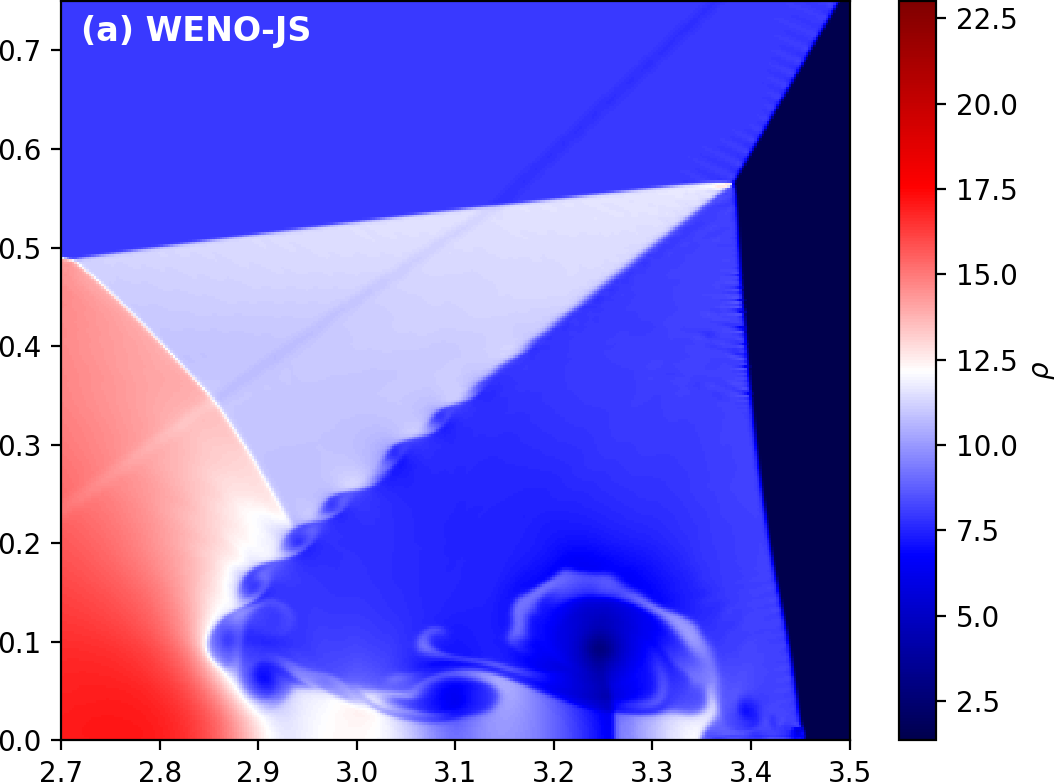}    
     \includegraphics[width=1.82in]    
    {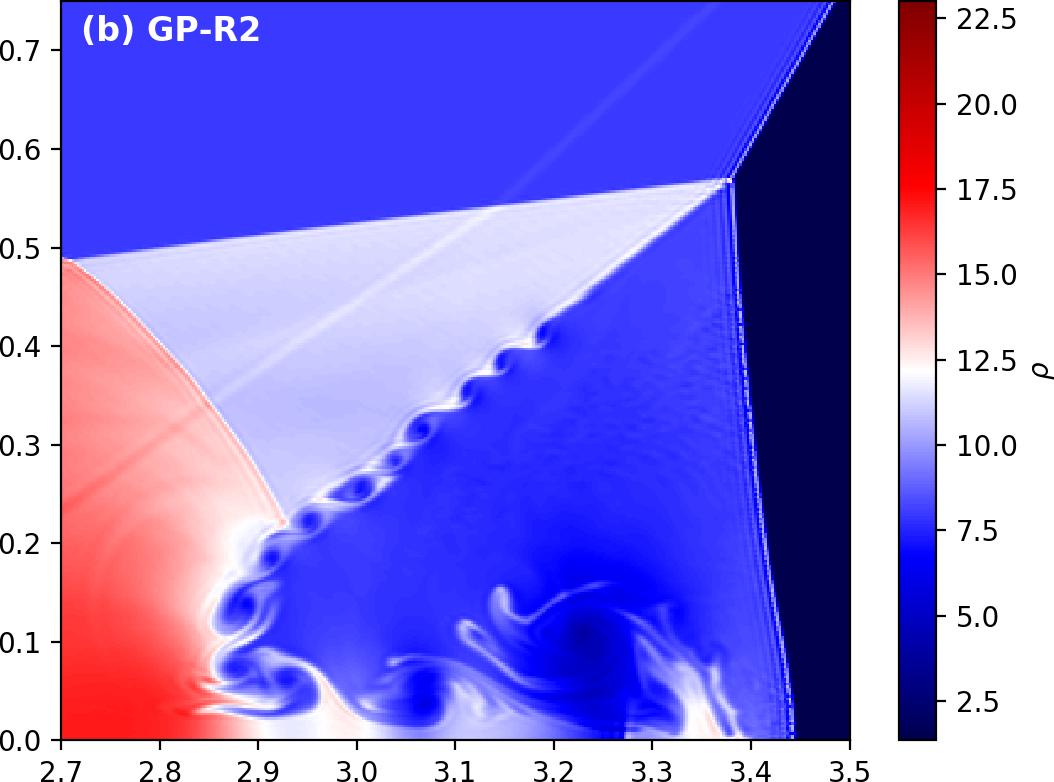} 
    \includegraphics[width=1.82in]    
    {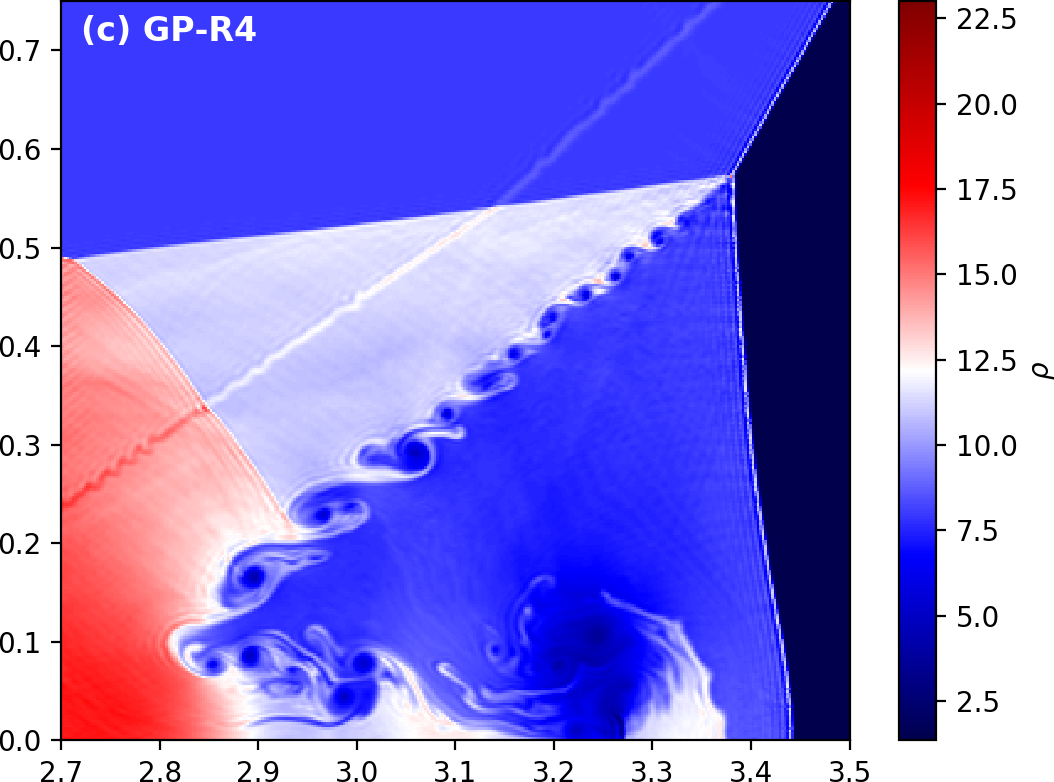}    
    \includegraphics[width=1.82in]    
    {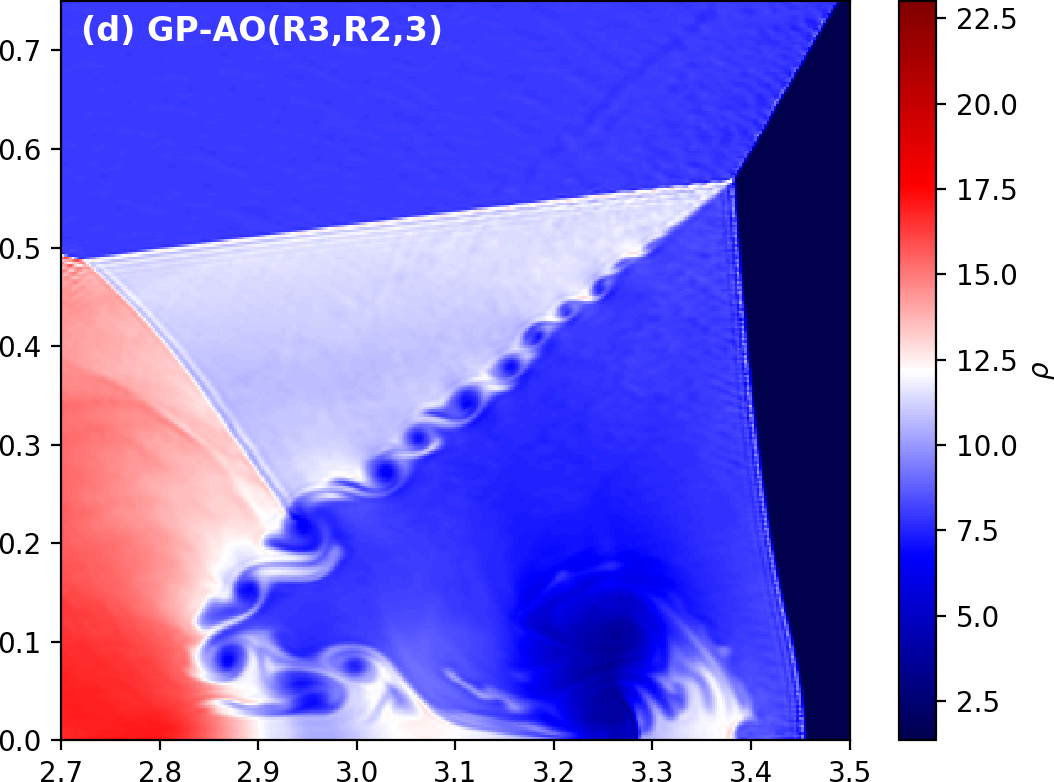}     
        
\caption{The density profiles of the DMR test performed on a high-resolution grid ($1600\times 400$). 
(a) WENO-JS, (b) GP-R2, and (c) GP-R4, and (d) GP-AO($R3$,$R2$,3), 
with $\sigma/\Delta=1.8$ and $\ell/\Delta=12$ for cases (b) to (d). 
The four bottom panels are close-ups of the top four panels near the triple point.
For all cases we used the HLLC Riemann solver, RK3, and CFL=0.4.}
\label{fig:DMR_highRes_zoomed_1600x400}
\end{sidewaysfigure}

Our last 2D test is the double Mach reflection problem introduced by Woodward and Colella
\cite{woodward_numerical_1984}. 
This test problem consists of a strong Mach 10 shock
that is incident on a reflecting wedge that forms a $60^\circ$ angle
with the plane of the shock. Fig.~\ref{fig:DMR} shows density profiles 
from $800\times 200$ grid resolution runs, for a variety of GP stencil radii ($R=1,2,3,4$),
as well as for the GP adaptive order (AO) hybridization \cite{balsara_efficient_2016} 
(detailed in \ref{sec:append-adapt-order}).
We present these GP solutions together with a 5th-order WENO-JS solution for comparison.
The reflection of the shock at the wall forms two Mach stems and two
contact discontinuities. The contact discontinuity that emanates from
the roll-up of the stronger shocks is known to be Kelvin-Helmholtz
unstable provided there is sufficiently low numerical dissipation in
the method. 
Thus, the problem quantifies the method's numerical dissipation by the number of
Kelvin-Helmholtz vortices present at the end of the run. 
%

The presence of a highly supersonic shock is a stringent test for the stability of the algorithm.
We find that the GP solutions remain stable for small values of $\sigma/\Delta$. 
The choice of $\ell/\Delta$ does not appear to considerably affect stability and thus
can be set to relatively larger values than $\sigma/\Delta$. 
All GP runs successfully reach $t=0.25$ for $\sigma/\Delta=1.8$ and $\ell/\Delta=12$
on two different resolutions, $800\times 200$ in Fig.~\ref{fig:DMR} and 
$1600\times 400$ in Fig.~\ref{fig:DMR_highRes_zoomed_1600x400}.

Close-ups in Fig.~\ref{fig:DMR_zoomed} reveal that 
GP-WENO is significantly better at capturing the development of Kelvin-Helmholtz vortices in the
Mach stem than the WENO-JS method. 
More specifically, the 5th order GP-R2 scheme is less dissipative than the WENO-JS scheme, which
in turn is less dissipative than the 3rd order GP-R1 scheme. 
Both 7th order GP-R3 and 9th order GP-R4 schemes are able to resolve more vortices.
While the GP-AO($R3,R2,3$) scheme is less dissipative than the GP-R2, it does not 
capture as many features as the GP-R3 scheme.  This is despite the fact that both GP-AO($R3,R2,3$) and GP-R3 
are of the same formal order of accuracy in smooth flows.

In Fig.~\ref{fig:DMR_highRes_zoomed_1600x400} we provide results for double the grid resolution, $1600\times400$.
The ranking derived from the lower resolution solutions still holds and the reduced dissipation of the
GP-R2 scheme over the 5th order WENO-JS is more evident. Further, the GP-R3 on the $800\times200$ grid
in Fig.~\ref{fig:DMR_zoomed}(d) captures more vortices than WENO-JS on the $1600\times 400$ grid
in Fig.~\ref{fig:DMR_highRes_zoomed_1600x400}(a).
Our results from Fig.~\ref{fig:DMR_highRes_zoomed_1600x400} can be directly compared to
one of the most recent finite difference WENO-AO solutions by Balsara \textit{et al.} \cite{balsara_efficient_2016}
(see their Fig. 7).

\subsection{3D Explosion}
\label{sec:3d-explosion}

This 3D explosion test problem was introduced by Boscheri and  Dumbser
\cite{noauthor_direct_2014} as a three-dimensional extension of
the Sod problem \cite{sod1978survey}. The test is set up on a $[-1,1]\times[-1,1]\times[-1,1]$
domain with outflow boundary conditions.  The initial condition is given by 
\begin{equation}
  \label{eq:3d-exp}
  \bV(\bfx) = \left \{
      \begin{array}{clrrl}
        \bV_\text{i} &= \; (1,0,0,0,1) & \text{if } ||\bfx|| \le R, \\
        \bV_\text{o} &= \; (0.125,0,0,0,0.1) & \text{if } ||\bfx|| > R,
      \end{array}
    \right .
  \end{equation}
  where $\bV=(\rho,u,v,w,P)$ and $R=0.5$. The ratio of specific heats is 
  $\gamma=1.4$ and the simulation completes at $t=0.25$.

\begin{figure}[h!]
  \centering
  \subfloat[][]{
    \includegraphics[width=.45\textwidth]{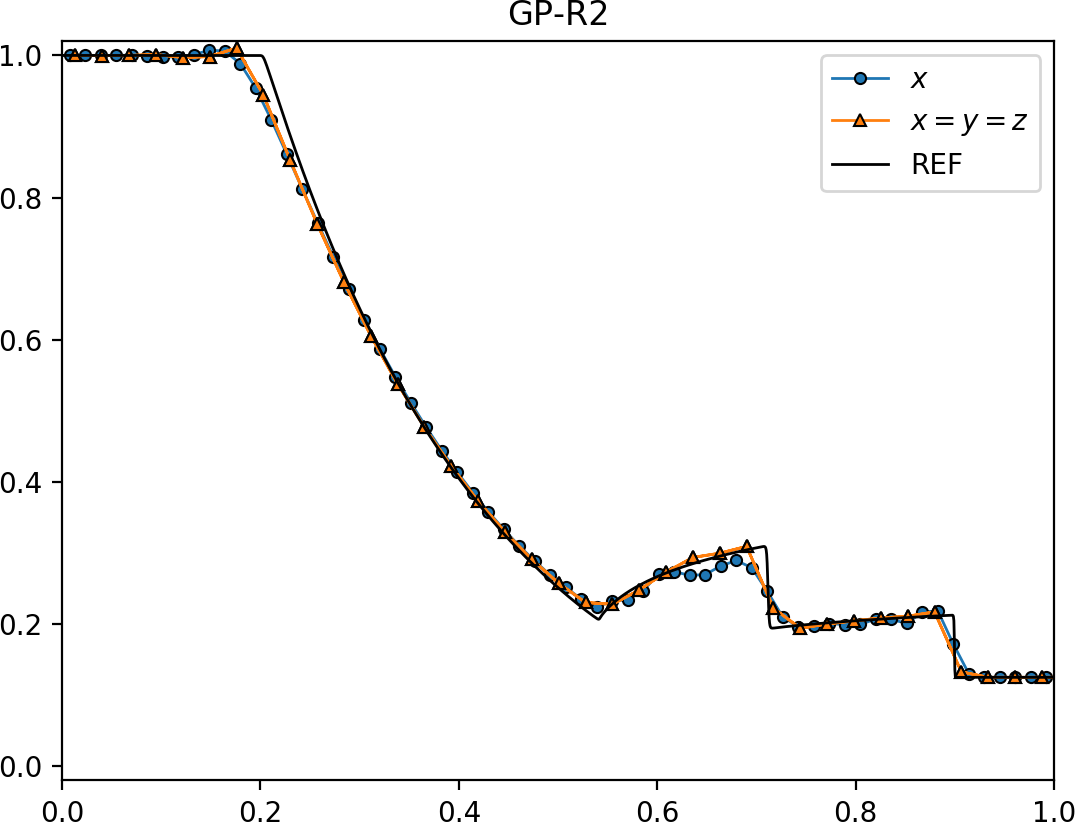}
    \label{subfig:exp-gp}
  }
  \subfloat[][]{
    \includegraphics[width=.45\textwidth]{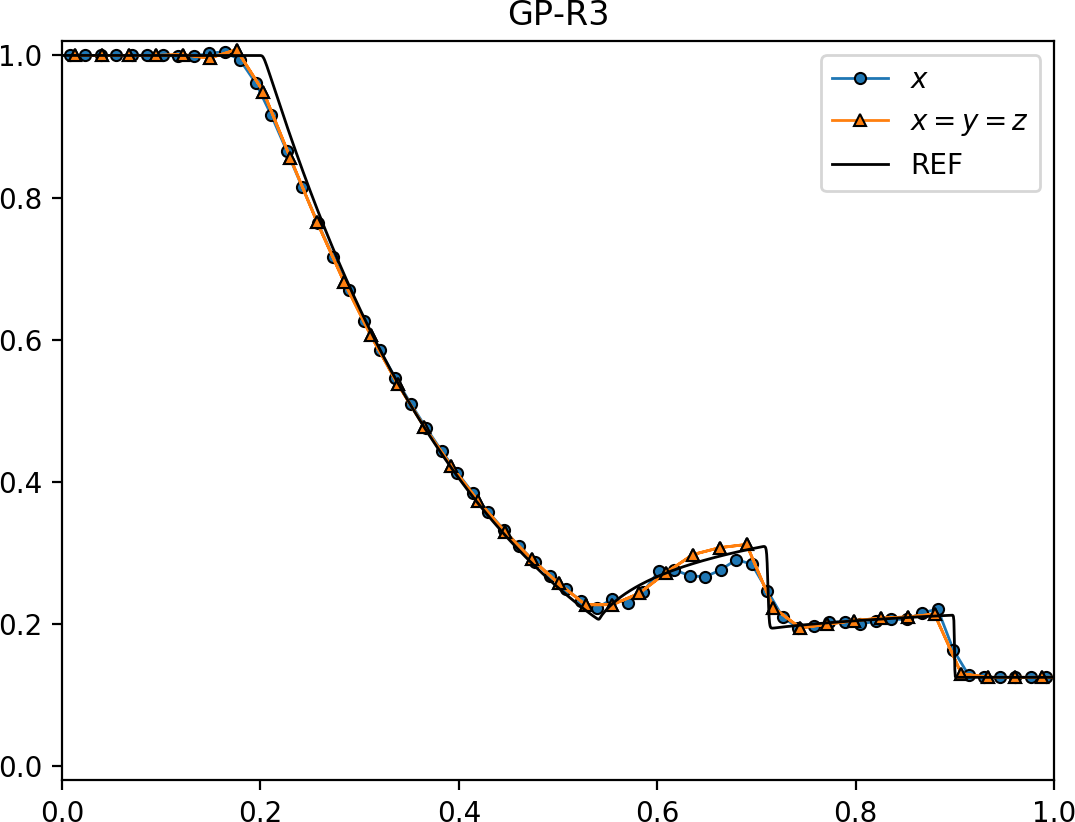}
    \label{subfig:exp-gpr3}
  }
  
  \centering
  \subfloat[][]{
    \includegraphics[width=.532 \textwidth]{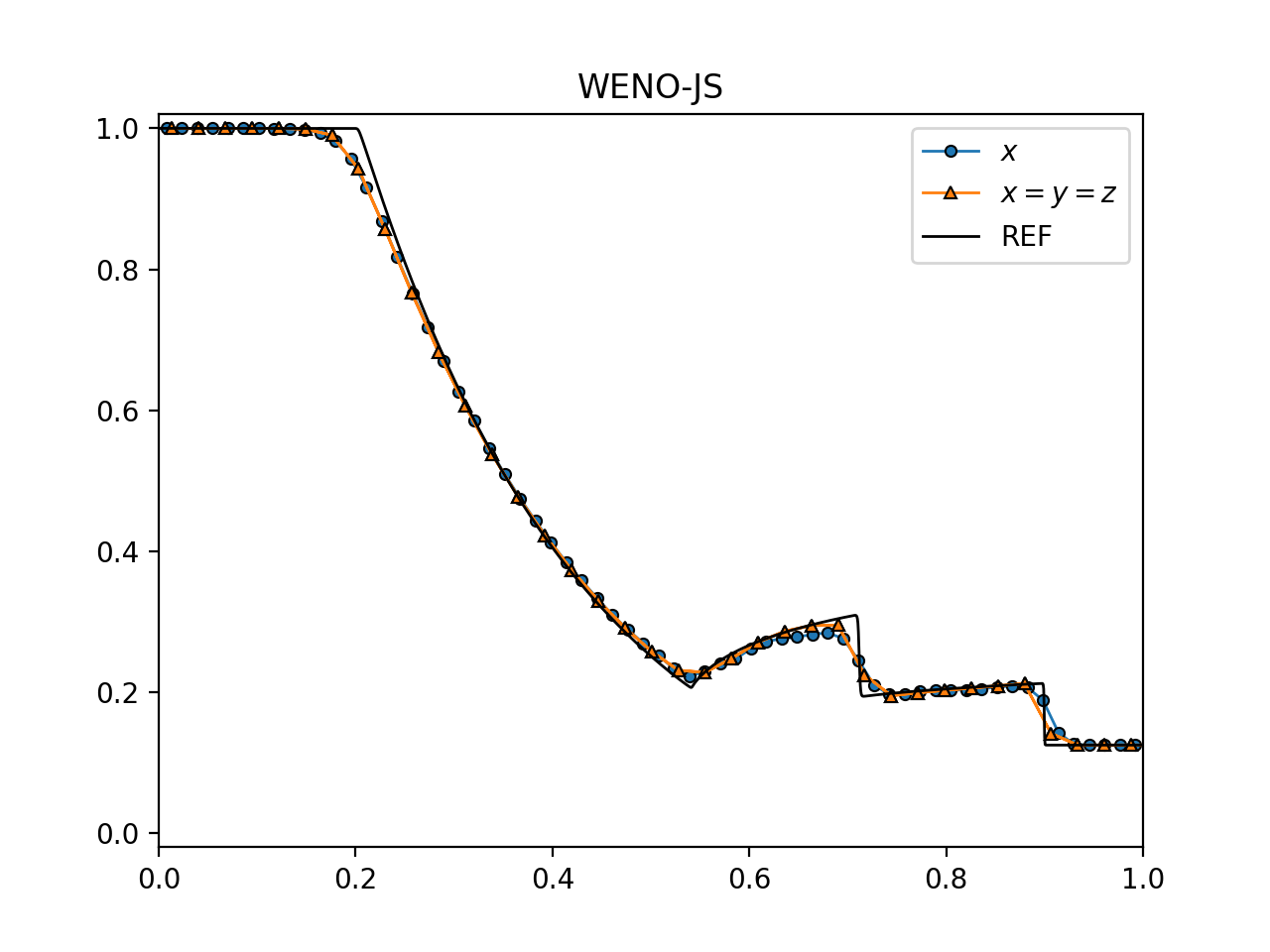}
    \label{subfig:exp-weno}
    }
  \caption{Radial density profiles along the $x$-axis and the diagonal
    $x=y=z$, for GP-R2, GP-R3 and WENO-JS on a $128\times128\times128$ grid, using a CFL
    of 0.3, HLLC, and RK3. GP simulations use $\ell/\Delta = 12$ and $\sigma/\Delta=3$.}
  \label{fig:explosion}
\end{figure}

  Fig.~\ref{fig:explosion} shows the resulting density profiles for
  the GP-R2 and WENO-JS schemes, along with a reference solution. The
  latter is obtained by considering the simplification
  of the spherically symmetric Euler equations as a 1D system with
  geometric source terms on 2056 grid points. Both methods
  produce well-acceptable solutions. We observe that 
  the GP-R2 solution produces some minor under- and over-shooting along
  the coordinate axes, which however did not affect the stability of the calculation.

\subsection{3D Riemann Problem}
\label{sec:3d-riemann-problem}

Finally, we consider the first configuration of the 3D Riemann problem
presented in \cite{balsara_three_2015}. The problem consists of eight
constant initial states in one octant of the $[-1,1]\times[-1,1]\times[-1,1]$
computational domain, with outflow boundary conditions. This initial condition
provides a set of 2D Riemann problems on each 
face of the domain, as well as on its diagonal planes.

\begin{figure}[h!]
  \centering
  \subfloat[]{
    \includegraphics[width=0.4\textwidth]{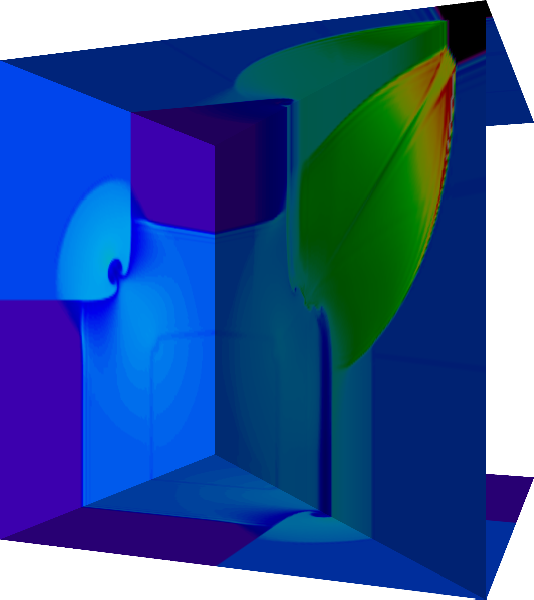}
    \label{subfig:3drp-gp-cube}
  }
  \subfloat[]{
    \includegraphics[width=0.54\textwidth]{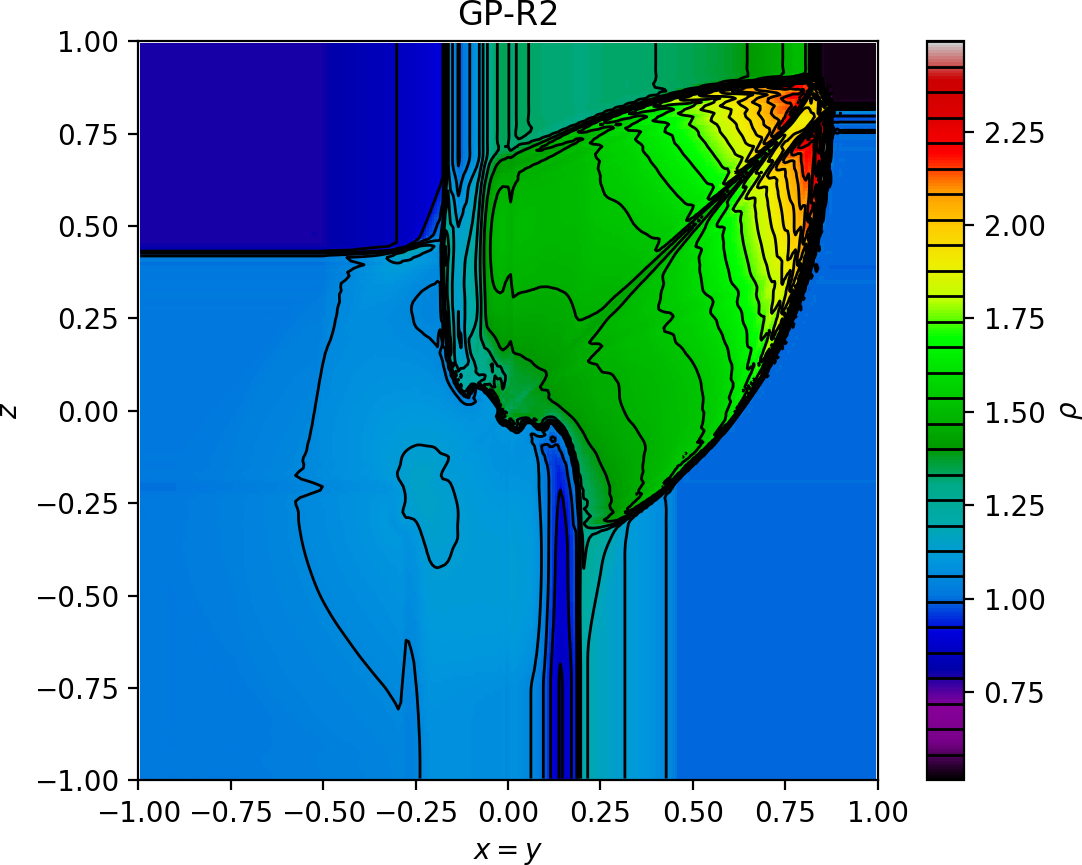}
    \label{subfig:3drp-gp-cut}
    }
  \caption{
  (a) 3D Riemann problem density profiles on the domain faces 
    and on the plane in the (1,1,0) direction using GP-R2. We use $\ell/\Delta=12$,
    $\sigma/\Delta=3$, a grid resolution of $200\times200\times200$, and a CFL of $0.3$. 
    (b) Density profile on the plane in the (1,1,0) direction. }

  \label{fig:3dRP-gp}
\end{figure}

\begin{figure}[h!]
  \centering
  \subfloat[]{
    \includegraphics[width=0.4\textwidth]{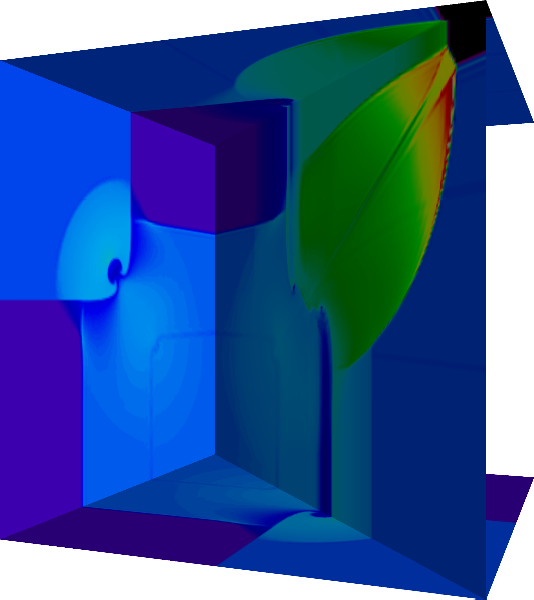}
    \label{subfig:3drp-gpr3-cube}
  }
  \subfloat[]{
    \includegraphics[width=0.54\textwidth]{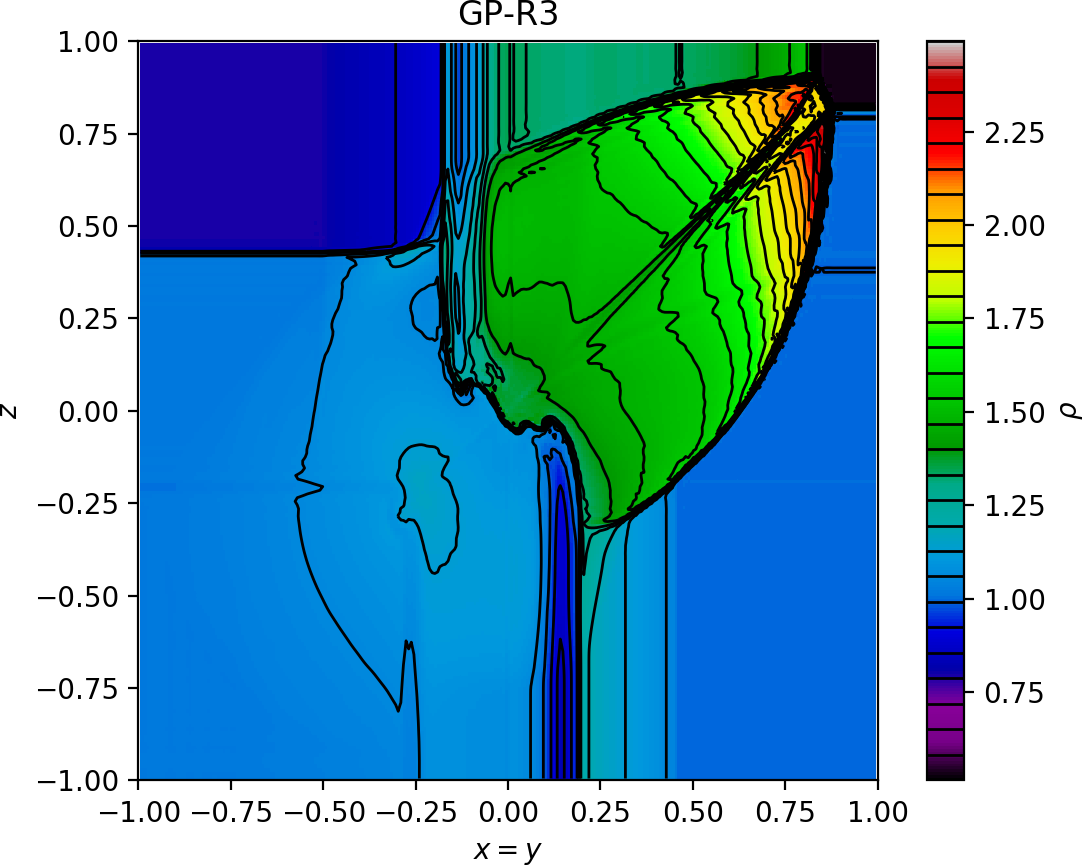}
    \label{subfig:3drp-gpr3-cut}
    }
  \caption{
  (a) Same as Fig.~\ref{subfig:3drp-gp-cube} but for GP-R3. 
    (b) Same as Fig.~\ref{subfig:3drp-gp-cut} but for GP-R3. }

  \label{fig:3dRP-gpr3}
\end{figure}

\begin{figure}[h!]
  \centering
  \subfloat[]{
    \includegraphics[width=0.4\textwidth]{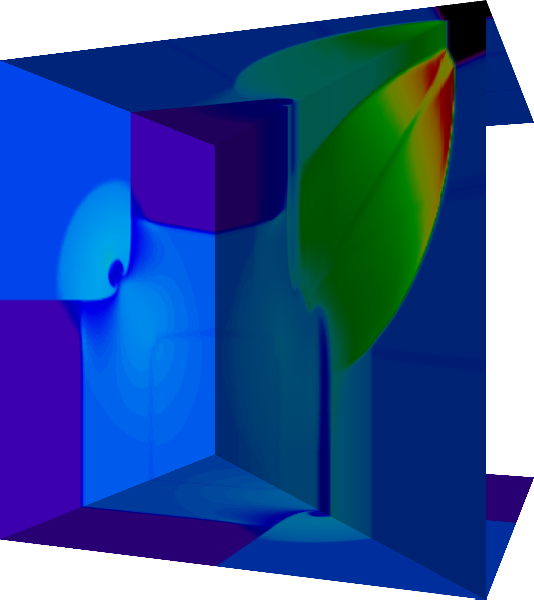}
    \label{subfig:3drp-w-cube}
  }
  \subfloat[]{
    \includegraphics[width=0.54\textwidth]{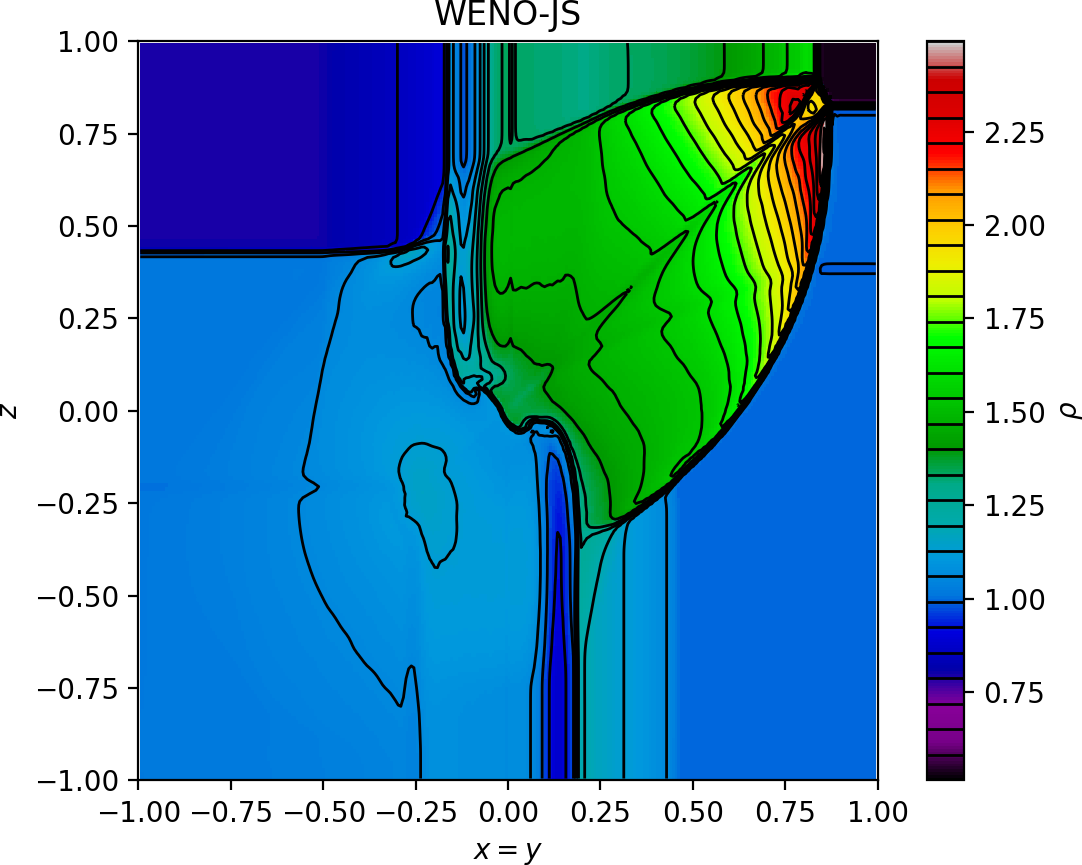}
    \label{subfig:3drp-w-cut}
    }
  \caption{
  (a) Same as Fig.~\ref{subfig:3drp-gp-cube} but for WENO-JS.
    (b) Same as Fig.~\ref{subfig:3drp-gp-cut} but for WENO-JS. }
  \label{fig:3dRP-weno}
\end{figure}

The results for the GP-R2, GP-R3, and WENO-JS schemes are
shown respectively in Figs.~\ref{fig:3dRP-gp}, \ref{fig:3dRP-gpr3}, and \ref{fig:3dRP-weno}.
In the left panels of the figures we discern three of the 2D Riemann problems, on the visible faces
of the domain and on the plane in the $(1,1,0)$ direction. The latter is also shown
in the right panels of figures, along with
contour lines. The GP methods are able to adequately capture 
the important features of the 2D Riemann problems to approximately the same extent.
Both GP solutions are better at resolving the contact discontinuity 
along the diagonal, when compared to the WENO-JS solution (see  right panels of
Figs.~\ref{fig:3dRP-gp}-\ref{fig:3dRP-weno}).
All calculations here employed the HLLC Riemann solver, RK3, and a CFL of 0.3.

\section{Conclusion}
\label{sec:conclusion}

In this paper we have extended the 1D GP-WENO FVM reconstruction from
\cite{reyes_new_2016} to operate as a high order interpolant for the
FD-Prim method for solving hyperbolic systems of full three-dimensional conservation
laws. 
To better capture shocks and discontinuities, we have refined
the GP-WENO smoothness indicators, based on the GP log-likelihood,
to use a separate length hyperparameter $\sigma$ from the actual GP
data interpolation.

The use of GP interpolation, along with GP-based smoothness indicators,
has shown fast rates of convergence in solution accuracy for smooth problems,
which can be variably controlled by the parameter $R$.
We demonstrated that the order of solution accuracy of the GP-WENO FD-Prim method 
varies linearly as $(2R+1)$ on smooth flows within one single algorithmic framework.
By being a polynomial-free method, GP-WENO does not require the formulation or implementation of
separate numerical algorithms as the order of accuracy varies.
This is not the case for most conventional polynomial-based approaches as one
needs to formulate a different piecewise polynomial algorithm for a given order,
introduce additional basis functions, and approach it in terms of the grid $\Delta$
and the local stencil size.
In this context, the GP-WENO formulation is more flexible and less
deterministic in providing a variable range of solution accuracies,
since the model is specified by its kernel function rather than explicit polynomial basis functions.
This novel, variable, high-order scheme allows for flexible algorithmic design
with minimal complexity and, at the same time, is able 
to captures shocks and discontinuities in a non-oscillatory, stable manner.

Moreover, our GP-based smoothness indicators are able to provide higher solution
accuracy relative to the standard WENO-based smoothness indicators with the same
weighting scheme. 
The solution accuracy can be tuned further using the kernel hyperparameter $\ell$,
which provides an additional knob for increased accuracy that is not available in traditional
polynomial-based methods.

Finally, the GP-based smoothness indicators can be implemented as a standalone numerical algorithm
to be integrated in any conventional polynomial-based WENO scheme, enhancing its solution accuracy.

\section{Acknowledgements}
This work was supported in part by
the National Science Foundation under grant AST-0909132,
in part by the U.S. DOE NNSA ASC through the Argonne
Institute for Computing in Science under field work proposal
57789, and in part by the Hellman Fellowship Program at UC Santa Cruz.
We acknowledge support from the U.S. DOE Office of Science under grant
DE-SC0016566 and the National Science Foundation under grant PHY-1619573.
The authors also thank anonymous referees for their helpful suggestions and comments
on our manuscript, which led us to improve the paper during the review process.

\appendix
\section{Adaptive Order (AO) GP}
\label{sec:append-adapt-order}

In this Appendix, we provide a hybrid scheme for handling discontinuities
in a non-oscillatory fashion, while keeping the desired high-order of accuracy away from shocks and
discontinuities. The approach we describe here is called the adaptive order (AO) approach
recently proposed by Balsara \textit{et al.} \cite{balsara_efficient_2016}.
%
The main idea behind the adaptive order approach is to nonlinearly
hybridize a lower-order interpolation that favors stability over
accuracy with a higher-order interpolation
on a large stencil. The hybridization is performed in such a way that
it biases towards the stable lower-order interpolation in the presence of
discontinuities, while the higher-order one is preferred in smooth
regions.

To begin, we start with a high-order GP interpolation of a Riemann
state $q^R_*$, given by Eq.~(\ref{eq:q-intrp}) on a stencil $S_R$ of radius
$R>1$. The stencil $S_R$ also comes with a smoothness indicator
$\beta^R$,  given by the log-likelihood in Eq.~(\ref{eq:log-like}). 
It is important to note that, 
for the AO formulation, we will be comparing stencils of different size. Thus, 
unlike the GP-WENO formulation discussed in Section~\ref{sec:gp-weno-interp}, 
it will be necessary to
retain the normalization and complexity penalty terms (i.e., the first and the second terms) 
that are in
Eq.~(\ref{eq:log-like}). We aim to nonlinearly hybridize the
high-order interpolation $q_*^R$, with a lower-order, stable
interpolation which we will call $q_*^C$.

We now describe how to obtain the lower-order stable
interpolation $q_*^C$. The choice of linear weights given by
Eq.~(\ref{eq:gammas}) or in the original WENO-JS scheme
\cite{jiang1996efficient} are designed to be \textit{optimal} only in
the sense of accuracy. Other types of WENO schemes can be formed that
prefer stability over accuracy, such as the central WENO (CWENO) schemes
\cite{cravero2016accuracy,semplice2016adaptive,levy2000compact}. Following
the AO approach of Balsara \textit{et al.} \cite{balsara_efficient_2016}, we adopt the
same set of three, three-point stencils of the 5th order WENO-JS
scheme \cite{jiang1996efficient}, the CWENO scheme
\cite{levy2000compact}, and that of the GP-WENO scheme with $R=2$ (or GP-R2) as our
candidate stencils,
\begin{equation}
  \label{eq:cweno_stencil}
  \begin{array}{lcr}
       S_1 = \{I_{i-2}, I_{i-1}, I_{i} \}, 
    & S_2 = \{I_{i-1}, I_{i},I_{i+1} \}, 
    & S_3 = \{I_{i}, I_{i+1}, I_{i+2} \}.
  \end{array}
\end{equation}
From each $m$-th stencil $S_m$, $m=1,2,3$, 
we obtain an interpolation of the Riemann states $q^m_*$ at the
point $x_*$, given by Eq.~(\ref{eq:candidate_interp}), as well as
smoothness indicators $\beta_m$, given by the GP log-likelihood
(Eq.~(\ref{eq:log-like})). As in the GP-WENO method, we form our GP-CWENO
interpolation as the convex combination of the $q_*^m$'s,
\begin{equation}
  \label{eq:CWENO}
  q_*^{C} = \sum_{m=1}^{3}\omega_mq_*^m,
\end{equation}
where the nonlinear weights (i.e., $\tilde{\omega}^R$ and $\tilde{\omega}_m$) 
as well as the associated normalized weights (i.e., ${\omega}^R$ and ${\omega}_m$) 
are given by
%
\begin{equation}
  \label{eq:nonlin-AO}
    \scalemath{1.2}{	
  \begin{array}{lr}
    \omega^R = \frac{\tilde{\omega}^R}{\tilde{\omega}^R+\sum_{m=1}^3\tilde{\omega}_m},
   & \tilde{\omega}^R = \frac{\gamma_\text{HI}}{(\epsilon + \beta^R)^p},
    \\ & \\
    \omega_m = \frac{\tilde{\omega}_m}{\tilde{\omega}^R+\sum_{m=1}^3\tilde{\omega}_m},
   & \tilde{\omega}_m = \frac{\gamma_\text{HI}}{(\epsilon + \beta_m)^p}.
  \end{array}
  }
\end{equation}

The linear weights $\gamma_m$ are
chosen to be
\begin{equation}
  \label{eq:CWENO-lin}
  \begin{array}{lr}
    \gamma_1=\gamma_3 = \inv{2}(1-\gamma_\text{HI})(1-\gamma_\text{LO}), &
                                                                     \gamma_2
                                                                     = (1-\gamma_\text{HI})\gamma_\text{LO}.
  \end{array}
\end{equation}
Here $\gamma_\text{HI}$ and $\gamma_\text{LO}$ are constants,
$\gamma_\text{HI},\gamma_\text{LO}\in[0.85,0.95]$ and are typically
chosen as $\gamma_\text{HI}=\gamma_\text{LO}=0.85$. Further, $\gamma_\text{LO}$
serves to bias the CWENO interpolation towards the centered 3-point
third-order stencil $S_2$, while $\gamma_\text{HI}$ serves to bias the
GP-AO interpolation towards the high-order stencil, as will be made
apparent shortly.

We may now hybridize the high-order interpolation $q_*^R$ with the
3rd-order stable $q_*^C$ to form the GP-AO interpolation
$q_*^{\text{AO}(R,3)}$ as,
\begin{equation}
  \label{eq:GP-AO}
  q_*^{\text{AO}(R,3)} = \frac{\omega^R}{\gamma_{\text{HI}}}q_*^R  +
  \sum_{m=1}^3\left(\omega_m - \frac{\omega^R}{\gamma_\text{HI}}\gamma_m\right)q_*^m .
\end{equation}

It becomes now apparent how the AO scheme operates: In the absence of
discontinuities on the high-order stencil $S_R$, all the
smoothness indicators will be roughly equal, i.e., $\beta^R \approx \beta_m$
for all $m=1,2,3$. In this case 
$\frac{\omega^R}{\gamma_\text{HI}} \approx 1$ 
and $\omega_m\approx\gamma_m$, which makes the term in the sum of
Eq.~(\ref{eq:GP-AO}) effectively zero so that the interpolation is
originates entirely from $q^R_*$, i.e., the high-order interpolation.
On the other hand, if a discontinuity is present on $S_R$ but, quite possibly,
absent on at least one of $S_m$, we get $\frac{\omega^R}{\gamma_\text{HI}} \approx 0$
so that the GP-AO interpolation in Eq.~(\ref{eq:GP-AO})
reduces to $q^C_*$ in Eq.~(\ref{eq:CWENO}),
which is the 3rd-order GP-CWENO interpolation.

Finally, we complete the discussion of the GP-AO with a description of the
recursive hybridization strategy, also presented in
\cite{balsara_efficient_2016}. In the current context we wish to
hybridize two GP-AO interpolations from Eq.~(\ref{eq:GP-AO}) of
different stencil radii, say $\text{GP-AO}(R,3)$ and
$\text{GP-AO}(R',3)$, to produce the GP-AO scheme which we denote as
$\text{GP-AO}(R',R,3)$. For example, taking $R'=3$ and $R=2$ would be similar to
the $\text{AO}(7,5,3)$ scheme presented in
\cite{balsara_efficient_2016}.

The recursive hybridization $\text{GP-AO}(R',R,3)$ is formed by
defining the nonlinear recursive weights $\nu^{R'}$ and $\nu^R$,
\begin{equation}
  \label{eq:recursive-weights}
  \scalemath{1.25}{
    \begin{array}{lr}
      \tilde{\nu}^{R'} =
      \frac{\gamma_\text{HI}}{(\epsilon+\beta^{R'})^p}, 
      & \tilde{\nu}^R = \frac{1-\gamma_{HI}}{(\epsilon+\beta^R)^p},
      \\
      \nu^{R'} = \frac{\tilde{\nu}^{R'}}{\tilde{\nu}^{R'}+\tilde{\nu}^R}, 
      & \nu^{R} = \frac{\tilde{\nu}^{R}}{\tilde{\nu}^{R'}+\tilde{\nu}^R}.
    \end{array}
  }
\end{equation}
The $\text{GP-AO}(R',R,3)$ interpolation $q_*^{\text{AO}(R',R,3)}$
is then
\begin{equation}
  \label{eq:AO-recursive}
  q_*^{\text{AO}(R',R,3)} =
  \frac{\nu^{R'}}{\gamma_{\text{HI}}}q_*^{\text{AO}(R',3)} + 
  \left(\nu^R - \frac{\nu^{R'}}{\gamma_\text{HI}}(1-\gamma_\text{HI}) \right)
  q_*^{\text{AO}(R,3)}.
\end{equation}
Note the similarity of equations (\ref{eq:AO-recursive}) and
(\ref{eq:GP-AO}). The factor of $\frac{\nu^{R'}}{\gamma_{\text{HI}}}$
acts exactly as $\frac{\omega^{R}}{\gamma_{\text{HI}}}$, effectively
being the switch between the higher ($S_{R'}$) and lower ($S_R$)
order stencils. In the case that there are no discontinuities on the
stencil $S_{R'}$ and $S_R$, both $q_*^{\text{AO}(R',3)}$ and
$q_*^{\text{AO}(R,3)}$ revert to the higher-order approximations
  $q^{R'}_*$ and $q^R_*$, respectively, and
  $\frac{\nu^{R'}}{\gamma_{\text{HI}}}\approx 1$ making the right hand
  side of Eq.~(\ref{eq:AO-recursive}) equal to
  $\frac{\nu^{R'}}{\gamma_{\text{HI}}}$. In the event that there is a
  discontinuity on $S_{R'}$ but not on $S_R$, we have
  $\frac{\nu^{R'}}{\gamma_{\text{HI}}}\approx 0$, which reduces $q_*^{\text{AO}(R',R,3)}$ 
  in Eq.~(\ref{eq:AO-recursive}) to $q_*^{\text{AO}(R,3)}$. 
  When there is a discontinuity on both
  $S_{R'}$ and $S_R$, both $q_*^{\text{AO}(R',3)}$ and
  $q_*^{\text{AO}(R,3)}$ will reduce to the centrally stable
  third-order interpolation as before.

\label{sec:references}

\bibliographystyle{model1-num-names}
\bibliography{mybibfile_cleaned}

\end{document}